\RequirePackage{fix-cm}
\documentclass[natbib,smallextended]{svjour3}       
\smartqed  
\usepackage{graphicx}
\usepackage{amsmath}
\usepackage{amssymb}
\usepackage{color}
\usepackage[citecolor=blue,urlcolor=blue,breaklinks]{hyperref}
\newcommand{\be}{\begin{equation}}
\newcommand{\ee}{\end{equation}}
\journalname{Living Rev. Relativ.}
\newcommand       \bea          {\begin{eqnarray}}
\newcommand       \eea          {\end{eqnarray}}

\newcommand       \apj          {ApJ}
\newcommand       \apjl         {ApJL}
\newcommand       \aap          {A\&A}
\newcommand       \nat          {Nature}
\newcommand       \mnras        {MNRAS}

\newcommand       \prl      {PRL}

\newcommand       \prd      {Phys.~Rev.~D.~}

\newcommand       \araa      {ARA\&A}

\newcommand       \pasj   {PASJ}

\newcommand      \physrep {Physics Reports}

\newcommand       \apss          {APSS}

\newcommand      \prc {PRC}

\newcommand      \pasa {PASA}

\newcommand      \LIGO {GW170817}

\begin{document}

\title{Kilonovae}

\author{Brian D.~Metzger}

\institute{B.~D.\ Metzger%
\at
Department of Physics\\
Columbia Astrophysics Laboratory\\
Columbia University\\
New York, NY 10027, USA\\
\email{bdm2129@columbia.edu}
}

\date{Received: date / Accepted: date}

\maketitle

\begin{abstract}
The coalescence of double neutron star (NS-NS) and black hole (BH)-NS binaries are prime sources of gravitational waves (GW) for Advanced LIGO/Virgo and future ground-based detectors.  Neutron-rich matter released from such events undergo rapid neutron capture ($r$-process) nucleosynthesis as it decompresses into space, enriching our universe with rare heavy elements like gold and platinum.  Radioactive decay of these unstable nuclei powers a rapidly evolving, approximately isotropic thermal transient known as a ``kilonova', which probes the physical conditions during the merger and its aftermath.  Here I review the history and physics of kilonovae, leading to the current paradigm of day-timescale emission at optical wavelengths from lanthanide-free components of the ejecta, followed by week-long emission with a spectral peak in the near-infrared (NIR).  These theoretical predictions, as compiled in the original version of this review, were largely confirmed by the transient optical/NIR counterpart discovered to the first NS-NS merger, GW170817, discovered by LIGO/Virgo.  Using a simple light curve model to illustrate the essential physical processes and their application to GW170817, I then introduce important variations about the standard picture which may be observable in future mergers.  These include $\sim$hour-long UV precursor emission, powered by the decay of free neutrons in the outermost ejecta layers or shock-heating of the ejecta by a delayed ultra-relativistic outflow; and enhancement of the luminosity from a long-lived central engine, such as an accreting BH or millisecond magnetar.  Joint GW and kilonova observations of GW170817 and future events provide a new avenue to constrain the astrophysical origin of the $r$-process elements and the equation of state of dense nuclear matter.

\keywords{gravitational waves, neutron stars, nucleosynthesis, black holes, radiative transfer}
\end{abstract}

\newpage

\setcounter{tocdepth}{3}
\tableofcontents


\section{Electromagnetic Counterparts of Binary Neutron Star Mergers}
\label{sec:intro}

The discovery in 2015 of gravitational waves (GW) from the inspiral and coalescence of binary black holes (BH) by the Laser Interferometer Gravitational Wave Observatory (LIGO) and later with its partner observatory, Virgo, has opened an entirely new window on the cosmos \citep{LIGO+16}.  This modest, but rapidly-expanding, sample\footnote{\url{https://dcc.ligo.org/LIGO-P1800307/public}} of BH-BH merger events \citep{LIGO+18CATALOG} is already being used to place constraints on the formation channels of compact binary systems (e.g.~\citealt{LIGO+16CHANNELS}), as well as fundamental tests of general relativity in the previously inaccessible strong-field regime (e.g.~\citealt{Miller16,LIGO+19GR}).   We are fortunate witnesses to the birth of a new field of research: Gravitational Wave Astronomy.  

On August 17, 2017, near the end of their second observing run, Advanced LIGO/Virgo detected its first merger of a double NS binary \citep{LIGO+17DISCOVERY}.  This event, like other GW detections, was dubbed GW170817 based on its date of discovery.  The individual masses of the binary components measured by the GW signal, $M_{1}, M_{2} \approx 1.16-1.60M_{\odot}$ (under the assumption of low NS spin) and the precisely-measured chirp mass 
\be \mathcal{M}_c \equiv \frac{(M_1M_2)^{3/5}}{(M_1+M_2)^{1/5}} \underset{\rm GW170817}\simeq 1.118M_{\odot}, \label{eq:Mchirp} \ee
are fully consistent with being drawn from the known population of Galactic binary NSs \citep{LIGO+19PARAMS}, particularly those with GW merger times less than the age of the universe \citep{Zhao&Lattimer18}.  The lack of evidence for tidal interaction between the merging objects in the inspiral GW signal allowed for stringent upper limits to be placed on the tidal deformability and radii of NSs \citep{LIGO+18EOS,DeSoumi+18}, properties closely related to the pressure of neutron-rich matter above nuclear saturation density (see \citealt{Horowitz+18} for a recent review).  

Beyond information encoded in the GW strain data, the discovery of electromagnetic (EM) emission accompanying the GW chirp has the potential to reveal a much richer picture of these events \citep{Bloom+09GW}.  By identifying the host galaxies of the merging systems, and their precise locations within or around their hosts, we obtain valuable information on the binary formation channels, age of the stellar population, evidence for dynamical formation channels in dense stellar systems, or displacement due to supernova (SN) birth kicks, in a manner similar to techniques long applied to gamma-ray bursts (GRBs) and supernovae (e.g.~\citealt{Fruchter+06, Fong&Berger13}).  From the host galaxy redshifts, we obtain independent distance estimates to the sources, thus reducing degeneracies in the GW parameter estimation, especially of the binary inclination with respect to the line of sight (e.g.~\citealt{Cantiello+18,Chen+18}). 
Redshift measurements also enable the use of GW events as standard rulers to measure the Hubble constant, or more generally, probe the cosmic expansion history \citep{Schutz86,Holz&Hughes05, Nissanke+13}.    Remarkably, all of these these opportunities, and many others to be discussed later in this review, became reality with GW170817.

Except in rare circumstances, the mergers of stellar-mass BH-BH binaries are not expected to produce luminous EM emission due to the absence of significant matter surrounding these systems at the time of coalescence.  Fruitful synthesis of the GW and EM skies will therefore most likely first be achieved from NS-NS and BH-NS mergers.  Given the discovery of a single NS-NS merger in the O1 and O2 observing runs, LIGO/Virgo infer a volumetric rate of $110-3840$ Gpc$^{-3}$ yr$^{-1}$ \citep{LIGO+18CATALOG}, corresponding to an expected NS-NS rate of $\approx  6-120$ yr$^{-1}$ once Advanced LIGO/Virgo reach design sensitivity by the early 2020s \citep{LIGO+17FERMI}.  The O1/O2 upper limit on the NS-BH merger rate is $\lesssim 600$ Gpc$^{-3}$ yr$^{-1}$ \citep{LIGO+18CATALOG}.  This range is broadly consistent with theoretical expectations on the rates (e.g.~population synthesis models of field binaries; e.g.~\citealt{Dominik+14}) as well as those derived empirically from the known population of Galactic double NS systems (\citealt{Phinney91,Kalogera+04,Kim+15}; see \citealt{Abadie+10} for a review of rate predictions).

Among the greatest challenge to the joint EM/GW endeavor are the large uncertainties in the measured sky positions of the GW sources, which are primarily determined by triangulating the GW arrival times with an array of interferometers.  When a detection is made by just the two North American LIGO facilities, sky error regions are very large (e.g.~$\approx 850\mathrm{\ deg}^{2}$ for the BH-BH merger GW150914, though later improved to $\approx 250 \mathrm{\ deg}^{2}$; \citealt{Abbott+16, Abbott+16PRX}).  However, with the addition of the Virgo detector in Italy, and eventually KAGRA in Japan \citep{KAGRA} and LIGO-India, these can be reduced to more manageable values of $\sim$ 10\,--\,100 deg$^{2}$ or less \citep{Fairhurst11,Nissanke+13,Rodriguez+14}.  Indeed, information from Virgo proved crucial in reducing the sky error region of GW170817 to 30 deg$^{2}$ \citep{LIGO+17CAPSTONE}, greatly facilitating the discovery of its optical counterpart.  Nevertheless, even in the best of cases, the GW-measured sky areas still greatly exceed that covered in a single pointing by most radio, optical, and X-ray telescopes, especially those with the required sensitivity to detect the potentially dim EM counterparts of NS-NS and BH-NS mergers \citep{Metzger&Berger12}.     

\begin{figure}[!t]
\includegraphics[width=1.0\textwidth]{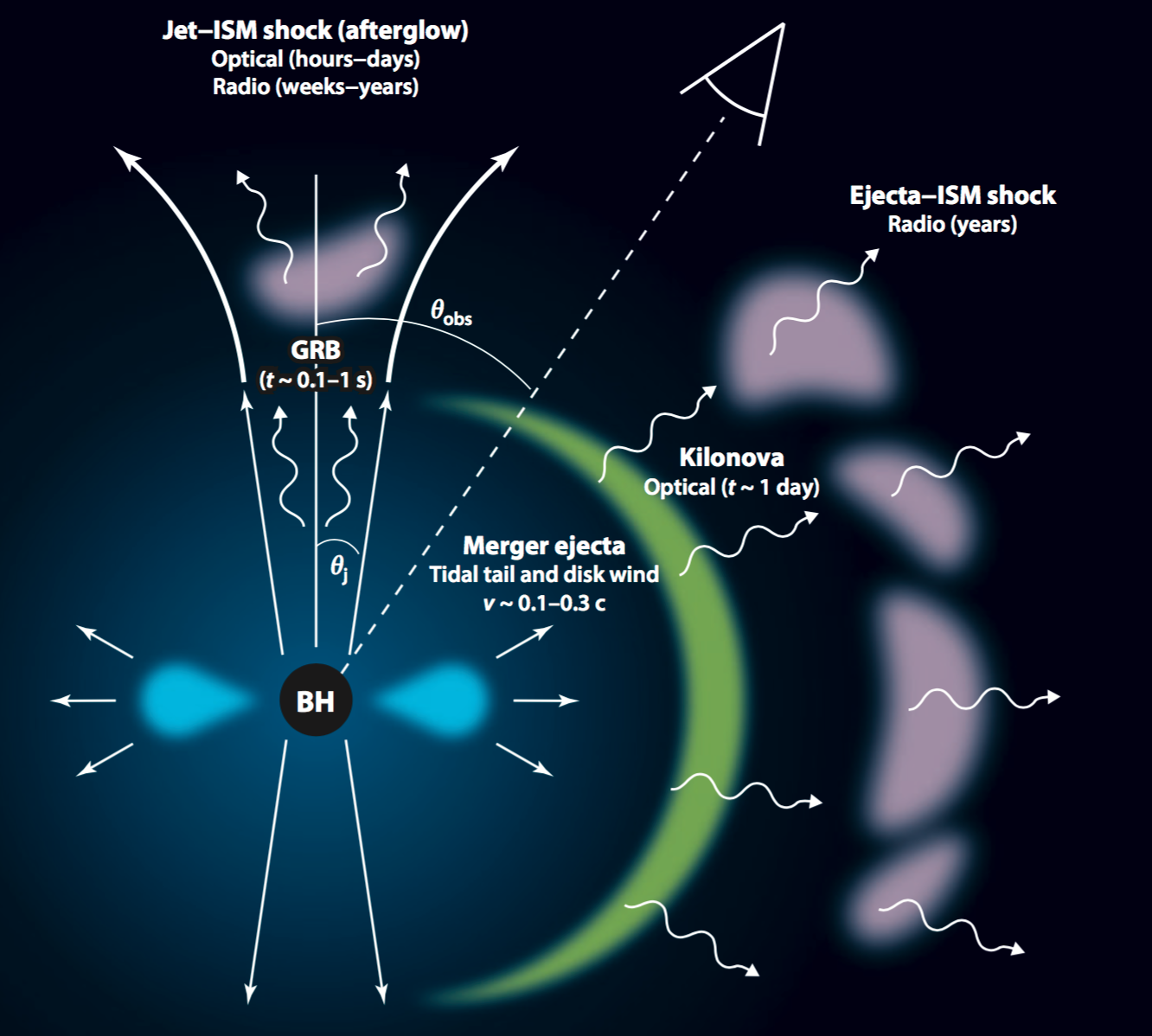}
\caption{Summary of the electromagnetic counterparts of NS-NS and BH-NS mergers and their dependence on the viewing angle with respect to the axis of the GRB jet.  The kilonova, in contrast to the GRB and its afterglow, is relatively isotropic and thus represents the most promising counterpart for the majority of GW-detected mergers.  Modified from an original version from \citep{Metzger&Berger12}.  Reproduced with permission from \citep{Berger14}, copyright of the author. }
\label{fig:MB12}
\end{figure}

Figure \ref{fig:MB12} summarizes the EM counterparts of NS-NS and BH-NS mergers as a function of the observer viewing angle relative to the binary axis.  Multiple lines of evidence, both observational (e.g.~\citealt{Fong+13}) and theoretical\footnote{One argument linking short GRBs to NS-NS/BH-NS mergers is the lack of viable alternative models.  Accretion-induced collapse of a NS to form a BH was once considered an alternative short GRB model \citep{MacFadyen+05,Dermer&Atoyan06}.  However,  \citet{Margalit+15} showed that even a maximally-spinning NS rotating as a solid body (the expected configuration in such aged systems) will not produce upon collapse a sufficiently massive accretion disk around the newly-formed BH given extant constraints on NS properties (see also \citealt{Shibata03,Camelio+18}).} \citep{Eichler+89,Narayan+92}, support an association between NS-NS/BH-NS mergers and the ``short duration'' class of GRBs.  The latter are those bursts with durations in the gamma-ray band less than about 2 seconds \citep{Nakar07,Berger14}, in contrast to the longer lasting bursts of duration $\gtrsim$ 2 seconds which are instead associated with the core collapse of very massive stars.  For a typical LIGO/Virgo source distance of $\lesssim 200$ Mpc, any gamma-ray transient with properties matching those of the well-characterized cosmological population of GRBs would easily be detected by the \textit{Fermi}, \textit{Swift} or {\it Integral} satellites within their fields of view, or even with the less sensitive but all-sky Interplanetary Network of gamma-ray telescopes \citep{Hurley10}.  

The tightly collimated, relativistic outflows responsible for short GRBs are commonly believed to be powered by the accretion of a massive remnant disk onto the compact BH or NS remnant following the merger (e.g.~\citealt{Narayan+92}).  This is expected to occur within seconds of the merger, making their temporal association with the termination of the GW chirp unambiguous (the gamma-ray sky is otherwise quiet).  Once a GRB is detected, its associated afterglow can in many cases be identified by promptly slewing a sensitive X-ray telescope to the location of the burst.  This exercise is now routine with \textit{Swift}, but may become less so in the future without a suitable replacement mission.  Although gamma-ray detectors themselves typically provide poor sky localization, the higher angular resolution of the X-ray telescope allows for the discovery of the optical or radio afterglow; this in turn provides an even more precise position, which can help to identify the host galaxy.  

A prompt burst of gamma-ray emission was detected from GW170817 by the {\it Fermi} and {\it Integral} satellites with a delay of $\approx 1.7$ seconds from the end of the inspiral \citep{LIGO+17FERMI,Goldstein+17,Savchenko+17}.  However, rapid localization of the event was not possible, for two reasons: (1) the merger was outside the field-of-view of the {\it Swift} BAT gamma-ray detector and therefore a relatively precise sky position was not immediately available; (2) even if rapidly slewing of the X-ray telescope had been made, the X-ray afterglow may not have been detectable at such early times.  Deep upper limits on the X-ray luminosity of GW170817 at $t = 2.3$ days \citep{Margutti+17} reveal a much dimmer event than expected for a cosmological GRB placed at the same distance at a similar epoch.  As we discuss below, the delayed rise and low luminosity of the synchrotron afterglow were the result of our viewing angle being far outside the core of the ultra-relativistic GRB jet, unlike the nearly on-axis orientation from which cosmological GRBs are typically viewed (e.g.~\citealt{Ryan+15}).

Although short GRBs are probably the cleanest EM counterparts, their measured rate within the Advanced LIGO detection volume, based on observations prior to GW170817, was expected to low, probably less than once per year to decade \citep{Metzger&Berger12}.  The measured volumetric rate of short GRBs in the local universe of $\mathcal{R}_{\rm SGRB} \sim 5$ Gpc$^{-3}$ yr$^{-1}$ \citep{Wanderman&Piran15} can be reconciled with the much higher NS-NS merger rate $\mathcal{R}_{\rm BNS} \sim 10^{3}$ Gpc$^{-3}$ yr$^{-1}$ \citep{LIGO+18CATALOG} if the gamma-ray emission is beamed into a narrow solid angle $\ll 4\pi$ by the bulk relativistic motion of the GRB jet \citep{Fong+15,Troja+16}.  Given a typical GRB jet opening angle of $\theta_{\rm jet} \approx 0.1$ radians, the resulting beaming fraction of $f_{\rm b}^{-1} = \theta_{\rm jet}^{2}/2 \sim 1/200 \sim \mathcal{R}_{\rm SGRB}/\mathcal{R}_{\rm BNS}$, is consistent with most or all short GRBs arising from NS-NS mergers (though uncertainties remain large enough that a contribution from other channels, such as BH-NS mergers, cannot yet be excluded). 

While the discovery of gamma-rays from the first GW-detected NS-NS merger came as a surprise to most, its properties were also highly unusual.  The isotropic luminosity of the burst was $\sim 10^{3}$ times smaller than the known population of cosmological short GRBs on which prior rate estimates had been based.  A similar burst would have been challenging to detect at even twice the distance of GW170817.  Several different theoretical models were proposed to explain the origin of gamma-ray signal from GW170817 (e.g.~\citealt{Granot+17,Fraija+17,Gottlieb+18b,Beloborodov+18}), but in all cases its low luminosity and  unusual spectral properties are related to our large viewing angle $\theta_{\rm obs} \approx 0.4$ relative to the binary angular momentum (\citealt{Finstad+18,LIGO+19PARAMS}) being $\gtrsim 4$ times larger than the opening angle of the jet core, $\theta_{\rm jet} \lesssim 0.1$ (e.g.~\citealt{Fong+17,Mooley+18,Beniamini+19}).  Given that the majority of future GW-detections will occur at greater distances and from even larger $\theta_{\rm obs}$ (for which the prompt jet emission is likely to be less luminous) than for GW170817, it remains true that only a fraction of NS-NS mergers detected are likely to be accompanied by detectable gamma-rays (e.g.~\citealt{Mandhai+18,Howell+19}).  Nevertheless, given the unique information encoded in the prompt gamma-ray emission when available (e.g.~on the properties of the earliest ejecta and the timing of jet formation relative to binary coalescence), every effort should be made to guarantee the presence of a wide-field gamma-ray telescopes in space throughout the next decades.

For the majority of GW-detected mergers viewed at $\theta_{\rm obs} \gg \theta_{\rm jet}$, the most luminous GRB emission will be beamed away from our line of sight by the bulk relativistic motion of the emitting jet material.  However, as the relativistic jet slows down by colliding with and shocking the interstellar medium, even off-axis viewers enter the causal emission region of the synchrotron afterglow (e.g.~\citealt{Totani&Panaitescu02}).  Such a delayed off-axis non-thermal afterglow emission was observed from GW170817 at X-ray (e.g.~\citealt{Troja+17,Margutti+17,Haggard+17}), radio (e.g.~\citealt{Hallinan+17,Alexander+17}), and optical frequencies (after the kilonova faded; e.g.~\citealt{Lyman+18}).  The afterglow light curve, which showed a gradual rise to a peak at $t \sim 200$ days followed by extremely rapid fading, reveals details on the angular structure of the jet (e.g.~\citealt{Perna+03,Lamb&Kobayashi17,Lazzati+18,Xie&MacFadyen19}).  The latter structured may have been imprinted by the relativistic GRB jet as it pierced through the merger ejecta \citep{Nakar&Piran16,Lazzati+17}.  We return later to an additional possible signature of the ejecta being shock-heated by the jet on the early-time kilonova emission.

In addition to the beamed GRB and its afterglow, the merger of NS-NS and BH-NS binaries are also accompanied by a more isotropic counterpart, commonly known as a `kilonova' (or, less commonly, `macronova').  Kilonovae are thermal supernova-like transients lasting days to weeks, which are powered by the radioactive decay of heavy neutron-rich elements synthesized in the expanding merger ejecta \citep{Li&Paczynski98}.  They provide both a robust EM counterpart to the GW chirp, which is expected to accompany a fraction of BH-NS mergers and essentially all NS-NS mergers, as well as a direct probe of the unknown astrophysical origin of the heaviest elements \citep{Metzger+10}.  

This article provides a pedagogical review of kilonovae, including a brief historical background and recent developments in this rapidly evolving field (Sect.~\ref{sec:history}).  Sect.~\ref{sec:basics} describes the basic physical ingredients, including the key input from numerical relativity simulations of the merger and its aftermath.  For pedagogical reasons, the discussion is organized around a simple toy model for the kilonova light curve (Sect.~\ref{sec:model}), which synthesizes most of the key ingredients in a common and easy-to-interpret framework.  My goal is to make the basic results accessible to anyone with the ability to solve a set of coupled ordinary differential equations.  

I begin by introducing the simplest model of lanthanide-rich ejecta heated by radioactivity, which produces a week-long near-infrared (NIR) transient (`red kilonova'; Sect.~\ref{sec:red}) which is proceeded in at least some cases by $\sim$ day-long UV/optical-wavelength emission (`blue kilonova') arising from Lanthanide-free components of the ejecta (Sect.~\ref{sec:blue}).  Sect.~\ref{sec:170817} describes observations of the thermal UVOIR kilonova emission observed following GW170817 and its theoretical interpretation within this largely pre-existing theoretical framework.  I also summarize the lessons GW170817 has provided, about the origin of $r$-process elements, the equation of state of neutron stars, and the final fate of the merger remnant.  

Sect.~$\ref{sec:variations}$ explores several variations on this canonical picture, some of which were not possible to test in the case of GW170817 and some which are ruled out in that event but could be relevant to future mergers, e.g. with different ingoing binary parameters.  These include early ($\sim$ hours-long) `precursor' emission at ultra-violet wavelengths (UV), which is powered either by the decay of free neutrons in the outermost layers of the ejecta (Sect.~\ref{sec:neutrons}) or prompt shock heating of the ejecta by a relativistic outflow such as the GRB jet (Sect~\ref{sec:cocoon}).  In Sect.~\ref{sec:engine} we consider the impact on the kilonova signal of energy input from a long-lived accreting BH or magnetar central engine.  In Sect.~\ref{sec:discussion}, I assess the prospects for discovering kilonovae following short GRBs and for future GW-triggers of NS-NS/BH-NS mergers.  I use this opportunity to make predictions for the diversity of kilonova signals with GW-measured properties of the binary, which will become testable once EM observations are routinely conducted in coincidence with a large sample of GW-detected merger events.  I conclude with some personal thoughts in Sect.~\ref{sec:conclusions}. 

Although I have attempted to make this review self-contained, the material covered is necessarily limited in scope and reflects my own opinions and biases.  I refer the reader to a number of other excellent recent reviews, which cover some of the topics discussed briefly here in greater detail: \citep{Nakar07,Faber&Rasio12,Berger14,Rosswog15,Fan&Hendry15,Baiotti&Rezzolla16,Baiotti19}, including other short reviews dedicated exclusively to kilonovae \citep{Tanaka16,Yu19}.  I encourage the reader to consult \cite{Fernandez&Metzger16} for a review of the broader range of EM counterparts of NS-NS/BH-NS mergers.  A few complementary reviews has appeared since GW170817 overviewing the interpretation of this event (e.g.~\citealt{Miller17,Bloom17,Metzger17,Siegel19}) or its constraints on the nuclear EOS (e.g.~\citealt{Raithel19}).  I also encourage the reader to consult the initial version of this review, written the year prior to the discovery of GW170817.

\begin{table}
\caption{Timeline of major developments in kilonova research}
\centering
\begin{tabular}{r || l}
1974 & Lattimer \& Schramm: $r$-process from BH-NS mergers \\
1975 & Hulse \& Taylor: discovery of binary pulsar system PSR 1913+16\\
1982 & Symbalisty \& Schramm: $r$-process from NS-NS mergers\\
1989 & Eichler et al.: GRBs from NS-NS mergers \\
1994 & Davies et al.: first numerical simulation of mass ejection from NS-NS mergers\\
1998 & Li \& Paczynski: first kilonova model, with parametrized heating \\
1999 & Freiburghaus et al.: NS-NS dynamical ejecta $\Rightarrow$ r-process abundances \\
2005 & Kulkarni: kilonova powered by free neutron-decay (``macronova"), central engine\\
2009 & Perley et al.: optical kilonova candidate following GRB 080503\\
2010 & Metzger et al., Roberts et al., Goriely et al.: ``kilonova" powered by $r$-process heating\\
2013 & Barnes \& Kasen, Tanaka \& Hotokezaka: La/Ac opacities $\Rightarrow$ NIR spectral peak\\
2013 & Tanvir et al., Berger et al.: NIR kilonova candidate following GRB 130603B\\
2013 & Yu, Zhang, Gao: magnetar-boosted kilonova (``merger-nova'')\\
2014 & Metzger \& Fernandez: blue kilonova from post-merger remnant disk winds\\
2017 & Coulter et al.: kilonova detected from NS-NS merger following GW-trigger\\
\end{tabular}
\end{table}


\section{Historical Background }  
\label{sec:history}

\subsection{NS mergers as sources of the $r$-process}
\label{sec:rprocess}

\cite{Burbidge+57} and \cite{Cameron57} realized that approximately half of the elements heavier than iron are synthesized via the capture of neutrons onto lighter seed nuclei like iron) in a dense neutron-rich environment in which the timescale for neutron capture is shorter than the $\beta$-decay timescale.  This `rapid neutron-capture process', or $r$-process, occurs along a nuclear path which resides far on the neutron-rich side of the valley of stable isotopes.  Despite these works occurring over 70 years ago, the astrophysical environments giving rise to the $r$-process remains an enduring mystery, among the greatest in nuclear astrophysics \citep[e.g.,][for contemporary reviews]{Qian&Wasserburg07,Arnould+07,Thielemann+11,Cowan+19}.  

Among the most critical quantities which characterize the viability of a potential $r$-process event is the {\it electron fraction} of the ejecta,
\be
Y_e \equiv \frac{n_p}{n_n + n_p},
\ee
where $n_p$ and $n_n$ are the densities of protons and neutrons, respectively.  Ordinary stellar material usually has more protons than neutrons ($Y_e \ge 0.5$), while matter with a neutron excess ($Y_e < 0.5$) is typically required for the $r$-process.

Core collapse supernovae have long been considered promising $r$-process sources.  This is in part due to their short delays following star formation, which allows even the earliest generations of metal-poor stars in our Galaxy to be polluted with $r$-process elements (e.g.~\citealt{Mathews+92,Sneden+08}).  Throughout the 1990s, the high entropy\footnote{A high entropy (low density) results in an $\alpha$-rich freeze-out of the 3-body and effective 4-body reactions responsible for forming seed nuclei in the wind, similar to big bang nucleosynthesis.  The resulting higher ratio of neutrons to seed nuclei (because the protons are trapped in $\alpha$ particles) then allows the $r$-process to proceed to heavier elements.} neutrino-heated winds from proto-neutron stars \citep{Duncan+86,Qian&Woosley96}, which emerge seconds after a successful explosion, were considered the most likely $r$-process site\footnote{Another $r$-process mechanism in the core collapse environment results from $\nu-$induced spallation in the He shell \citep{Banerjee+11}.  This channel is limited to very low metallicity $Z \lesssim 10^{-3}$ and thus cannot represent the dominant $r$-process source over the age of the galaxy (though it could be important for the first generations of stars).  } within the core collapse environment \citep{Woosley+94, Takahashi+94}.  However, more detailed calculations of the wind properties \citep{Thompson+01, Arcones+07, Fischer+10, Hudepohl+10, Roberts+10, MartinezPinedo+12, Roberts+12} later showed that the requisite combination of neutron-rich conditions ($Y_e \lesssim  0.5$) and high entropy were unlikely to obtain.  Possible exceptions include the rare case of a very massive proto-NS \citep{Cardall&Fuller97}, or in the presence of non-standard physics such as an eV-mass sterile neutrino \citep{Tamborra+12,Wu+14}.  

Another exception to this canonical picture may occur if the NS is formed rotating rapidly and is endowed with an ultra-strong ordered magnetic field $B \gtrsim 10^{14}\mbox{\,--\,}10^{15}$ G, similar to those which characterize Galactic magnetars.  Magneto-centrifugal acceleration within the wind of such a ``millisecond magnetar''  to relativistic velocities can act to lower its electron fraction or reduce the number of seed nuclei formed through rapid expansion \citep{Thompson+04}.  This could occur during the early supernova explosion phase \citep{Winteler+12,Nishimura+15} as well as during the subsequent cooling phase of the proto-NS over several seconds \citep{Thompson03,Metzger+07,Vlasov+14}.  Despite the promise of such models, numerical simulations of MHD supernovae are still in a preliminary state, especially when it comes to the accurate neutrino transport need to determine the ejecta $Y_e$ and the high resolution three-dimensional grid needed to capturing the growth of non-axisymmetric (magnetic kink or sausage mode) instabilities.  The latter can disrupt and slow the expansion rate of jet-like structures \citep{Mosta+14}, rendering the creation of the heaviest (third abundance-peak) $r$-process elements challenging to obtain \citep{Halevi+18}.

The observed rate of hyper-energetic supernovae (the only {\it bona fide} MHD-powered explosions largely agreed upon to exist in nature) is only $\sim 1/1000$ of the total core collapse supernova rate (e.g.~\citealt{Podsiadlowski+04}).  Therefore, a higher $r$-process yield per event $\gtrsim 10^{-2}M_{\odot}$ is required to explain a significant fraction of the Galactic abundances through this channel.  However, in scenarios where the $r$-process takes place in a prompt jet during the supernova explosion, it is inevitable that the $r$-process material will mix into the outer layers of the supernova ejecta along with the shock-synthesized $^{56}$Ni, the latter being responsible for powering the supernova's optical luminosity.  As we shall discuss later in the context of kilonovae, such a large abundance of lanthanide elements mixed into the outer ejecta layers would substantially redden the observed colors of the supernova light in a way incompatible with observed hyper-energetic (MHD) supernovae \citep*{Siegel+19}.\footnote{Nevertheless, a more deeply embedded source of heavy $r$-process ejecta would be less conspicuous, as the characteristic signatures of lanthanide elements would only appear well after the supernova's optical peak.  Promising in this regard are outflows from a fall-back accretion disk around the central BH or NS, such as those which may be responsible for powering long-duration GRBs (\citealt{Pruet+04}; \citealt*{Siegel+19}).}

Contemporaneously with the discovery of the first binary pulsar \citep{Hulse&Taylor75}, \cite{Lattimer&Schramm74, Lattimer&Schramm76} proposed that the merger of compact star binaries---in particular the collision of BH-NS systems---could give rise to the $r$-process by the decompression of highly neutron-rich ejecta \citep{Meyer89}.  \cite{Symbalisty&Schramm82} were the first to suggest NS-NS mergers as the site of the $r$-process.  \cite{Blinnikov+84} and \cite{Paczynski86} first suggested a connection between NS-NS mergers and GRBs.  \cite{Eichler+89} presented a more detailed model for how this environment could give rise to a GRB, albeit one which differs significantly from the current view.  \cite{Davies+94} performed the first numerical simulations of mass ejection from merging neutron stars, finding that $\sim 2\%$ of the binary mass was unbound during the process.  \cite{Freiburghaus+99} presented the first explicit calculations showing that the ejecta properties extracted from a hydrodynamical simulation of a NS-NS merger \citep{Rosswog+99b} indeed produces abundance patterns in basic accord with the solar system $r$-process.    

The neutrino-driven wind following a supernova explosion accelerates matter from the proto-NS surface relatively gradually, in which case neutrino absorption reactions on nucleons (particularly $\nu_e + n \rightarrow p + e^{-}$) have time to appreciably raise the electron fraction of the wind from its initial low value near the NS surface.  By contrast, in NS-NS/BH-NS mergers the  different geometry and more dynamical nature of the system allows at least a fraction of the unbound ejecta (tidal tails and disk winds) to avoid strong neutrino irradiation, maintaining a much lower value of $Y_e \lesssim 0.2$ (Sect.~\ref{sec:ejecta}).  

When averaged over the age of the Galaxy, the required production rate of heavy $r$-process nuclei of mass number $A > 140$ is $\sim 2\times 10^{-7} M_{\odot}$ yr$^{-1}$ \citep{Qian00}.  Given a rate $R_{\rm NS-NS}$ of detection of NS-NS mergers by Advanced LIGO/Virgo at design sensitivity, the required $r$-process mass yield per merger event to explain the entire Galactic abundances is very approximately given by \citep[e.g.,][]{Metzger+09,Vangioni+16}
\be
\langle M_{r} \rangle \sim 10^{-2}M_{\odot}\left(\frac{R_{\rm NS-NS}}{10\,{\rm yr^{-1}}}\right)^{-1}.
\label{eq:Mr}
\ee
As described in Sect.~\ref{sec:ejecta}, numerical simulations of NS-NS/BH-NS mergers find a range of total ejecta masses of $\langle M_{r} \rangle \sim 10^{-3}-10^{-1} M_{\odot}$, while $\langle M_{r} \rangle\approx 0.03-0.06M_{\odot}$ was inferred from the kilonova of GW170817 (Sect.~\ref{sec:170817}).  Although large uncertainties remain, it is safe to conclude that NS mergers are likely major, if not the dominant, sources of the $r$-process in the universe.  

Several additional lines of evidence support `high yield' $r$-process events consistent with NS-NS/BH-NS mergers being common in our Galaxy, both now and in its early history.  These include the detection of $^{244}$Pu on the ocean floor at abundances roughly two orders lower than that expected if the currently active $r$-process source were frequent, low-yield events like ordinary core collapse supernovae \citep{Wallner+15,Hotokezaka+15}.  Similar arguments show that actinide abundances in the primordial solar system require a rare source for the heaviest $r$-process elements (\citealt{Bartos&Marka19,Cote+19b}).  A fraction of the stars in the ultra-faint dwarf galaxy Reticulum II are highly enriched in $r$-process elements, indicating that this galaxy was polluted early in its history by a single rare $r$-process event \citep{Ji+16}.  Similarly, the large spread seen in the $r$-process abundances in many Globular Clusters (e.g.~\citealt{Roederer11}) may also indicate a rare source acting at low metallicity.  Reasonable variations in the ejecta properties of NS mergers could in principle explain the observed variability in the actinide abundances of metal-poor stars \citep{Holmbeck+19}.

Nevertheless, NS mergers are challenged by some observations, which may point to alternative $r$-process sites.  Given the low escape speeds of dwarf galaxies of $\sim 10\mathrm{\ km\ s}^{-1}$, even moderate velocity kicks to binaries from the process of NS formation would remove the binaries from the galaxy prior to merger (and thus preventing the merger ejecta from polluting the next generation of stars).  Although a sizable fraction of the Galactic NS-NS binaries have low proper motions and are indeed inferred to have experienced very low supernova kicks \citep{Beniamini+16}, even relatively modest spatial offsets of the merger events from the core of the galaxies make it challenging to retain enough $r$-process enhanced gas \citep{Bonetti+19}.  Another challenge to NS mergers are the short delay times $\lesssim 10-100$ Myr between star formation and merger which are required to explain stellar populations in the low-metallicity Galactic halo (\citealt{Safarzadeh+19}) and Globular Clusters (\citealt{Zevin+19}).  Depending on the efficiency of compositional mixing between the merger ejecta and the ISM of the Galaxy, realistic delay time distributions for NS-NS/NS-BH mergers within a consistent picture of structure formation via hierarchical growth \citep{Kelley+10} were argued to produce chemical evolution histories consistent with observations of the abundances of $r$-process elements in metal-poor halo stars as a function of their iron abundance (\citealt{Shen+15,Ramirez-Ruiz+15,vandeVoort+15}).  However, \citealt{Safarzadeh+19} comes to a different conclusion, while \citet{vandeVoort+19} found that $r$-process production by rare supernovae better fit the abundances of metal-poor stars than NS mergers.  Due to the fact that the delay time distribution of NS mergers at late times after star formation is expected to similar to that of Type Ia supernova (which generate most of the iron in the Galaxy), NS mergers are also challenged to explain the observed decrease of [Eu/Fe] with increasing [Fe] at later times in the chemical evolution history of the Galaxy \citep{Hotokezaka+18,Cote+19a}.  

Together, these deficiencies may hint at the existence of an additional high-yield $r$-process channel beyond NS-NS mergers which can operate at low metallicities with short delay times.  The collapse of massive, rotating stars (``collapsars"), the central engines of long-duration gamma-ray bursts (which are directly observed to occur in dwarf low-metallicity galaxies), are among the most promising contenders (\citealt{Pruet+04}; \citealt{Fryer+06}; \citealt*{Siegel+19}).  As we shall discuss, the physical conditions of hyper-accreting disks and their outflows following NS mergers, as probed by their kilonova emission, offers an indirect probe of the broadly similar physical conditions which characterize the outflows generated in collapsars.  Evidence for $r$-process in the outflows of NS merger accretion flows would indirectly support an $r$-process occurring in collapsars as well.

\subsection{A Brief History of Kilonovae}

\begin{figure}[!t]
\includegraphics[width=1.0\textwidth]{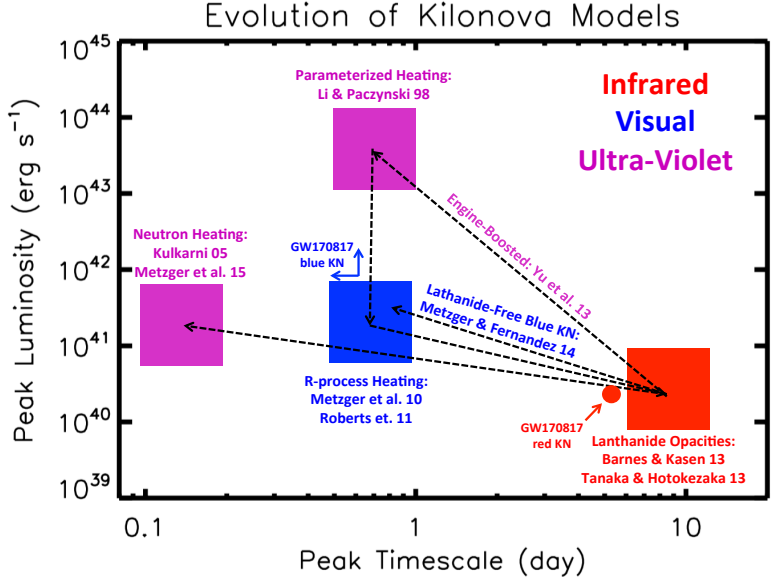}
\caption{Schematic timeline of the development kilonova models in the space of peak luminosity and peak timescale.  The wavelength of the predicted spectral peak are indicated by color as marked in the figure. Shown for comparison are the approximate properties of the ``red" and ``blue" kilonova emission components observed following GW170817 (e.g.~\citealt{Cowperthwaite+17,Villar+17}).}
\label{fig:timeline}
\end{figure}

\cite{Li&Paczynski98} first argued that the radioactive ejecta from a NS-NS or BH-NS merger provides a source for powering thermal transient emission, in analogy with supernovae.  They developed a toy model for the light curve, similar to that we describe in Sect.~\ref{sec:model}.  Given the low mass and high velocity of the ejecta from a NS-NS/BH-NS merger, they concluded that the ejecta will become transparent to its own radiation quickly, producing emission which peaks on a timescale of about one day, much faster than for normal supernovae (which instead peak on a timescale of weeks or longer).  

Lacking a model for the nucleosynthesis (the word ``$r$-process" does not appear in their work), \cite{Li&Paczynski98} parametrized the radioactive heating rate of the ejecta at time $t$ after the merger according to the following prescription,
\be
\dot{Q}_{\rm LP} = \frac{f M c^{2}}{t},
\label{eq:LP98}
\ee
where $M$ is the ejecta mass and $f$ was a free parameter.  The $\propto 1/t$ time dependence was motivated by the total heating rate which results from the sum of the radioactive decay heating rate $\dot{Q}_i \propto \exp(-t/\tau_i)$ of a large number of isotopes $i$, under the assumption that their half-lives $\tau_i$ are distributed equally per logarithmic time (at any time $t$, the heating rate is dominated by isotopes with half-lives $\tau_i \sim t$).  Contemporary models, which process the thermodynamic history of the expanding ejecta based on numerical simulations of the merger through a detailed nuclear reaction network, show that the heating rate at late times actually approaches a steeper power law decay $\propto t^{-\alpha}$, with $\alpha \approx 1.1\mbox{\,--\,}1.4$ \citep{Metzger+10, Roberts+11, Korobkin+12}, similar to that found for the decay rate of terrestrial radioactive waste \citep{Way&Wigner48}.  \citet{Metzger+10} and \citet{Hotokezaka+17} describe how this power-law decay can be understood from the basic physics of $\beta$-decay and the properties of nuclei on the neutron-rich valley of stability.

\cite{Li&Paczynski98} left the normalization of the heating rate $f$, to which the peak luminosity of the kilonova is linearly proportional, as a free parameter, considering a range of models with different values of $f = 10^{-5}\mbox{\,--\,}10^{-3}$.  More recent calculations, described below, show that such high heating rates are extremely optimistic, leading to predicted peak luminosities $\gtrsim 10^{43}\mbox{\,--\,}10^{44}\mathrm{\ erg\ s}^{-1}$ \citep[][their Fig.~2]{Li&Paczynski98} which exceed even those of supernovae.  These over-predictions leaked to other works throughout the next decade; for instance, \cite{Rosswog05} predicted that BH-NS mergers are accompanied by transients of luminosity $\gtrsim 10^{44}\mathrm{\ erg\ s}^{-1}$, which would rival the most luminous transients ever discovered.  This unclear theoretical situation led to observational searches for kilonovae following short GRBs which were inconclusive since they were forced to parametrized their results (usually non-detections) in terms of the allowed range of $f$ \citep{Bloom+06,Kocevski+10} instead of in terms of more meaningful constraints on the ejecta properties such as its mass.

\cite{Metzger+10} were the first to determine the true luminosity scale of the radioactively-powered transients of NS mergers by calculating light curve models using radioactive heating rates derived self-consistently from a nuclear reaction network calculation of the $r$-process, based on the dynamical ejecta trajectories of \cite{Freiburghaus+99}.\footnote{As a student entering this field in the mid 2000s, it was clear to me that the optical transients proposed by \cite{Li&Paczynski98} were not connected in most people's mind with the $r$-process.  \cite{Rosswog05} in principle had all the information needed to calculate the radioactive heating rate of the ejecta based on the earlier \cite{Freiburghaus+99} calculations, and thus to determine the true luminosity scale of these merger transients well before \cite{Metzger+10}.  I make this point not to cast blame, but simply to point out that the concept, now taken for granted, that the radioactive heating rate was something that could actually be calculated with any precision, came as a revelation, at least to a student of the available literature.}  Based on their derived peak luminosities being approximately one thousand times brighter than a nova, \cite{Metzger+10} introduced the term `kilonova' to describe the EM counterparts of NS mergers powered by the decay of $r$-process nuclei.  They showed that the radioactive heating rate was relatively insensitive to the precise electron fraction of the ejecta and to the assumed nuclear mass model, and they were the first to consider how efficiently the decay products thermalize their energy in the ejecta.  Their work highlighted the critical four-way connection, now taken for granted, between kilonovae, short GRBs, GWs from NS-NS/BH-NS mergers, and the astrophysical origin of the $r$-process.

Prior to \citet{Metzger+10}, it was commonly believed that kilonovae were in fact brighter, or much brighter, than supernovae \citep{Li&Paczynski98,Rosswog05}.  One exception is \cite{Kulkarni05}, who assumed that the radioactive power was supplied by the decay of $^{56}$Ni or free neutrons.  However, $^{56}$Ni cannot be produced in the neutron-rich ejecta of a NS merger, while all initially free neutrons are captured into seed nuclei during the $r$-process (except perhaps in the very outermost, fastest expanding layers of the ejecta; see Sect.~\ref{sec:neutrons}).  Kulkarni introduced the term ``macronovae'' for such Nickel/neutron-powered events.  Despite its inauspicious physical motivation and limited use in the literature until well after the term kilonova was already in use, many authors continue to use the macronova terminology, in part because this name is not tied to a particular luminosity scale (which may change as our physical models evolve).    

Once the radioactive heating rate was determined, attention turned to the yet thornier issue of the ejecta opacity.  The latter is crucial since it determines at what time and wavelength the ejecta becomes transparent and the light curve peaks.  Given the general lack of experimental data or theoretical models for the opacity of heavy $r$-process elements, especially in the first and second ionization states of greatest relevance, \cite{Metzger+10,Roberts+11} adopted grey opacities appropriate to the Fe-rich ejecta in Type Ia supernovae.  However, \cite{Kasen+13} showed that the opacity of $r$-process elements can be significantly higher than that of Fe, due to the high density of line transitions associated with the complex atomic structures of some lanthanide and actinide elements (Sect.~\ref{sec:opacity}).  This finding was subsequently confirmed by \cite{Tanaka&Hotokezaka13}.  As compared to the earlier predictions \citep{Metzger+10}, these higher opacities push the bolometric light curve to peak later in time ($\sim 1$ week instead of a $\sim 1$ day timescale), and at a lower luminosity \citep{Barnes&Kasen13}.  More importantly, the enormous optical wavelength opacity caused by line blanketing moved the spectral peak from optical/UV frequencies to the near-infrared (NIR).  Later that year, \cite{Tanvir+13} and \cite{Berger+13} presented evidence for excess infrared emission following the short GRB 130603B on a timescale of about one week using the \textit{Hubble Space Telescope}.   

However, not all of the merger ejecta necessarily will contain lanthanide elements with such a high optical opacity (e.g.~\citealt*{Metzger+08c}).  While ejecta with a relatively high electron fraction $0.25 \lesssim Y_{e} \lesssim 0.4$ has enough neutrons to synthesize radioactive $r$-process nuclei, the ratio of neutrons to lighter seed nuclei is insufficient to reach the relatively heavy lanthanide elements of atomic mass number $A \gtrsim 140$ which (if not blocked by high-opacity low-$Y_e$ material further out) produces emission with more rapid evolution and bluer colors, similar to those predicted by the original models \citep{Metzger+10,Roberts+11}.  \citet{Metzger&Fernandez14} dubbed the emission from high-$Y_e$, lanthanide-poor ejecta a ``blue'' kilonova, in contrast to ``red'' kilonova emission originating from low-$Y_e$, lanthanide-rich portions of the ejecta \citep{Barnes&Kasen13}.  They further argued that both ``blue'' and ``red'' kilonova emission, arising from different components of the merger ejecta, could be seen in the same merger event, at least for some observing geometries.  As will be discussed in $\S\ref{sec:170817}$, such hybrid ``blue'' + ``red'' kilonova models came to play an important role in the interpretation of GW170817.  Fig.~\ref{fig:timeline} is a timeline of theoretical predictions for the peak luminosities, timescales, and spectral peak of the kilonova emission.

\section{Basic Ingredients}
\label{sec:basics}

The physics of kilonovae can be understood from basic considerations.  Consider the merger ejecta of total mass $M$, which is expanding at a constant mean velocity $v$, such that its mean radius is $R \approx vt$ after a time $t$ following the merger.  Perhaps surprisingly, it is not unreasonable to assume spherical symmetry to first order because the ejecta will have a chance to expand laterally over the many orders of magnitude in scale from the merging binary ($R_{0} \sim 10^{6}$ cm) to the much larger radius ($R_{\rm peak} \sim 10^{15}$ cm) at which the kilonova emission peaks \citep{Roberts+11,Grossman+14,Rosswog+14}.  

The ejecta is extremely hot immediately after being ejected from the viscinity of the merger (Sect.~\ref{sec:ejecta}).  This thermal energy cannot, however, initially escape as radiation because of its high optical depth at early times,
\be
\tau \simeq \rho \kappa R = \frac{3M\kappa}{4\pi R^{2}} \simeq 70\left(\frac{M}{10^{-2}M_{\odot}}\right)\left(\frac{\kappa}{\rm 1\,cm^{2}\,g^{-1}}\right)\left(\frac{v}{0.1c}\right)^{-2}\left(\frac{t}{\rm 1\,day}\right)^{-2},
\label{eq:tau}
\ee
and the correspondingly long photon diffusion timescale through the ejecta,
\be
t_{\rm diff} \simeq \frac{R}{c}\tau =  \frac{3M\kappa}{4\pi c R} = \frac{3M\kappa}{4\pi c vt},
\label{eq:tdiff}
\ee
where $\rho = 3M/(4\pi R^{3})$ is the mean density and $\kappa$ is the opacity (cross section per unit mass).  As the ejecta expands, the diffusion time decreases with time $t_{\rm diff} \propto t^{-1}$, until eventually radiation can escape on the expansion timescale, as occurs once $t_{\rm diff} = t$ \citep{Arnett82}.  This condition determines the characteristic timescale at which the light curve peaks,
\be
t_{\rm peak} \equiv \left(\frac{3 M \kappa }{4\pi \beta v c}\right)^{1/2} \approx 1.6\,{\rm d}\,\,\left(\frac{M}{10^{-2}M_{\odot}}\right)^{1/2}\left(\frac{v}{0.1c}\right)^{-1/2}\left(\frac{\kappa }{1\,{\rm cm^{2}\,g^{-1}}}\right)^{1/2},
\label{eq:tpeak}
\ee
where the constant $\beta \approx 3$ depends on the precise density profile of the ejecta (see Sect.~\ref{sec:model}).  For values of the opacity $\kappa \sim 0.5\mbox{\,--\,}30\mathrm{\ cm^{2}\ g^{-1}}$ which characterize the range from Lanthanide-free and Lanthanide-rich matter (\citealt{Tanaka+19}; Table \ref{table:opacity}), respectively, Eq.~(\ref{eq:tpeak}) predicts characteristic durations $\sim 1$ day\,--\,1 week.

The temperature of matter freshly ejected at the radius of the merger $R_0 \lesssim 10^{6}$ cm generally exceed $10^{9}-10^{10}$ K.  However, absent a source of persistent heating, this matter will cool through adiabatic expansion, losing all but a fraction $\sim R_0/R_{\rm peak} \sim 10^{-9}$ of its initial thermal energy before reaching the radius $R_{\rm peak} = vt_{\rm peak}$ at which the ejecta becomes transparent (Eq.~\ref{eq:tpeak}).  Such adiabatic losses would leave the ejecta so cold as to be effectively invisible at large distances.  

\begin{figure}[!t]
\includegraphics[width=0.5\textwidth]{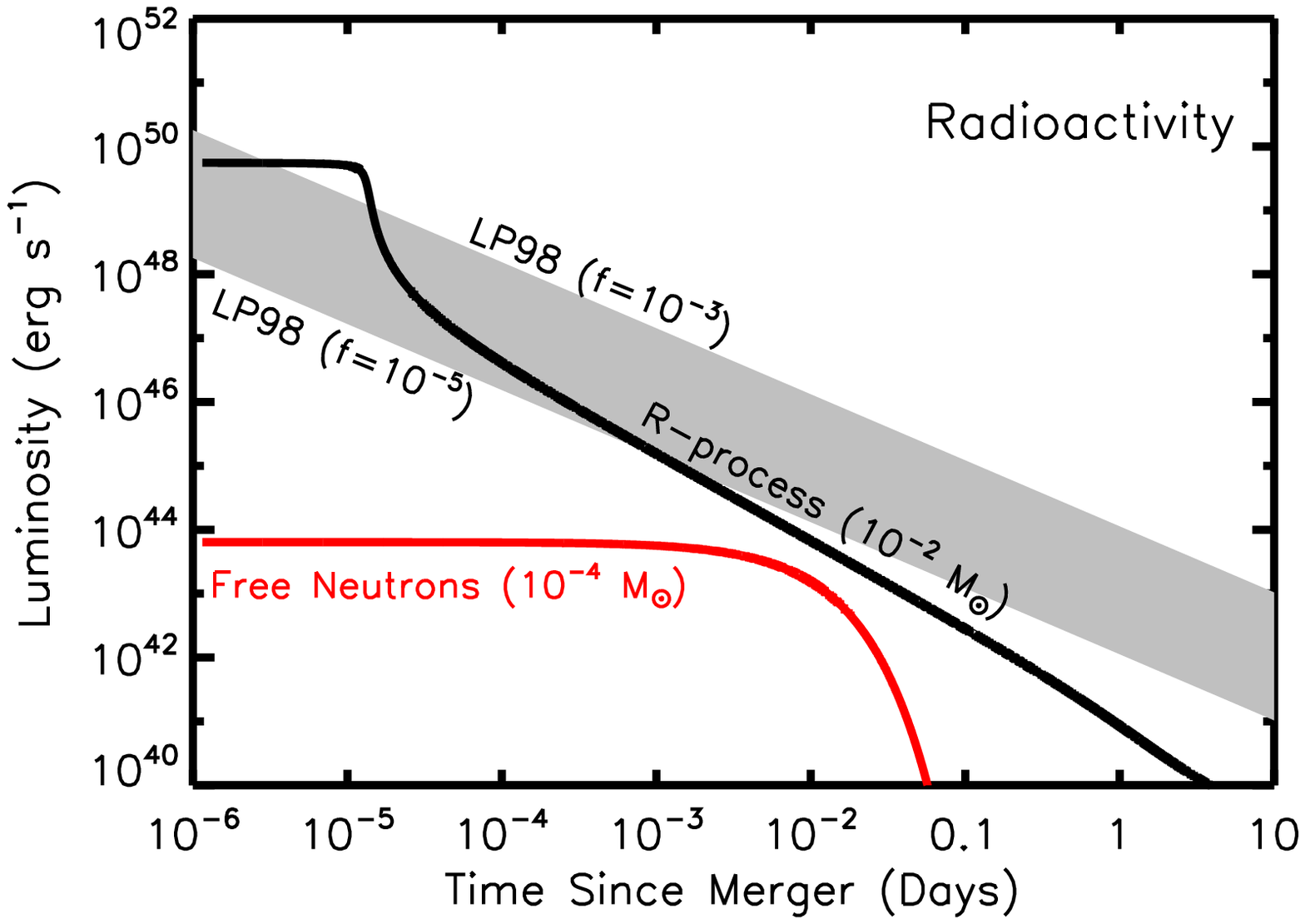}
\includegraphics[width=0.5\textwidth]{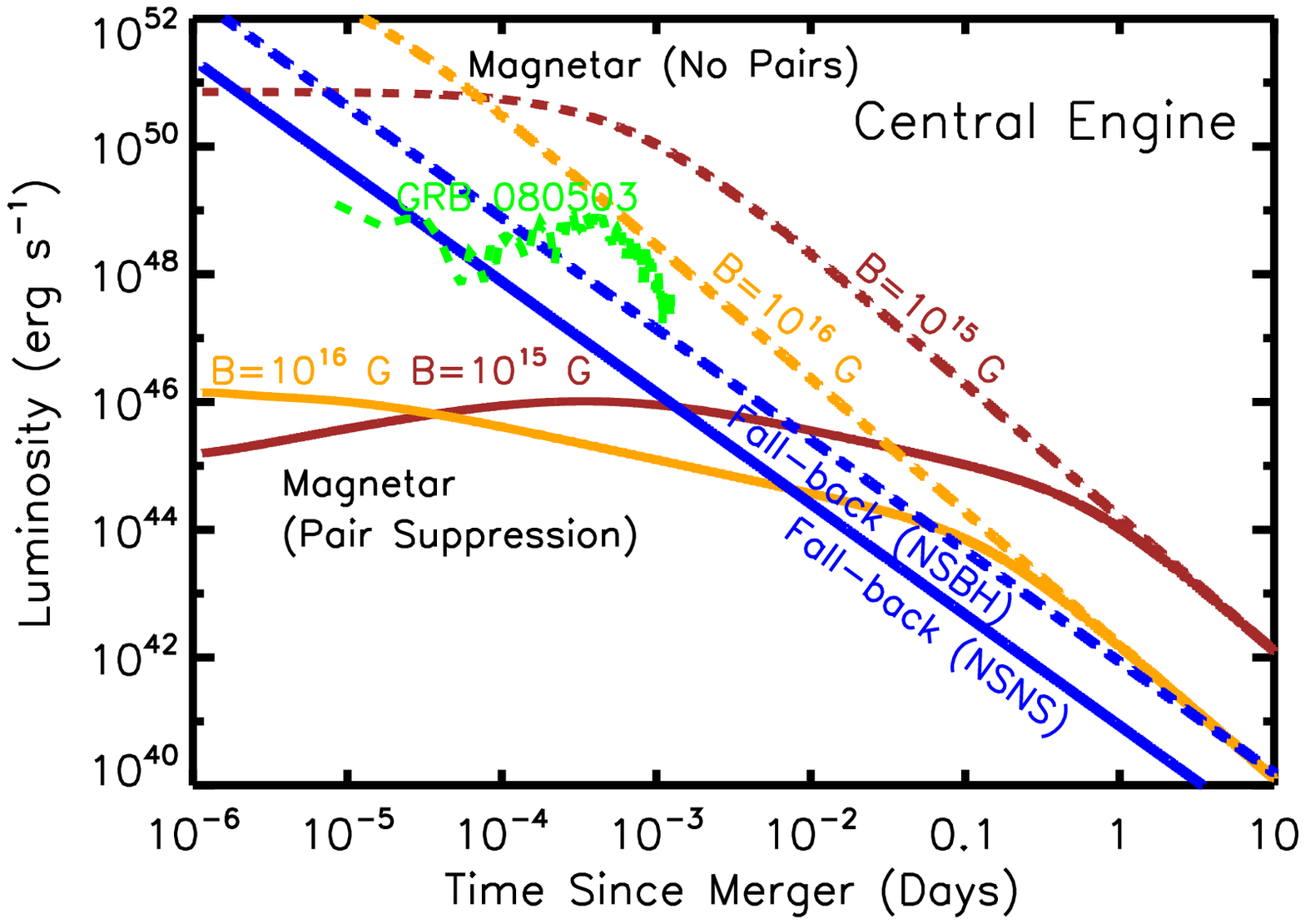}
\caption{Luminosity versus time after the merger of a range of heating sources relevant to powering kilonovae.  {\bf Left}: Sources of radioactive heating include the decay of $\sim 10^{-2}M_{\odot}$ of r-process nuclei, as first modeled in a parametrized way by \cite{Li&Paczynski98} (Eq.~\ref{eq:LP98}, grey band) and then by \cite{Metzger+10} using a full reaction network, plotted here using the analytic fit of \cite{Korobkin+12} (Eq.~\ref{eq:edotr}, black line) and including the thermalization efficiency of \cite{Barnes+16} (Eq.~\ref{eq:eth}).  The outermost layers of the ejecta may contain $\sim 10^{-4}M_{\odot}$ free neutrons (red line), which due to their comparatively long half-life can enhance the kilonova emission during the first few hours if present in the outermost layers of the ejecta due to premature freeze-out of the $r$-process (Sect.~\ref{sec:neutrons}).  {\bf Right}: Heating sources from a central engine.  These include fall-back accretion (blue lines), shown separately for NS-NS (solid line) and BH-NS (dashed line) mergers, based on results by \cite{Rosswog07} for an assumed jet efficiency $\epsilon_j = 0.1$ (Eq.~\ref{eq:Lxfb}).  Also shown is the rotational energy input from the magnetic dipole spin-down of a stable magnetar remnant with an initial spin period of $P = 0.7$ ms and dipole field strengths of $B = 10^{15}$ G (brown lines) and $10^{16}$ G (orange lines).  Dashed lines show the total spin-down luminosity $L_{\rm sd}$ (Eq.~\ref{eq:Lsd}), while solid lines show the effective luminosity available to power optical/X-ray emission once accounting for suppression of the efficiency of thermalization due to the high scattering opacity of $e^{\pm}$ pairs in the nebula (Eq.~\ref{eq:Lobs}; \citealp{Metzger&Piro14}).  The isotropic luminosity of the temporally-extended X-ray emission observed following the short GRB 080503 is shown with a green line (for an assumed source redshift $z = 0.3$; \citealt{Perley+09}).}
\label{fig:heating}
\end{figure}

In a realistic situation, the ejecta will be continuously heated, by a combination of sources, at a total rate $\dot{Q}(t)$ (Fig.~\ref{fig:heating}).  At a minimum, this heating includes contributions from radioactivity due to $r$-process nuclei and, possibly at early times, free neutrons.  More speculatively, the ejecta can also be heated from within by a central engine, such as the emergence of the GRB jet or over longer timescales by the rotational energy of a magnetar remnant.  In most cases of relevance, $\dot{Q}(t)$ is constant or decreasing with time less steeply than $\propto t^{-2}$.  The peak luminosity of the observed emission then equals the heating rate at the peak time ($t = t_{\rm peak}$), i.e.,
\be
L_{\rm peak} \approx \dot{Q}(t_{\rm peak}),
\label{eq:Arnett}
\ee 
a result commonly known as ``Arnett's Law'' \citep{Arnett82}.  

Equations (\ref{eq:tpeak}) and (\ref{eq:Arnett}) make clear that, in order to quantify the key observables of kilonovae (peak timescale, luminosity, and effective temperature), we must understand three key ingredients: 
\begin{itemize}
\item{The mass and velocity of the ejecta from NS-NS/BH-NS mergers, as comprised by several distinct components.}
\item{The opacity $\kappa$ of expanding neutron-rich matter.}
\item{The variety of sources contributing to the ejecta heating $\dot{Q}(t)$, particularly on timescales of $t_{\rm peak}$, when the ejecta first becomes transparent.}
\end{itemize}
 The remainder of this section addresses the first two issues.  The range of different heating sources, which can give rise to different types of kilonovae, are covered in Sect.~\ref{sec:model}.

\subsection{Sources of Neutron-Rich Ejecta}
\label{sec:ejecta}

\begin{table}[!t]
\caption{Sources of Ejecta in NS-NS Mergers \label{table:ejecta}}

\begin{tabular}{ccccccc}

Ejecta Type & $M_{\rm ej} (M_{\odot})$ & $v_{\rm ej}(c)$ & $Y_e$ & $M_{\rm ej}$ decreases with  \\

\hline

Tidal Tails$\dagger$ & $10^{-4}-10^{-2}$ & $0.15-0.35$ & $\lesssim 0.2$ & $q = M_{2}/M_{1} < 1$ & &  \\
Polar Shocked & $10^{-4}-10^{-2}$ & $0.15-0.35$ & $0.25-0.45$ & $M_{\rm tot}/M_{\rm TOV}, R_{\rm ns}$ &  \\
Magnetar Wind & $10^{-2}$ & $0.2-1$ & $0.25-0.45$  & $M_{\rm tot}/M_{\rm TOV}$ &  \\
Disk Outflows$\dagger$ & $10^{-3}-0.1$ & $0.03-0.1$ & $0.1-0.4$ & $M_{\rm tot}/M_{\rm TOV}$ &  \\
\hline \\
\end{tabular}
\\
$\dagger$Present in NS-BH mergers.
\end{table}


Two broad sources of ejecta characterize NS-NS and BH-NS mergers \citep[see][for recent reviews]{Fernandez&Metzger16,Shibata&Hotokezaka19}.  First, there is matter ejected on the dynamical timescale of milliseconds, either by tidal forces or due to compression-induced heating at the interface between merging bodies (Sect.~\ref{sec:dynamical}).  Debris from the merger, which is not immediately unbound or incorporated into the central compact object, can possess enough angular momentum to circularize into an accretion disk around the central remnant.  A disk can also be generated by outwards transport of angular momentum and mass during the post-merger evolution of the central NS remnant prior to BH formation.  Outflows from this remnant disk, taking place on longer timescales of up to seconds, provide a second important source of  ejecta (Sect.~\ref{sec:diskejecta}).

In BH-NS mergers, significant mass ejection and disk formation occurs only if the BH has a low mass $M_{\bullet}$ and is rapidly spinning; in such cases, the NS is tidally disrupted during the very final stages of the inspiral instead of being swallowed whole (giving effectively zero mass ejection).  Roughly speaking, the condition for the latter is that the tidal radius of the NS, $R_{\rm t} \propto M_{\bullet}^{1/3}$, exceed the innermost stable circular orbit of the BH, $R_{\rm isco} \propto M_{\bullet}$ (see \citealt{Foucart12,Foucart+18} for a more precise criterion for mass ejection, calibrated to GR merger simulations).  For a NS of radius $12$ km and mass $1.4M_{\odot}$, this requires a BH of mass $\lesssim 4(12)M_{\odot}$ for a BH Kerr spin parameter of $\chi_{\rm BH} = 0.7(0.95)$.  For slowly-spinning BHs (as appears to characterize most of LIGO/Virgo's BH-BH systems), the BH mass range giving rise to tidal disruption---and hence a kilonova or GRB---could be very small.

In the case of a NS-NS merger, the ejecta properties depend sensitively on the fate of the massive NS remnant which is created by the coalescence event.  The latter in turn depends sensitively on the total mass of the original NS-NS binary, $M_{\rm tot}$ \citep{shibata2000,Shibata&Taniguchi06}.  For $M_{\rm tot}$ above a threshold mass of $M_{\rm crit} \sim 2.6-3.9M_\odot$ (covering a range of soft and stiff nuclear-theory based equations of state [EOS], respectively), the remnant collapses to a BH essentially immediately, on the dynamical time of milliseconds or less \citep{Hotokezaka+11,Bauswein+13b}.  \citet{Bauswein+13b} present an empirical fitting formula for the value of $M_{\rm crit}$ in terms of the maximum mass $M_{\rm TOV}$ of a non-rotating NS (the Tolman-Oppenheimer-Volkoff [TOV] mass) and the NS compactness (see also \citealt{Koppel+19}), which they find is insensitive to the binary mass ratio $q = M_{2}/M_{1}$ for $q\gtrsim 0.7$ (however, see \citealt{Kiuchi+19}).  

Mergers that do not undergo prompt collapse ($M_{\rm tot} < M_{\rm crit}$) typically result in the formation of rapidly-spinning NS remnant of mass $\sim M_{\rm tot}$ (after subtracting mass lost through neutrino and GW emission and in the dynamical ejecta), which is at least temporarily stable against gravitational collapse to a BH.  The maximum stable mass of a NS exceeds its non-rotating value, $M_{\rm TOV}$, if the NS is rapidly spinning close to the break-up velocity \citep{Baumgarte+00,Ozel+10,Kaplan+14}.  

A massive NS remnant, which is supported exclusively by its differential rotation, is known as a \emph{hypermassive} NS (HMNS).  A somewhat less massive NS, which can be supported even by its solid body rotation (i.e.~after differential rotation has been removed), is known as a \emph{supramassive} NS (SMNS).  A HMNS is unlikely to survive for more than a few tens to hundreds of milliseconds after the merger, before collapsing to a BH due to the loss of differential rotation and accretion of mass by internal hydro-magnetic torques and gravitational wave radiation \citep{Shibata&Taniguchi06,Duez+06,Siegel+13}. In contrast, SMNS remnants must spin-down to the point of collapse through the global loss of angular momentum.  The latter must take place through less efficient processes, such as magnetic dipole radiation or GW emission arising from small non-axisymmetric distortions of the NS, and hence such objects can in principle survive for much longer before collapsing.  Finally, the merger of a particularly low mass binary, which leaves a remnant mass less than $M_{\rm TOV}$, will produce an indefinitely stable remnant \citep{Metzger+08b,Giacomazzo&Perna13}, from which a BH can never form, even once its angular momentum has been entirely removed.  Such cases are likely very rare.

\begin{table}[!t]
\caption{Remnants of NS-NS Mergers \label{table:remnants}}

\begin{tabular}{ccccccc}

Remnant & Binary Mass Range & NS Lifetime ($t_{\rm collapse}$) & $\%$ of mergers$^{(a)}$ \\
\hline

Prompt BH & $M_{\rm tot} \gtrsim M_{\rm th}^{\dagger} \sim 1.3-1.6M_{\rm TOV}$ & $\lesssim 1$ ms & $\sim 0-32\%$ \\
HMNS & $\sim 1.2M_{\rm TOV} \lesssim M_{\rm tot} \lesssim M_{\rm th}$ & $\sim 30-300$ ms & $\sim 0-79\%$ \\ 
SMNS & $M_{\rm TOV} \lesssim M_{\rm tot} \lesssim 1.2M_{\rm TOV}$ & $\gg 300$ ms & $\sim 18-65\%$ \\ 
Stable NS & $M_{\rm tot} < M_{\rm TOV}$ & $\infty$ & $\lesssim 3\%$ \\
\hline \\
\end{tabular}
\\
$^{(a)}$Percentage of mergers allowed by current EOS constraints on NS radii and $M_{\rm TOV}$ (Sect.~\ref{sec:EOS}) assuming the merging extragalactic NS-NS binary population is identical to the known Galactic NS-NS binaries (from \citealt{Margalit&Metzger19}).  $^{\dagger}$The prompt collapse threshold, $M_{\rm th}$, depends on both $M_{\rm TOV}$ and the NS compactness/radius (see text).
\end{table}

\begin{figure}[!t]
\includegraphics[width=0.5\textwidth]{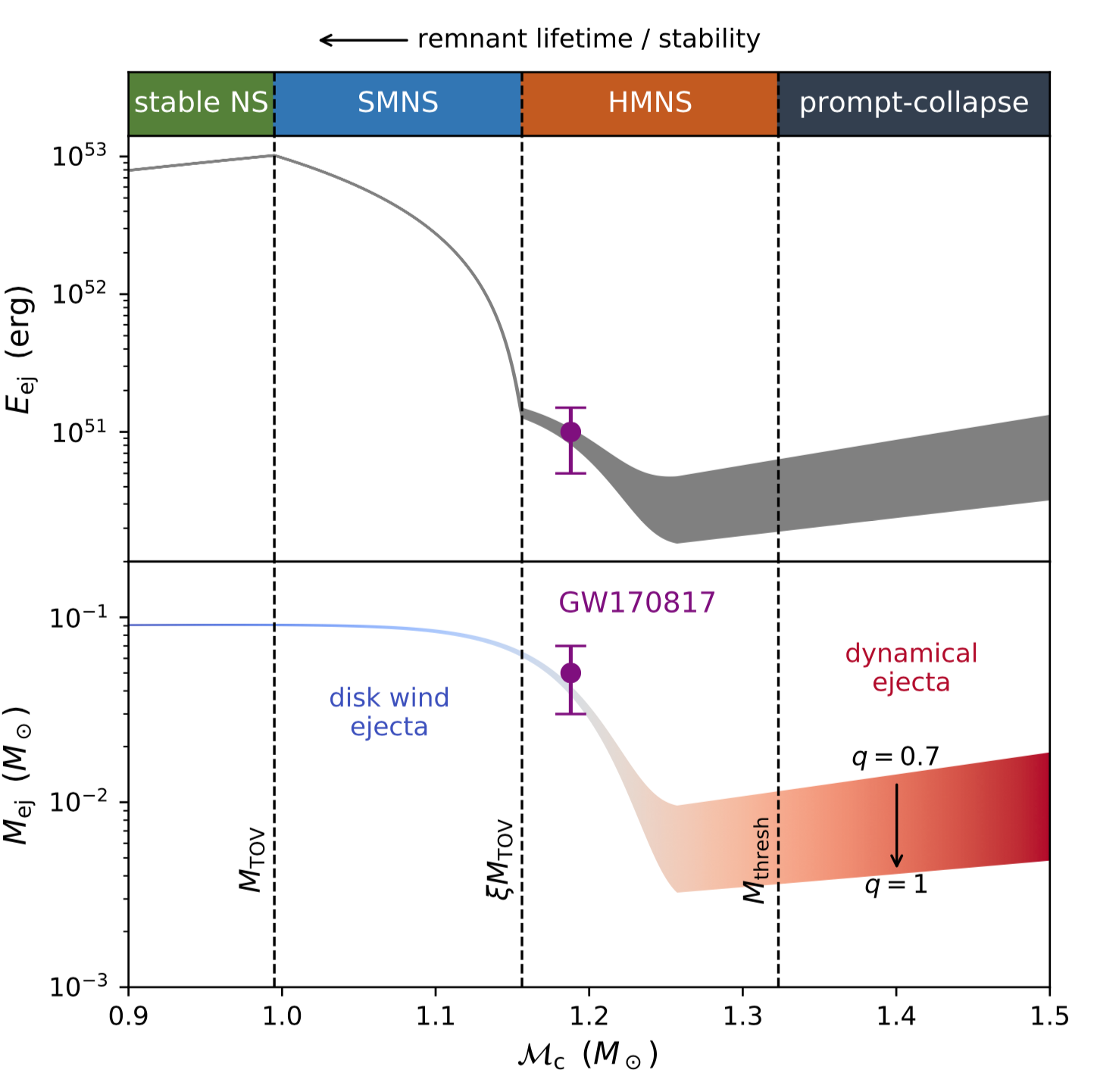}
\includegraphics[width=0.5\textwidth]{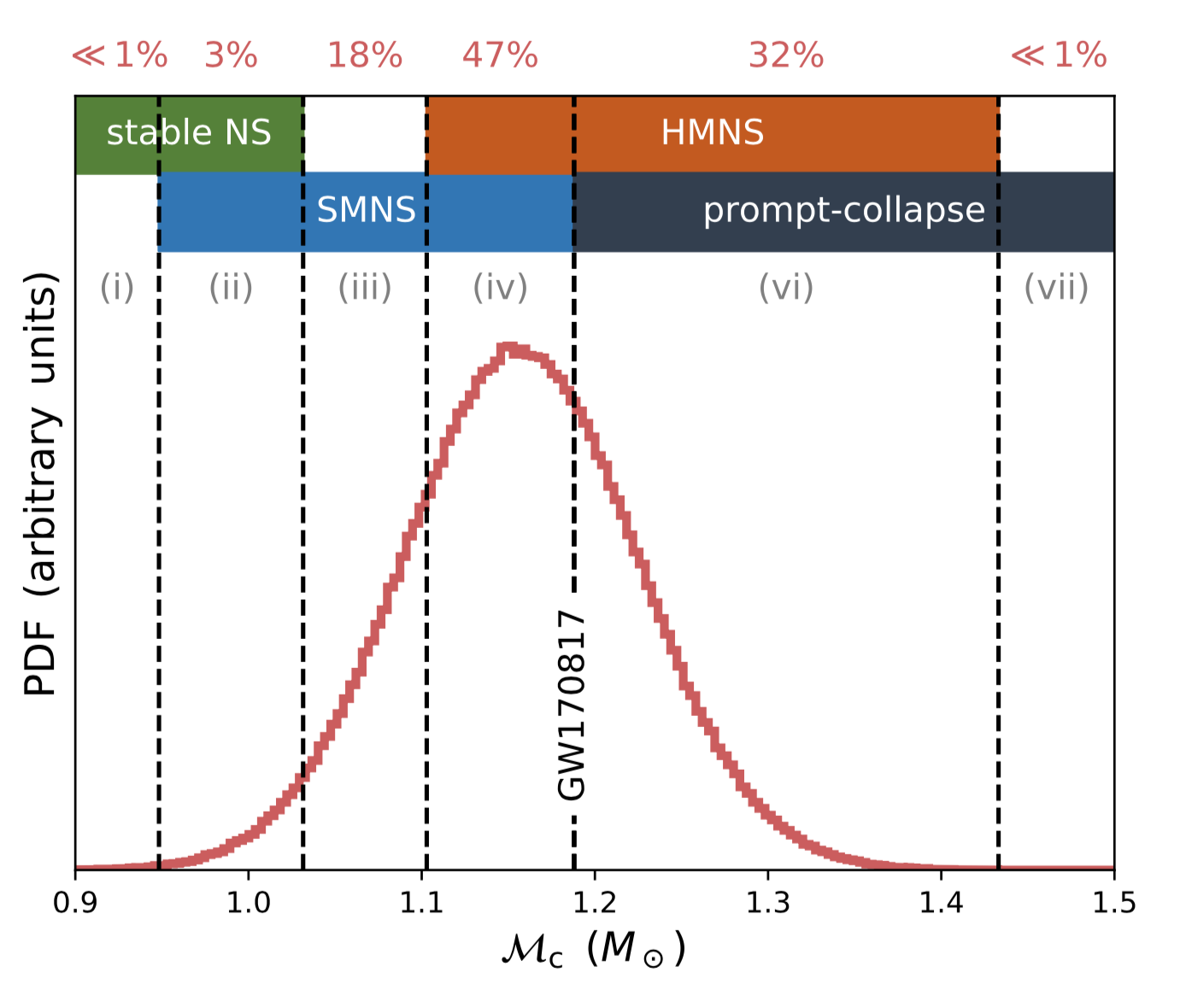}
\caption{{\bf Left:} Properties of the merger ejecta which affect the EM emission as a function of the binary chirp mass $\mathcal{M}_{\rm c}$ (eq.~\ref{eq:Mchirp}), taken here as a proxy for the total binary mass $M_{\rm tot}$. Vertical dashed lines delineate the threshold masses for different merger remnants as marked, for an example EOS with $M_{\rm TOV} = 2.1M_{\odot}$ and radius $R_{1.6} = 12$ km of a 1.6$M_{\odot}$ NS. The top panel shows the ejecta kinetic energy, which we take to be the sum of the initial kinetic energy of the ejecta (estimated using fits to numerical relativity simulations; \citealt{Coughlin+18a,Coughlin+18b}) and, in the case of stable or SMNSs, the rotational energy which can be extracted from the remnant before forming a BH \citep{Margalit&Metzger17}. The bottom panel shows the ejecta mass, both dynamical and disk wind ejecta, estimated as in \citet{Coughlin+18b}, where 50\% of the disk mass is assumed to be ejected at $v = 0.15$ c (e.g. \citealt{Siegel&Metzger17}).  The finite width of the lines results from a range of binary mass ratio $q = 0.7-1$, to which the tidal dynamical ejecta is most sensitive. The ejecta mass line is colored qualitatively according to the dominant color of the kilonova emission, which becomes redder for more massive binaries (with shorter-lived remnants) due to their more neutron-rich ejecta (\citealt{Metzger&Fernandez14}). {\bf Right:}  Distribution of BNS merger chirp masses drawn from a NS population representative of Galactic double NSs (\citealt{Kiziltan+13}).  Dashed vertical curves separate the $\mathcal{M}_{\rm c}$ parameter space based on the possible merger outcomes in each region. The fraction of mergers expected to occur in each region (the integral over the PDF within this region) is stated above the region in red (see also Table \ref{table:remnants}).  Figures reproduced with permission from \citet{Margalit&Metzger19}, copyright of the authors. }
\label{fig:MM}
\end{figure}

Table \ref{table:remnants} summarizes the four possible outcomes of a NS-NS merger and estimates of their expected rates.
  The left panel of Fig.~\ref{fig:MM} from \citet{Margalit&Metzger19} illustrates these mass divisions in terms of the chirp mass $\mathcal{M}_{\rm c} \simeq 0.435M_{\rm tot}$ (eq.~\ref{eq:Mchirp}; under the assumption of an equal mass binary $M_{1} = M_{2}$) and taking as an example EOS one which predicts a $1.6M_{\odot}$ NS radius $R_{1.6} = 12$ km and TOV mass $M_{\rm TOV} \approx 2.1M_{\odot}$.  The latter is consistent with the lower limit of $M_{\rm TOV} \gtrsim 2-2.1M_{\odot}$ set by the discovery of pulsars with similar masses \citep{Demorest+10, Antoniadis+13,Cromartie+19} and the upper limit $M_{\rm TOV} \lesssim 2.16M_{\odot}$ supported by joint EM/GW observations of GW170817 (Sect.~\ref{sec:EOS}).  

The right panel of Fig.~\ref{fig:MM} shows the chirp mass distribution of known Galactic NS-NS binaries \citep{Kiziltan+13} compared to the allowed ranges in the binary mass thresholds separating different remnant classes (stable NS, SMNS, HMNS, prompt collapse) given current EOS constraints.  The chirp mass of GW170817 is fully consistent with being drawn from the Galactic NS-NS population, while indications from the EM observations suggest that a HMNS remnant formed in this event (Sect.~\ref{sec:170817}).  If the extra-galactic population of merging NS-NS binaries is indeed similar to the known Galactic population, \citet{Margalit&Metzger19} predict that 18$\% -65\%$ of mergers would result in SMNS remnants, while only a small fraction $< 3\%$ would produce indefinitely stable NS remnants (Table \ref{table:remnants}).  As we discuss in Sect.~\ref{sec:magnetar}, additional energy input from a long-lived magnetar remnant could substantially boost the kilonova emission.  The fractions of mergers leading to a prompt BH collapse, and relatively little ejecta or disk, ranges from tens of percent to extremely infrequently.

\subsubsection{Dynamical Ejecta}
\label{sec:dynamical}

NS-NS mergers eject unbound matter through processes that operate on the dynamical time, and which depend primarily on the total  binary mass, the mass ratio, and the EOS.  Total dynamical ejecta masses typically lie in the range $10^{-4}\mbox{\,--\,}10^{-2}M_\odot$ for NS-NS mergers (e.g.~\citealt{Hotokezaka+13,radice2016,Bovard+17}), with velocities $0.1-0.3$ c.   For BH-NS mergers, the ejecta mass can be up to $\sim 0.1M_\odot$ with similar velocities as in the NS-NS case \citep{Kyutoku+13,Kyutoku+15,Foucart+17}.  The ejecta mass is typically greater for eccentric binaries \citep{East+12,Gold+12}, although the dynamical interactions giving rise to eccentric mergers require high stellar densities, probably making them rare events compared to circular inspirals \citep{Tsang13}.  Very high NS spin can also enhance the quantity of dynamical ejecta (e.g.~\citealt{Dietrich+17b,East+19,Most+19}).

\begin{figure}[!t]
\includegraphics[width=1.0\textwidth]{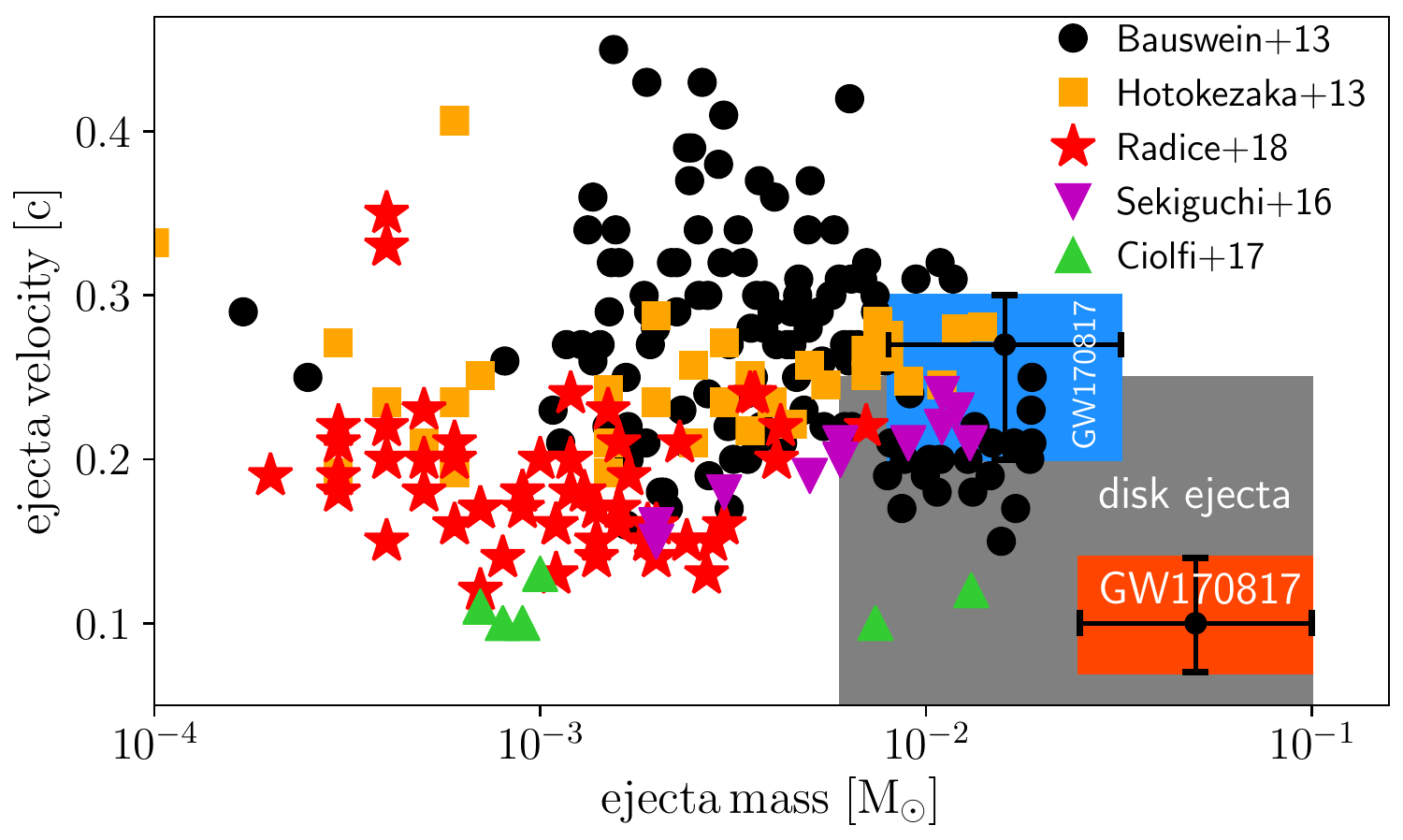}
\caption{Dynamical ejecta masses and velocities from a range of binary neutron star merger simulations encompassing different numerical techniques, various equations of state, binary binary mass ratios $q = 0.65-1$, effects of neutrinos and magnetic fields, together with the corresponding ejecta parameters inferred from the ‘blue’ and ‘red’ kilonova of GW170817.  Figure from \citet{Siegel19}.}
\label{fig:Siegel}
\end{figure}

Two main ejection processes operate in NS-NS mergers. First, material at the contact interface between the merging stars is squeezed out by hydrodynamic forces and is subsequently expelled by quasi-radial pulsations of the remnant \citep{Oechslin+07,Bauswein+13,Hotokezaka+13}, ejecting shock-heated matter in a broad range of angular directions.  The second process involves spiral arms from tidal interactions during the merger, which expand outwards in the equatorial plane due to angular momentum transport by hydrodynamic processes.  The relative importance of these mechanisms depends on the EOS and the binary mass ratio $q$, with lower values of $q \ll 1$ (asymmetric) binaries ejecting greater quantities of mass  \citep{bauswein2013,Lehner+16}.  The ejecta mass also depends on the BH formation timescale; for the prompt collapses which characterize massive binaries, mass ejection from the contact interface is suppressed due to prompt swallowing of this region.  Figure \ref{fig:Siegel} shows the total ejecta mass and mean velocity of the dynamical ejecta inferred from a range of NS-NS simulations compiled from the literature \citep{bauswein2013,Hotokezaka+13,Radice+18,Sekiguchi+16,Ciolfi+17}.

In BH-NS mergers, mass is ejected primarily by tidal forces that disrupt the NS, with the matter emerging primarily in the equatorial plane \citep{Kawaguchi+15}.  The ejecta from BH-NS mergers also often covers only part of the azimuthal range \citep{Kyutoku+15}, which may introduce a stronger viewing angle dependence on the kilonova emission than for NS-NS mergers. 

Another key property of the dynamical ejecta, in addition to the mass and velocity, is its electron fraction, $Y_e$.  Simulations that do not account for weak interactions find the ejecta from NS-NS mergers to be highly neutron-rich, with an electron fraction $Y_e \lesssim 0.1$, sufficiently low to produce a robust\footnote{This robustness is rooted in `fission recycling' \citep{Goriely+05}: the low initial $Y_e$ results in a large neutron-to-seed ratio, allowing the nuclear flow to reach heavy nuclei for which fission is possible ($A \sim 250$). The fission fragments are then subject to additional neutron captures, generating more heavy nuclei and closing the cycle.} abundance pattern for heavy nuclei with $A \gtrsim 130$ \citep{Goriely+11,Korobkin+12,Bauswein+13,Mendoza-Temis.Wu.ea:2015}.  More recent merger calculations that include the effects of $e^\pm$ captures and neutrino irradiation in full general-relativity have shown that the dynamical ejecta may have a wider electron fraction distribution ($Y_e \sim 0.1-0.4$) than models which neglect weak interactions \citep{sekiguchi2015,radice2016}.  As a result, lighter $r$-process elements with $90 \lesssim A \lesssim 130$ are synthesized in addition to third-peak elements \citep{wanajo2014}.  These high-$Y_e$ ejecta components are distributed in a relatively spherically-symmetric geometry, while the primarily tidally-ejected, lower-$Y_e$ matter is concentrated closer to the equatorial plane (Fig.~\ref{fig:schematic}).

\subsubsection{Disk Outflow Ejecta}
\label{sec:diskejecta}

\begin{figure}[!t]
\includegraphics[width=1.0\textwidth]{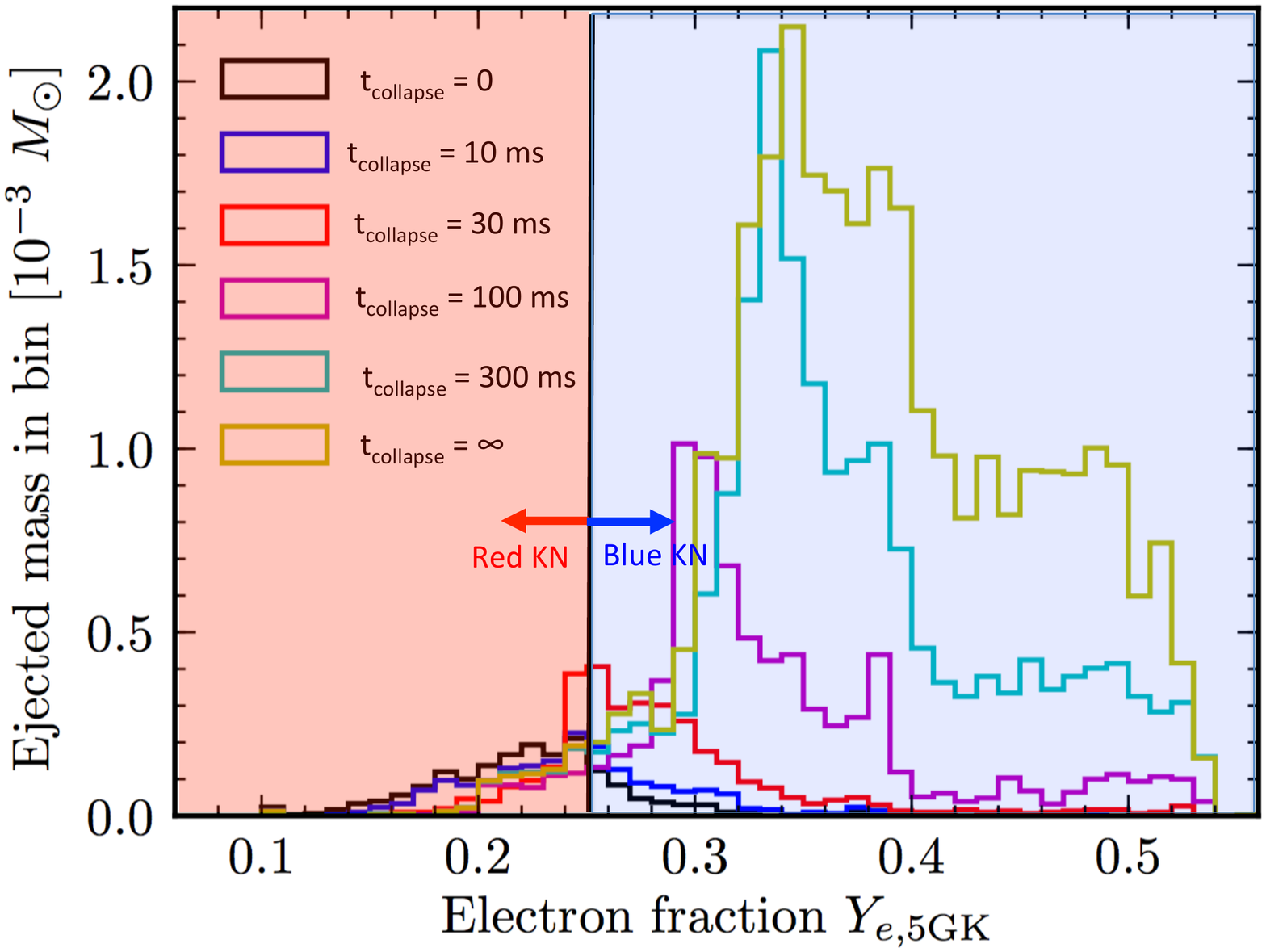}
\caption{Longer-lived remnants produce higher $Y_e$ disk wind ejecta and bluer kilonovae.  Shown here is the mass distribution by electron fraction $Y_e$ of the disk wind ejecta, calculated for different assumptions about the lifetime, $t_{\rm collapse}$, of the central NS remnant prior to BH formation, from the axisymmetric $\alpha$-viscosity hydrodynamical calculations of \citet{Metzger&Fernandez14}.  A vertical line approximately delineates the ejecta with enough neutrons to synthesize lanthanide elements ($Y_e \lesssim 0.25$) generate a red kilonova from that with $Y_e \gtrsim 0.25$ which is lanthanide-poor and will generate blue kilonova emission.  The NS lifetime has a strong effect on the ejecta composition because it is a strong source of electron neutrinos, which convert neutrons in the disk to protons via the process $\nu_e + n \rightarrow p + e^{-}$.  This figure is modified from a version in \citet{Lippuner+17}.}
\label{fig:Lippuner}
\end{figure}

All NS-NS mergers, and those BH-NS mergers which end in NS tidal disruption outside the BH horizon, result in the formation of an accretion disk around the central NS or BH remnant.  The disk mass is typically $\sim 0.01-0.3M_{\odot}$, depending on the total mass and mass ratio of the binary, the spins of the binary components, and the NS EOS (e.g.~\citealt{Oechslin&Janka06}).  Relatively low disk masses are expected in the case of massive binaries that undergo prompt collapse to a BH, because the process of massive disk formation is intimately related to the internal redistribution of mass and angular momentum of the remnant as it evolves from a differentially rotating to solid body state (which has no time to occur in a prompt collapse).  Outflows from this disk, over a timescales of seconds or longer, represent an important source of ejecta mass which can often dominate that of the dynamical ejecta.

At early times after the disk forms, its mass accretion rate is high and the disk is a copious source of thermal neutrinos \citep{Popham+99}.  During this phase, mass loss is driven from the disk surface by neutrino heating, in a manner analogous to neutrino-driven proto-NS winds in core collapse supernovae \citep{Surman+08,Metzger+08a}.  Spiral density waves, which are excited in the disk by the oscillations of the central NS remnant, may also play a role in outwards angular momentum transport and mass ejection during this early phase \citep{Nedora+19}.  Time dependent models of the long-term evolution of these remnant tori, which include neutrino emission and absorption, indicate that when BH formation is prompt, the amount of mass ejected through this channel is small, contributing at most a few percent of the outflow, because the neutrino luminosity decreases rapidly in time \citep{Fernandez&Metzger13,Just+15}.  However, if the central NS remnant survives for longer than $\sim 50$~ms (as a HMNS or SMNS), then the larger neutrino luminosity from the NS remnant ejects a non-negligible amount of mass ($\sim 10^{-3}M_\odot$, primarily from the NS itself instead of the disk; \citealp{Dessart+09, Perego+14,Martin+15,Richers+15}).  As we discuss below, ejecta from the star could be substantially enhanced if the central remnant has a strong ordered magnetic field \citep{Metzger+18}.

The disk evolves in time due to the outwards transport of angular momentum, as mediated e.g.~by spiral density waves or (more generically) magnetic stresses created by MHD turbulence generated by the magneto-rotational instability.  Initial time-dependent calculations of this `viscous spreading' followed the disk evolution over several viscous times using one-zone \citep{Metzger+08c} and one-dimensional height-integrated \citep{Metzger+09} models.  These works showed that, as the disk evolves and its accretion rate decreases, the disk transitions from a neutrino-cooled state to a radiatively inefficient (geometrically thick disk) state as the temperature, and hence the neutrino cooling rate, decreases over a timescale of seconds \citep[see also][]{Lee+09,Beloborodov08}.  Significant outflows occur once during the radiatively inefficient phase, because viscous turbulent heating and nuclear recombination are unbalanced by neutrino cooling \citep{Kohri+05}.  This state transition is also accompanied by ``freeze-out''\footnote{A useful analogy can be drawn between weak freeze-out in the viscously-expanding accretion disk of a NS merger, and that which occurs in the expanding Universe during the first minutes following the Big Bang.  However, unlike a NS merger, the Universe freezes-out proton-rich, due to the lower densities (which favor proton-forming reactions over the neutron-forming ones instead favored under conditions of high electron degeneracy).} of weak interactions, leading to the winds being neutron-rich \citep{Metzger+08c,Metzger+09}.  Neutron-rich mater is shielded within the degenerate disk midplane, being ejected only once the disk radius has become large enough, and the neutrino luminosity low enough, that weak interactions no longer appreciably raise $Y_e$ in the outflow.  

These early estimates were followed by two-dimensional, axisymmetric hydrodynamical models of the disk evolution, which show that, in the case of prompt BH formation, the electron fraction of the disk outflows lies in the range $Y_e \sim 0.2\mbox{\,--\,}0.4$ \citep{Fernandez&Metzger13,Just+15}, sufficient to produce the entire mass range of $r$-process elements \citep{Just+15,Wu+16}.  The total fraction of the disk mass which is unbound by these ``viscously-driven'' winds ranges from $\sim 5\%$ for a slowly spinning BH, to $\sim 30\%$ for high BH spin $\chi_{\rm BH} \simeq 0.95$ \citep{Just+15,Fernandez+15b}; see also \cite{Kiuchi+15}, who simulated the long-term evolution of BH-NS disks but without following the electron fraction evolution.  These large disk ejecta fractions and neutron-rich ejecta were confirmed by the first 3D GRMHD simulations of the long-term disk evolution \citep{Siegel&Metzger17,Siegel&Metzger18,Fernandez+19}, with \citet{Siegel&Metzger17} finding that up to 40\% of the initial torus may be unbound.  The velocity and composition of magnetized disk outflows appears to be sensitive to the strength and geometry of the large-scale net magnetic flux threading the accretion disk \citep{Fernandez+19,Christie+19}.

An even larger fraction of the disk mass (up to $\sim 90\%$) is unbound when the central remnant is a long-lived hypermassive or supramassive NS instead of a BH, due to the presence of a hard surface and the higher level of neutrino irradiation from the central remnant \citep{Metzger&Fernandez14,Fahlman&Fernandez19}.  A longer-lived remnant also increases the electron fraction of the ejecta, which increases monotonically with the lifetime of the HMNS (Fig.~\ref{fig:Lippuner}).  Most of the ejecta is lanthanide-free ($Y_e \gtrsim 0.3$) if the NS survives longer than about 300 ms \citep{Metzger&Fernandez14,Kasen+15,Lippuner+17}.  Even when BH formation is prompt, simulations with Monte Carlo radiation transport included find that the earliest phases of disk evolution can produce at least a modest quantity of high-$Y_e$ material \citep{Miller+19}.

\begin{figure}[!t]

\includegraphics[width=0.5\textwidth]{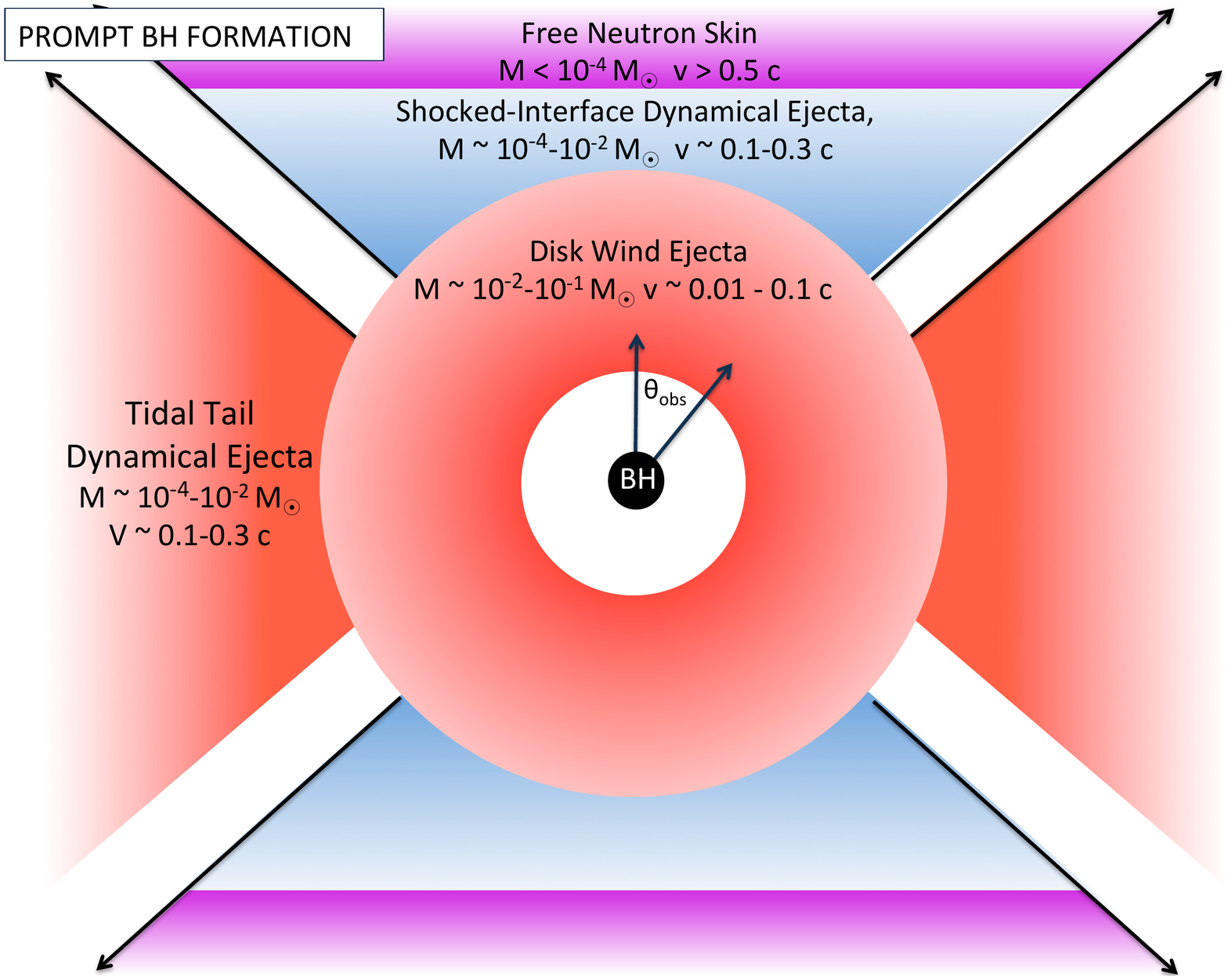}
\includegraphics[width=0.5\textwidth]{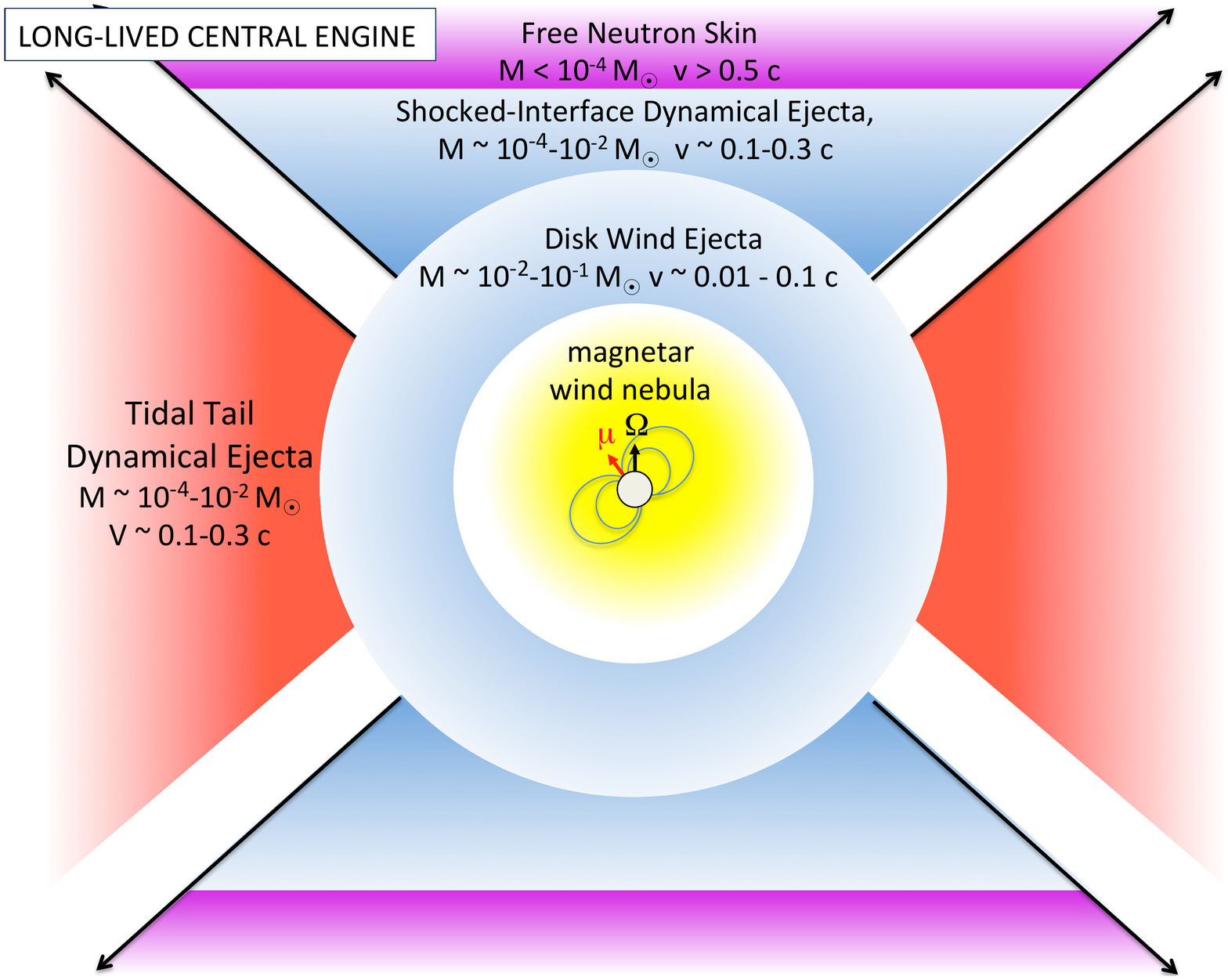}
\caption{Different components of the ejecta from NS-NS mergers and the possible dependence of their kilonova emission on the observer viewing angle, $\theta_{\rm obs}$, relative to the binary axis, in the case of a relatively prompt BH formation (left panel) and a long-lived magnetar remnant (right panel).  In both cases, the dynamical ejecta in the equatorial plane is highly neutron-rich ($Y_e \lesssim 0.1$), producing lanthanides and correspondingly ``red'' kilonova emission peaking at NIR wavelengths.  Mass ejected dynamically in the polar directions may be sufficiently neutron-poor ($Y_e \gtrsim 0.3$) to preclude Lanthanide production, powering ``blue'' kilonova emission at optical wavelengths (although this component may be suppressed if BH formation is extremely prompt).  The outermost layers of the polar ejecta may contain free neutrons, the decay of which powers a UV transient lasting a few hours following the merger (Sect.~\ref{sec:neutrons}).  Re-heating of the ejecta by a delayed relativistic outflow (e.g. the GRB jet or a wind from the magnetar remnant) may also contribute to early blue emission (Sect.~\ref{sec:cocoon}).  The innermost ejecta layers originate from accretion disk outflows, which may emerge more isotropically.  When BH formation is prompt, the disk wind ejecta is mainly neutron-rich, powering red kilonova emission \citep{Fernandez&Metzger13,Just+15,Wu+16,Siegel&Metzger17}.  If the NS remnant is instead long-lived relative to the disk lifetime, then neutrino emission can increase $Y_e$ sufficiently to suppress Lanthanide production and result in blue disk wind emission (Fig.~\ref{fig:Lippuner}; e.g.~\citealt{Metzger&Fernandez14,Perego+14}).  Energy input from the central accreting BH or magnetar remnant enhance the kilonova luminosity compared to that exclusively from radioactivity (Sect.~\ref{sec:engine}).    }
\label{fig:schematic}
\end{figure}

The mass ejected by the late disk wind can easily be comparable to, or larger than, that in the dynamical ejecta (e.g.~\citealt{Wu+16}, their Fig.~1).  Indeed, the total ejecta mass inferred for GW170817 greatly exceeds that of the dynamical ejecta found in merger simulations (Fig.~\ref{fig:Siegel}), but is consistent in both its mass and velocity with originating from a disk wind (e.g.~\citealt{Siegel&Metzger17}).  As the disk outflows emerge after the dynamical ejecta, the disk outflow material will be physically located behind the latter (Fig.~\ref{fig:schematic}).  

Beyond the dynamical and disk wind ejecta, other ejecta sources have been proposed, though these remain more speculative because the physical processes at work are less robust.  Mass loss may occur from the differentially rotating NS during the process of angular momentum redistribution \citep{Fujibayashi+18,Radice+18}.  However, the details of this mechanism and its predictions for the ejecta properties depend sensitively on the uncertain physical source and operation of the ``viscosity'' currently put into the simulations by hand; unlike the quasi-Keplerian accretion disk on larger radial scales, the inner regions of the NS remnant possess a positive shear profile $d \Omega/dr > 0$ are therefore not unstable to the magneto-rotational instability. 

 Outflows can also occur from the HMNS/SMNS or stable NS remnant as it undergoes Kelvin-Helmholtz contraction and neutrino cooling over a timescale of seconds.  At a minimum there will be outflows driven from the NS surface by neutrino heating \citep{Dessart+09}, which typically will possess a relatively low mass-loss rate $\lesssim 10^{-3}M_{\odot}$ s$^{-1}$ and low asymptotic velocity $\sim 0.1$ c.   
However, the NS remnant possesses an ordered magnetic field of strength $\sim 10^{14}-10^{15}$ G, then the mass-loss rate and velocity of such an outflow is substantially enhanced by the centrifugal force along the open magnetic field lines (e.g.~\citealt{Thompson+04}).  \citep*{Metzger+18} argue that such a magnetar wind, from a HMNS of lifetime $\sim 0.1-1$ s, was responsible for the fastest ejecta in GW170817 (Sect.~\ref{sec:170817}).  While the presence of an ordered magnetic field of this strength is physically reasonable, its generation from the smaller scale magnetic field generated during the merger process has yet to be conclusively demonstrated by numerical simulations.

\subsection{Ejecta Opacity}
\label{sec:opacity}

\begin{table}[!t]
\caption{Colors and Sources of Kilonova Ejecta in NS-NS Mergers \label{table:opacity}}

\begin{tabular}{ccccccc}

Color & $Y_e$ & $A_{\rm max}$$^{(a)}$ & $\kappa$$^{(b)}$ (cm$^{2}$ g$^{-1}$) & Main Sources  \\

\hline
Red & $\lesssim 0.2$ & $\sim 200$ & $\sim 30$ &  Tidal Tail Dynamical\\
& & & &  Disk Wind (Prompt BH/HMNS) \\
Blue/Purple & $0.25-0.35$ & $\sim$ 130 & $\sim 3$ & Shock-Heated Dynamical\\
&&&& Disk Wind (HMNS/SMNS) \\
&&&& Magnetar Wind\\
&&&& ``Viscous" Outflows/Spiral Arm \\
Blue  & $\sim0.40$ & $\lesssim$ 100 & $\sim 1$ & Disk Wind (SMNS/Stable NS) \\
\hline \\
\end{tabular}
\\
$^{(a)}$Maximum atomic mass of nuclei synthesized.  $^{(b)}$Effective gray opacity (\citealt{Tanaka+19}). 
\end{table}

\begin{figure}[!t]
\includegraphics[width=1.0\textwidth]{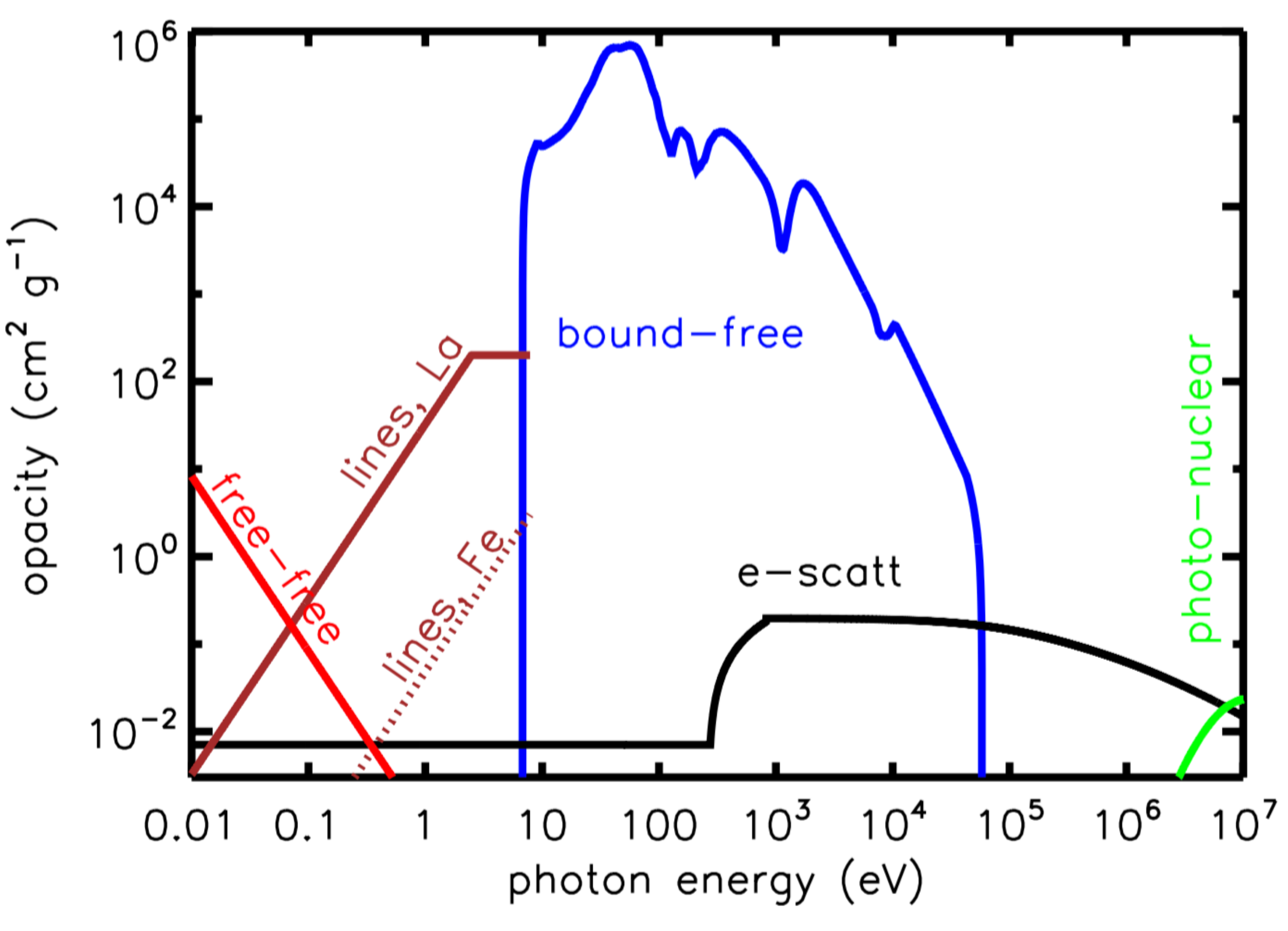}
\caption{Schematic illustration of the opacity of the NS merger ejecta as a function of photon energy at a fixed epoch near peak light.  The free-free opacity (red line) is calculated assuming singly-ionized ejecta of temperature $T = 2\times 10^{4}$ K and density $\rho = 10^{-14}$ g cm$^{-3}$, corresponding to the mean properties of 10$^{-2} M_{\odot}$ of ejecta expanding at $v = 0.1$ c at $t =$ 3 days.  Line opacities of iron-like elements and lanthanide-rich elements are approximated from Figures 3 and 7 of \cite{Kasen+13}.  Bound-free opacities are estimated as that of neutral iron \citep{Verner+96}, which should crudely approximate the behavior of heavier $r$-process elements.  Electron scattering opacity accounts for the Klein-Nishina suppression at energies $\gg m_e c^{2}$ and (very schematically) for the rise in opacity that occurs above the keV energy scale due to all electrons (including those bound in atoms) contributing to the scattering opacity when the photon wavelength is smaller than the atomic scale.  At the highest energies, opacity is dominated by pair creation by gamma-rays interacting with the electric fields of nuclei in the ejecta (shown schematically for Xenon, $A = 131$, $Z = 54$).  Not included are possible contributions from $r$-process dust; or $\gamma-\gamma$ pair creation opacity at photon energies $\gg m_e c^{2} \sim 10^{6}$ eV  (see Eq.~\ref{eq:compactness}).  }
\label{fig:opacities}
\end{figure}

It is no coincidence that kilonova emission is centered in the optical/IR band, as this is the first spectral window through which the expanding merger ejecta becomes transparent.  Figure \ref{fig:opacities} illustrates semi-quantitatively the opacity of NS merger ejecta near peak light as a function of photon energy. 

 At the lowest frequencies (radio and far-IR), free-free absorption from ionized gas dominates, as shown with a red line in Fig.~\ref{fig:opacities}, and calculated for the approximate ejecta conditions three days post merger.  As the ejecta expands, the free-free opacity will decrease rapidly due to the decreasing density $\propto \rho \propto t^{-3}$ and the fewer number of free electrons as the ejecta cools and recombines.  

At near-IR/optical frequencies, the dominant source of opacity is a dense forest of line (bound-bound) transitions.  The magnitude of this effective continuum opacity is determined by the strengths and wavelength density of the lines, which in turn depend sensitively on the ejecta composition.  If the ejecta contains elements with relatively simple valence electron shell structures, such as iron, then the resulting opacity is comparatively low (dashed brown line), only moderately higher than the Fe-rich ejecta in Type Ia supernovae \citep{Pinto&Eastman00}.  On the other hand, if the ejecta also contains even a modest fraction of elements with partially-filled f-shell valence shells, such as those in the lanthanide and actinide group, then the opacity can be an order of magnitude or more higher \citep{Kasen+13,Tanaka&Hotokezaka13,Fontes+16,Fontes+17,Even+19}.  In both cases, the opacity rises steeply from the optical into the UV, due to the increasing line density moving to higher frequencies.  

Considerable uncertainty remains in current calculations of the La/Ac opacities because the atomic states and line strengths of these complex elements are not measured experimentally.  Theoretically, such high$-Z$ atoms represent an unsolved problem in N-body quantum mechanics, with statistical models that must be calibrated to experimental data.  Beyond identifying the line transitions themselves, there is considerably uncertainty in how to translate these data into an effective opacity.  The commonly employed ``line expansion opacity'' formalism \citep{Pinto&Eastman00}, based on the Sobolev approximation and applied to kilonovae by \cite{Barnes&Kasen13} and \cite{Tanaka&Hotokezaka13}, may break down if the line density is sufficiently high that the wavelength spacing of strong lines becomes comparable to the intrinsic thermal) width of the lines \citep{Kasen+13,Fontes+16,Fontes+17}.  Nevertheless, the qualitative dichotomy between the opacity of La/Ac-free and La/Ac-bearing ejecta is likely to be robust and will imprint diversity in the kilonova color evolution (Sect.~\ref{sec:blue}).\footnote{Another uncertainty arises because, at low temperatures $\lesssim 10^{3}$ K, the ejecta may condense from gaseous to solid phase \citep{Takami+14,Gall+17}.  The formation of such $r$-process dust could act to either increase or decrease the optical/UV opacity, depending on uncertain details such as when the dust condenses and how quickly it grows.  Dust formation is already complex and poorly understood in less exotic astrophysical environments \citep{Cherneff&Dwek09,Lazzati&Heger16}.  }        

Despite the strong time- and wavelength-dependence of the opacity, for purposes of an estimate it is reasonable to model the kilonova using a constant effective ``grey'' (wavelength-independent) opacity, $\kappa$.  Including a large range of $r$-process nuclei in their analysis, \citet{Tanaka+19} found (for temperatures $T = 5-10\times 10^{3}$ K characteristic of the ejecta near the time of peak emission) values of $\kappa \lesssim 20-30$ cm$^{2}$ g$^{-1}$ for $Y_{e} \lesssim 0.2$ (sufficient neutrons for the $r$-process to extend up to or beyond the third abundance peak at $A \sim 195$ with a large lanthanide mass fraction), $\kappa \sim 3-5$ cm$^{2}$ g$^{-1}$ for $Y_{e} \approx 0.25-0.35$ ($r$-process extending only to the second abundance peak $A \sim 130$ with a small or zero lanthanide mass fraction) and $\kappa \sim 1$ cm$^{2}$ g$^{-1}$ for $Y_{e} \approx 0.40$ (mainly neutron-rich Fe-group nuclei and a weak $r$-process).  The approximate opacity range corresponding to different ejecta composition is summarized in Table \ref{table:opacity}.

 Throughout the far UV and X-ray bands, bound-free transitions of the partially neutral ejecta dominates the opacity (blue line in Fig.~\ref{fig:opacities}).  This prevents radiation from escaping the ejecta at these frequencies, unless non-thermal radiation from the central magnetar or BH remnant remains luminous enough to re-ionize the ejecta at late times (Sect.~\ref{sec:magnetar}).  
\citet{Margalit+18} find that an X-ray luminosity $L_{\rm X} \gtrsim 10^{42}$ erg/s would be required to ionize $M_{\rm ej} \sim 10^{-3}M_{\odot}$ of Fe-like ejecta expanding at $\approx 0.2$ c at $t \sim 1$ day after the merger.  However, a substantially higher luminosity $L_{\rm X} \gtrsim 10^{44}-10^{45}$ erg s$^{-1}$ would be needed\footnote{This is also consistent with upper limits on non-afterglow contributions to the X-ray emission from a central remnant in GW170817 (for which $M_{\rm ej} \gtrsim 0.03M_{\odot}$; e.g.~\citealt{Margutti+17}).} to ionize the greater expected quantity of disk wind ejecta $M_{\rm ej} \sim 0.01-0.1M_{\odot}$, especially considering that the bound-free opacity of $r$-process nuclei will be even higher than Fe.  Such high X-ray luminosities are too large to be powered by fall-back accretion onto the central remnant from the merger ejecta, but could be achievable from the rotational energy input from a long-lived magnetar remnant (Fig.~\ref{fig:heating}, right panel).  

At hard X-rays and gamma-ray energies, electron scattering, with Klein-Nishina corrections, provides an important opacity (which becomes highly inelastic at energies $\gtrsim m_e c^{2}$).  For gamma-ray energies $\gtrsim m_e c^{2}$, the opacity approaches a constant value $\kappa_{A\gamma} \approx \alpha_{\rm fs}\kappa_{T}(Z^{2}/A)$ due to electron/positron pair creation on nuclei, where $\alpha_{\rm fs} \simeq 1/137$, and $A$ and $Z$ are the nuclear mass and charge, respectively (e.g.~\citealt{Zdziarski&Svensson89}).  For $r$-process nuclei with $Z^{2}/A \gtrsim 10-20$ this dominates inelastic scattering at energies $\gtrsim 10$ MeV.   The low opacity $\lesssim 0.1$ cm$^{2}$ g$^{-1}$ in the $\sim$ MeV energy range implies that gamma-rays released by radioactive decay of $r$-process elements largely escape the ejecta prior to the optical/NIR peak without their energy being thermalized (Sect.~\ref{sec:rprocessheating}).  

Gamma-rays with very high energies $\gtrsim (m_e c^{2})^{2}/h\nu_{\rm s} \sim 0.3{\rm TeV}(h\nu_{\rm s}/1{\rm eV})^{-1}$ can also create electron/positron pairs by interacting with (more abundant) lower energy optical or X-ray photons of energy $h\nu_{\rm s} \ll m_e c^{2}$.  The $\gamma-\gamma$ pair creation optical depth through the ejecta of radius $R = vt$ is roughly given by
\begin{eqnarray}
\tau_{\gamma-\gamma} &\simeq& \frac{U_{\rm rad}\sigma_T R}{h\nu_{\rm opt}} \simeq \frac{L \sigma_T}{4\pi R  h\nu_{\rm opt} c}  \nonumber \\
 &\approx& 2\times 10^{3}\left(\frac{L}{10^{42}\,{\rm erg\,s^{-1}}}\right)\left(\frac{v}{0.2\,\rm c}\right)^{-1}\left(\frac{h\nu_{\rm s}}{\rm 1\,eV}\right)^{-1}\left(\frac{t}{\rm 1\,day}\right)^{-1},
\label{eq:compactness}
\end{eqnarray}
where $U_{\rm rad} \simeq L/(4\pi R^{2}c)$ is the energy density of the seed photons of luminosity $L$.  The fact that $\tau_{\gamma-\gamma} \gg 1$ for characteristic thermal luminosities $\sim 10^{40}-10^{42}$ erg s$^{-1}$ shows that $\sim$ GeV-TeV photons from a putative long-lived central engine (e.g. millisecond magnetar; Sect.~\ref{sec:magnetar}) will be trapped for days to weeks after the merger.  Prompt TeV emission from a NS-NS merger is thus unlikely to come from the merger remnant, but still could be generated within the relativistically-beamed GRB jet on much larger physical scales.

\section{Unified Toy Model}
\label{sec:model}

Thermal emission following a NS-NS or BH-NS merger (a ``kilonova'', broadly defined) can be powered by a variety of different energy sources, including radioactivity and central engine activity (see Fig.~\ref{fig:heating} for a summary of different heating sources).  This section describes a simple model for the evolution of the ejecta and its radiation, which we use to motivate the potential diversity of kilonova emission properties.  Though ultimately no substitute for full, multi-dimensional, multi-group radiative transfer, this 1D toy model does a reasonable job at the factor of a few level.  Some sacrifice in accuracy may be justified in order to facilitate a qualitative understanding, given the other uncertainties on the mass, heating rate, composition, and opacity of the ejecta.    

Following the merger, the ejecta velocity structure approaches one of homologous expansion, with the faster matter lying ahead of slower matter \citep{Rosswog+14}.  We approximate the distribution of mass with velocity greater than a value $v$ as a power-law,
\be
M_{v} = M(v/v_{\rm 0})^{-\beta},\,\,\,\, v \ge v_{0},
\label{eq:veldist}
\ee
where $M$ is the total mass, $v_{0} \approx 0.1$ c is the average ($\sim$ minimum) velocity.  We adopt a fiducial value of $\beta \approx 3$, motivated by a power-law fit to the dynamical ejecta in the numerical simulations of \citep{Bauswein+13}.  In general the velocity distribution derived from numerical simulations cannot be fit by a single power-law (e.g., Fig.~3 of \citealt{Piran+13}), but the following analysis can be readily extended to the case of an arbitrary velocity distribution.

In analogy with Eq.~(\ref{eq:tdiff}), radiation escapes from the mass layer $M_{v}$ on the diffusion timescale
\be
t_{d,v} \approx \frac{3 M_{v} \kappa_{v}}{4\pi \beta R_v c} \underset{R_v = vt}= \frac{M_{v}^{4/3}\kappa_{v}}{4\pi M^{1/3} v_{0} t c},
\label{eq:tdv}
\ee
where $\kappa_v$ is the opacity of the mass layer $v$ and in the second equality makes use of Eq.~(\ref{eq:veldist}) with $\beta = 3$.  Equating $t_{d,v} = t$ gives the mass depth from which radiation peaks for each time $t$,
\be
M_{v}(t)  = \left\{
\begin{array}{lr}
 M(t/t_{\rm peak})^{3/2}
, &
t < t_{\rm peak}\\
M &
t > t_{\rm peak} \\
\end{array}
\label{eq:Mv}
\right. ,
\ee
where $t_{\rm peak}$ is the peak time for diffusion out of the whole ejecta mass, e.g., Eq.~(\ref{eq:tpeak}) evaluated for $v = v_0$.  Emission from the outer layers (mass $M_v < M$) peaks first, while the luminosity of the innermost shell of mass $\sim M$ peaks at $t = t_{\rm peak}$.  The deepest layers usually set the peak luminosity of the total light curve, except when the heating rate and/or opacity are not constant with depth (e.g.~ if the outer layers are free neutrons instead of $r$-process nuclei; Sect.~\ref{sec:neutrons}).

As the ejecta expands, the radius of each layer of depth $M_{v}$ of mass $\delta M_{v}$ evolves according to 
\be \frac{dR_v}{dt} = v.
\ee
The thermal energy $\delta E_v$ of the layer evolves according to
\be \frac{d(\delta E_v)}{dt} = -\frac{\delta E_v}{R_v}\frac{dR_v}{dt} - L_v + \dot{Q},
\label{eq:dEdt}
\ee
where the first term accounts for losses due to PdV expansion in the radiation-dominated ejecta.  The second term in Eq.~(\ref{eq:dEdt}), 
\be
L_{v} = \frac{\delta E_v}{t_{d,v} + t_{lc,v}},
\ee 
accounts for radiative losses (the observed luminosity) and $t_{lc,v} = R_v/c$ limits the energy loss time to the light crossing time (this becomes important at late times when the layer is optically thin).  The third term in Eq.~(\ref{eq:dEdt}),
\be
\dot{Q}(t) = \dot{Q}_{r,v} + \dot{Q}_{\rm mag} + \dot{Q}_{\rm fb}
\ee
accounts for sources of heating, including radioactivity ($\dot{Q}_{r,v}$; Sect.~\ref{sec:rprocessheating}), a millisecond magnetar ($\dot{Q}_{\rm mag}$; Sect.~\ref{sec:magnetar}) or fall-back accretion ($\dot{Q}_{\rm fb}$; Sect.~\ref{sec:fallback}).  The radioactive heating rate, being intrinsic to the ejecta, will in general vary between different mass layers $v$.  In the case of magnetar or accretion heating, radiation must diffuse from the central cavity through the entire ejecta shell (Fig.~\ref{fig:schematic}, right panel).

One must in general also account for the evolution of the ejecta velocity (Eq.~\ref{eq:veldist}) due to acceleration by pressure forces.  For radioactive heating, the total energy input $\int \dot{Q}_{r,v}dt$ is less than the initial kinetic energy of the ejecta \citep{Metzger+10b,Rosswog+12,Desai+19}, in which case changes to the initial velocity distribution (Eq.~\ref{eq:veldist}) are safely ignored.  However, free expansion is not a good assumption when there is substantial energy input from a central engine.  In such cases, the velocity $v_0$ of the central shell (of mass $M$ and thermal energy $E_{v_0}$) is evolved separately according to
\be
\frac{d}{dt}\left(\frac{M v_0^{2}}{2}\right) = Mv_0 \frac{dv_0}{dt} = \frac{E_{v_{0}}}{R_0}\frac{dR_0}{dt},
\label{eq:dvdt}
\ee
where the source term on the right hand side balances the PdV \emph{loss} term in the thermal energy equation (\ref{eq:dEdt}), and $R_0$ is the radius of the inner mass shell.  Equation (\ref{eq:dvdt}) neglects two details: (1) special relativistic effects, which become important for low ejecta mass $\lesssim 10^{-2}M_{\odot}$ and the most energetic magnetar engines \citep{Zhang13,Gao+13,Siegel&Ciolfi16a,Siegel&Ciolfi16b}; (2) the secondary shock driven through the outer ejecta layers by the nebula inflated by a long-lived central engine (e.g.~\citealt{Kasen+16}) and its effects on the outer velocity distribution (e.g.~\citealt{Suzuki&Maeda17}).

Under the idealization of blackbody emission, the temperature of the thermal emission is 
\be T_{\rm eff} = \left(\frac{L_{\rm tot}}{4\pi \sigma R_{\rm ph}^{2}}\right)^{1/4},
\ee where $L_{\rm tot} = \int_{v_0}L_v \frac{dM_v}{dv}dv \simeq \sum_{v} (L_v \delta M_v)$ is the total luminosity (summed over all mass shells).  The radius of the photosphere $R_{\rm ph} (t)$ is defined as that of the mass shell at which the optical depth $\tau_v = \int_{v_0}\frac{dM_v}{dv}\frac{\kappa_v}{4\pi R_{\rm v}^2}dv \simeq \sum_{v} \left(\frac{\kappa_v \delta M_v}{4\pi R_v^{2}}\right) = 1$.  The flux density of the source at photon frequency $\nu$ is given by
\be
F_{\nu}(t) = \frac{2\pi h \nu^{3}}{c^{2}}\frac{1}{\exp\left[h\nu/kT_{\rm eff}(t)\right]-1}\frac{R_{\rm ph}^{2}(t)}{D^{2}},
\ee
where $D$ is the source luminosity distance.  We have neglected cosmological effects such as K-corrections, but these can be readily included.

For simplicity in what follows, we assume that the opacity $\kappa_v$ of each mass layer depends entirely on its composition, i.e. we adopt a temperature-independent grey opacity.  The most relevant feature of the composition is the mass fraction of lanthanide or actinide elements, which in turn depends most sensitively on the ejecta $Y_e$ (Sect.~\ref{sec:opacity}).  Following \citet[Table \ref{table:opacity}]{Tanaka+19}, one is motivated to take
\be
\kappa_v(Y_e) = \left\{
\begin{array}{lr}
 20-30\,{\rm cm^{2}\,g^{-1}} 
, &
Y_{e} \lesssim 0.2 \,\,\,\,\,({\rm Red}) \\
3-5\,{\rm cm^{2}\,g^{-1}}  &
Y_e \approx 0.25-0.35 \,\,\,\,\,({\rm Blue/Purple}) \\
1\,{\rm cm^{2}\,g^{-1}} &
Y_e \approx 0.4\,\,\,\,\, ({\rm Blue}) \\
\end{array}
\label{eq:kappav}
\right. ,
\ee
smoothly interpolating when necessary.  We caution that these values were calculated by \citep{Tanaka+19} for ejecta temperatures in the range $5-10\times 10^{3}$ K, i.e. similar to those obtained close to peak light, and therefore may not be appropriate much earlier (the first hours) or later (nebular phase).

The full emission properties are determined by solving Eq.~(\ref{eq:dEdt}) for $\delta E_v$, and hence $L_v$, for a densely sampled distribution of shells of mass $\delta M_v$ and velocity $v > v_0$.  When considering  radioactive heating acting alone, one can fix the velocity distribution (Eq.~\ref{eq:veldist}).  For an energetic engine, the velocity of the central shell is evolved simultaneously using Eq.~(\ref{eq:dvdt}).  

As initial conditions at the ejection radius $R(t = 0) \approx 10-100$ km, it is reasonable to assume that the initial thermal energy is comparable to its final kinetic energy, $\delta E_{v}(t = 0) \sim (1/2)\delta M_v v^2(t=0)$.  If the ejecta expands freely from the site of ejection, the predicted light curves are largely insensitive to the details of this assumption because the initial thermal energy is anyways quickly removed by adiabatic expansion.  Sect.~\ref{sec:cocoon} explores an exception to this rule when the ejecta is re-heated by shock interaction with a delayed outflow from the central engine.    

\subsection{$R$-Process Heating}
\label{sec:rprocessheating}
At a minimum, the ejecta is heated by the radioactive decay of heavy $r$-process nuclei.  This occurs at a rate
\be
\dot{Q}_{r,v} = \delta M_v X_{r,v} \dot{e}_r(t),
\label{eq:qdotr}
\ee
where $X_{r,v}$ is the $r$-process mass fraction in mass layer $M_v$ and $e_r$ is the specific heating rate.  For neutron-rich ejecta ($Y_e \lesssim 0.2$), the latter can be reasonably well approximated by the fitting formula \citep{Korobkin+12}
\be
\dot{e}_r  = 4\times 10^{18}\epsilon_{th,v} \left(0.5-\pi^{-1}\arctan[(t-t_0)/\sigma]\right)^{1.3}\,{\rm erg\,s^{-1}\,g^{-1}},
\label{eq:edotr}
\ee
where $t_0 = 1.3$ s and $\sigma = 0.11$ s are constants, and $\epsilon_{\rm th,m}$ is the thermalization efficiency (see below).  Equation (\ref{eq:edotr}) predicts a constant heating rate for the first second (while neutrons are being consumed during the $r$-process), followed by a power-law decay at later times as nuclei decay back to stability \citep{Metzger+10, Roberts+11}; see Figs.~\ref{fig:heating} and \ref{fig:wu}.  The latter is reasonably well approximated by the expression
\be
\dot{e}_r \underset{t \gg t_0}\approx 2\times 10^{10}\epsilon_{th,v} \left(\frac{t}{1\,{\rm day}}\right)^{-1.3}{\rm erg\,s^{-1}\,g^{-1}}
\label{eq:rshort}
\ee
Using equations (\ref{eq:tpeak}) and (\ref{eq:Arnett}) the peak luminosity can be estimated as
\begin{eqnarray}
L_{\rm peak} &\approx& M\dot{e}_r(t_{\rm peak}) \nonumber \\
& \approx& 10^{41}{\rm erg\,s^{-1}}\left(\frac{\epsilon_{th,v}}{0.5}\right)\left(\frac{M}{10^{-2}M_{\odot}}\right)^{0.35}\left(\frac{v}{0.1c}\right)^{0.65}\left(\frac{\kappa }{1\,{\rm cm^{2}\,g^{-1}}}\right)^{-0.65}.
\end{eqnarray}
Given the reasonably large range in values allowed in NS-NS or BH-NS mergers, $M \sim 10^{-3}-0.1M_{\odot}$, $\kappa \sim 0.5-30$ cm$^{2}$ g$^{-1}$, $v \sim 0.1-0.3$ c, one can have $L_{\rm peak} \approx 10^{39}-10^{42}$ erg s$^{-1}$.

The time dependence of $\dot{q}_r$ is more complicated for higher $0.2 \lesssim Y_e \lesssim 0.4$, with 'wiggles' caused by the heating rate being dominated by a few discrete nuclei instead of the large statistical ensemble present at low $Y_e$ \citep{Korobkin+12,Martin+15}.  However, when averaged over a realistic $Y_e$ distribution, the heating rate on timescales of days-weeks (of greatest relevance to the peak luminosity; Eq.~\ref{eq:Arnett}), is constant to within a factor of a few for $Y_e \lesssim 0.4$  (\citealt{Lippuner&Roberts15,Wu+19a}; see Fig.~\ref{fig:wu}).  The radioactive decay power is sensitive to various uncertainties in the assumed nuclear physics (nuclear masses, cross sections, and fission fragment distribution) at the factor of a few level (e.g.~\citealt{Wu+19a})\footnote{The $r$-process abundance pattern itself is much more sensitive to these nuclear physics uncertainties \citep{Eichler+15,Wu+16,Mumpower+16}.}, a point we shall return to when discussing GW170817 (Sect.~\ref{sec:170817}).

Radioactive heating occurs through a combination of $\beta$-decays, $\alpha$-decays, and fission \citep{Metzger+10,Barnes+16,Hotokezaka+16b}.  The thermalization efficiency, $\epsilon_{th,v}$, depends on how these decay products share their energies with the thermal plasma.  Neutrinos escape from the ejecta without interacting; $\sim$ MeV gamma-rays are trapped at early times ($\lesssim 1$ day), but leak out at later times given the low Klein-Nishina-suppressed opacity (Fig.~\ref{fig:opacities}; \citealp{Hotokezaka+16b,Barnes+16}).  $\beta$-decay electrons, $\alpha$-particles, and fission fragments share their kinetic energy effectively with the ejecta via Coulomb collisions \citep{Metzger+10} and by ionizing atoms \citep{Barnes+16}.  For a fixed energy release rate, the thermalization efficiency is smallest for $\beta$-decay, higher for $\alpha$-decay, and the highest for fission fragments.  The thermalization efficiency of charged particles also depends on the magnetic field orientation within the ejecta, since the particle Larmor radius is generally shorter than the mean free path for Coulomb interactions.  Because the actinide yield around mass number $A \sim 230$ varies significantly with the assumed nuclear mass model, \cite{Barnes+16}  finds that the effective heating rate including thermalization efficiency can vary by a factor of 2\,--\,6, depending on time. 

\cite{Barnes+16} find that the combined efficiency from all of these processes typically decreases from $\epsilon_{ th,v} \sim 0.5$ on a timescale of 1 day to $\sim 0.1$ at $t \sim 1$ week (their Fig.~13). In what follows, we adopt the fit provided in their Table 1,
\be
\epsilon_{th,v}(t) = 0.36\left[\exp(-a_v t_{\rm day})  + \frac{{\rm ln}(1+2b_v t_{\rm day}^{d_v})}{2b_v t_{\rm day}^{d_v}}\right],
\label{eq:eth}
\ee
where $t_{\rm day} = t/1$ day, and $\{a_v,b_v,d_v\}$ are constants that will in general depend on the mass and velocity of the layer under consideration.  For simplicity, we adopt fixed values of $a_v = 0.56, b_v = 0.17, c_v = 0.74$, corresponding to a layer with $M = 10^{-2}M_{\odot}$ and $v_0 = 0.1$ c.

As we shall discuss, the luminosity and color evolution of kilonovae encode information on the total quantity of $r$-process ejecta and, in principle, the abundance of lanthanide/actinide elements.  However, insofar as the lanthanides cover the atomic mass range $A \sim 140-175$, kilonova observations at peak light do not readily probe the creation of the heaviest elements, those near the third $r$-process peak ($A \sim 195$) and the transuranic elements ($A \gtrsim 240$).  

One avenue for probing the formation of ultra-heavy elements is by the light curve's decay at late times, weeks to months after maximum light.  At such late times the radioactive heating is often dominated by a few discrete isotopes with well-measured half-lifes (e.g.~$^{223}$Ra [$t_{1/2} = 11.4$ d], $^{225}$Ac [$t_{1/2} = 10.0$ d], $^{225}$Ra [$t_{1/2} = 14.9$ d], $^{254}$Cf [$t_{1/2} = 60.5$ d]) which could produce distinctive features (e.g.~bumps or exponential decay-like features) in the bolometric light curve of characteristic timescale $\sim t_{1/2}$ \citep{Zhu+18,Wu+19a}, much in the way that the half-life of $^{56}$Co is imprinted in the decay of Type Ia supernovae.  The ability in practice to identify individual isotopes through this method will depend on accurate models for the ejecta thermalization entering the nebular phase \citep{Kasen&Barnes19,Waxman+19} as well as dedicated broad-band, multi-epoch follow-up of nearby kilonovae in the NIR (where most the nebular emission likely emerges) with sensitive facilities like the James Webb Telescope \citep{Kasliwal+19,Villar+18}.  Table \ref{table:halflives} compiles all $r$-process isotopes with half-lives in the range 10-100 day \citep{Wu+19a}.  

Gamma-ray decay lines from the $r$-process element decays escape the ejecta within days or less of the merger and could in principle be directly observed from an extremely nearby event $\lesssim 3-10$ Mpc with future gamma-ray satellites \citep{Hotokezaka+16,Korobkin+19}.  A related, but potentially more promising near-term strategy is a gamma-ray search for remnants of past NS mergers in our Galaxy \citep{Wu+19b,Korobkin+19}.  Among the most promising isotopes for this purpose is $^{126}$Sn, which has several lines in the energy range 415$-$695 keV and resides close to the second $r$-process peak, because of its half-life $t_{1/2} = 2.3\times 10^{5}$ yr is comparable to the ages of the most recent Galactic merger(s).  \citet{Wu+19b} estimate that multiple remnants are detectable as individual sources by next-generation gamma-ray satellites with line sensitivities $\sim 10^{-6}-10^{-8}$ $\gamma$ cm$^{-2}$ s$^{-1}$.

\begin{table}[!t]
\caption{$r$-Process Nuclei with Half-lives $t_{1/2} = 10-100$ d  \label{table:halflives}}

\begin{tabular}{ccccccc}

Isotope & Decay Channel & $t_{1/2}$ (days)  \\

\hline
$^{225}$Ra & $\beta^-$ & 14.9(2)  \\ 
  $^{225}$Ac & $\alpha\beta^-$ to $^{209}$Bi &  10.0(1) \\ \hline
  $^{246}$Pu & $\beta^-$ to $^{246}$Cm & 10.84(2) \\ \hline
  $^{147}$Nd & $\beta^-$ & 10.98(1) \\ \hline
  $^{223}$Ra & $\alpha\beta^-$  to $^{207}$Pb & 11.43(5) \\ \hline
  $^{140}$Ba & $\beta^-$  to $^{140}$Ce &  12.7527(23) \\ \hline
  $^{143}$Pr & $\beta^-$ & 13.57(2) \\ \hline   
  $^{156}$Eu & $\beta^-$ & 15.19(8) \\ \hline
  $^{191}$Os & $\beta^-$ & 15.4(1)  \\ \hline
  $^{253}$Cf & $\beta^-$ & 17.81(8) \\ 
  $^{253}$Es & $\alpha$ & 20.47(3) \\ \hline
  $^{234}$Th & $\beta^-$ to $^{234}$U & 24.10(3) \\ \hline
  $^{233}$Pa & $\beta^-$ & 26.975(13)  \\ \hline
  $^{141}$Ce & $\beta^-$ & 32.511(13)  \\ \hline
  $^{103}$Ru & $\beta^-$ & 39.247(3) \\ \hline
  $^{255}$Es & $\alpha\beta^-$ to $^{251}$Cf & 39.8(12) \\ \hline
  $^{181}$Hf & $\beta^-$ & 42.39(6)  \\ \hline
  $^{203}$Hg & $\beta^-$ & 46.594(12) \\ \hline
  $^{89}$Sr & $\beta^-$ & 50.563(25)  \\ \hline
  $^{91}$Y  & $\beta^-$ & 58.51(6)  \\ \hline      
  $^{95}$Zr  & $\beta^-$ & 64.032(6)  \\ 
  $^{95}$Nb  & $\beta^-$ & 34.991(6)  \\ \hline      
  $^{188}$W  & $\beta^-$ to $^{188}$Os & 69.78(5)  \\ \hline      
  $^{185}$W  & $\beta^-$ & 75.1(3)   \\ \hline 
  $^{254}$Cf & Fission & 60.5(2)\\
\end{tabular}
\\
Modified from Table II in \citep{Wu+19a}.\\
\end{table}

\subsubsection{Red Kilonova: Lanthanide-Bearing Ejecta}
\label{sec:red}

All NS-NS mergers, and the fraction of BH-NS mergers in which the NS is tidally disrupted before being swallowed by the BH, will unbind at least some highly neutron-rich matter ($Y_e \lesssim 0.25$) capable of forming heavy $r$-process nuclei.  This Lanthanide-bearing high-opacity material resides within the equatorially-focused tidal tail, or in more spherical outflows from the accretion disk (Fig.~\ref{fig:schematic}, top panel).  The disk outflows will contain a greater abundance of low-$Y_e \lesssim 0.25$ material in NS-NS mergers if the BH formation is prompt or the HMNS phase short-lived (Fig.~\ref{fig:schematic}, top panel).

\begin{figure}[!t]
\includegraphics[width=0.5\textwidth]{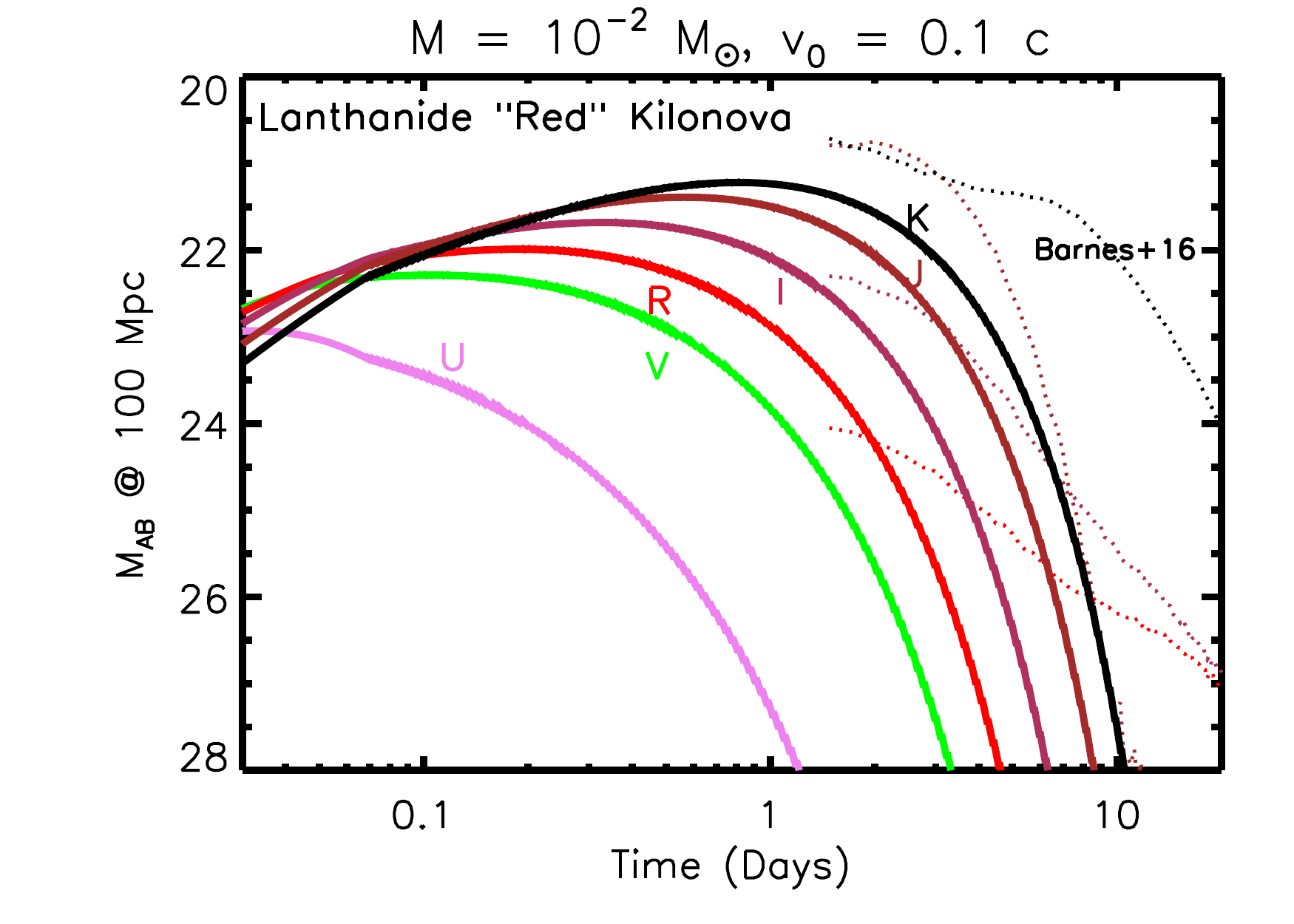}
\includegraphics[width=0.5\textwidth]{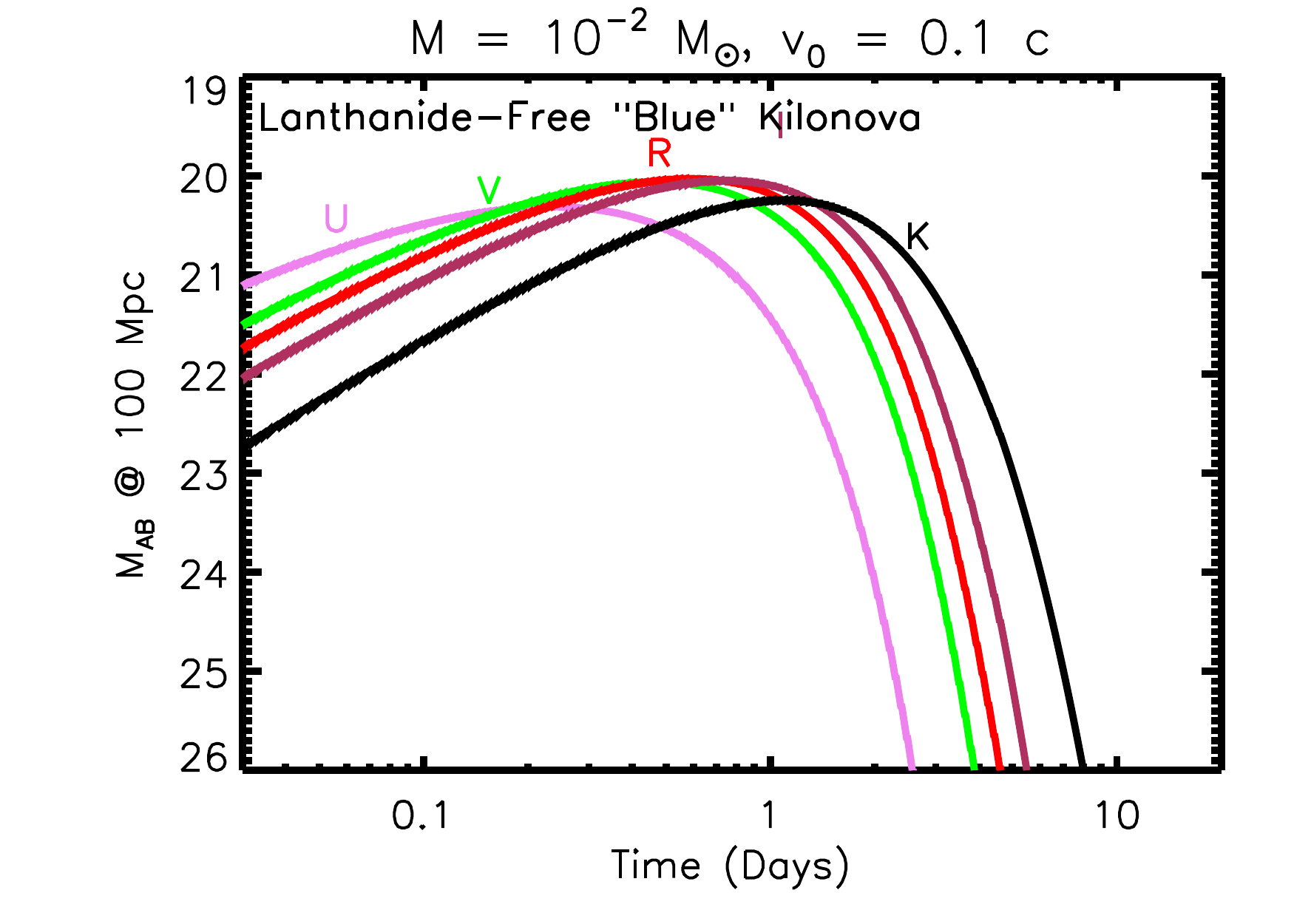}
\caption{Kilonova light curves in AB magnitudes for a source at 100 Mpc, calculated using the toy model presented in Sect.~\ref{sec:model}, assuming a total ejecta mass $M = 10^{-2}M_{\odot}$, minimum velocity $v_0 = 0.1$ c, and gray opacity $\kappa = 20$ cm$^{2}$ g$^{-1}$.  The left panel shows a standard ``red'' kilonova, corresponding to ejecta bearing Lanthanide elements, while the right panel shows a ``blue'' kilonova poor in Lanthanides ($\kappa = 1$ cm$^{2}$ g$^{-1}$).   Shown for comparison in the red kilonova case with dashed lines are models from \cite{Barnes+16} for $v = 0.1$ c and $M = 10^{-2}M_{\odot}$.  Depending on the relative speeds of the two components and the viewing angle of the observer, both red and blue emission components can be present in a single merger, originating from distinct portions of the ejecta (Fig.~\ref{fig:schematic}).}
\label{fig:vanilla}
\end{figure}

The left panel of Fig.~\ref{fig:vanilla} shows an example light curve of such a `red' kilonova, calculated using the toy model assuming an ejecta mass $M = 10^{-2}M_{\odot}$, opacity $\kappa = 20$ cm$^{2}$ g$^{-1}$, minimum velocity $v_0 = 0.1$ c, and velocity index $\beta =3$, at an assumed distance of 100 Mpc.  For comparison, dashed lines show light curves calculated from \cite{Barnes+16}, based on a full one-dimensional radiative transfer calculation, for similar parameters.  The emission is seen to peak at NIR wavelengths on a timescale of several days to a week at J and K bands (1.2 and 2.2 $\mu$m, respectively). 

One notable feature of these light curves is the significant suppression of the emission in the UV/optical wavebands $URV$ due to the high lanthanide opacity.  Here, the assumption of a gray opacity made in the toy model results in a substantial underestimation of the blueward suppression relative to that found by the full radiative transfer calculation \citep{Barnes+16}.  This difference results in part because the true line opacity increases strongly moving to higher frequencies due to the higher density of lines in the UV (Fig.~\ref{fig:opacities}).

\subsubsection{Blue Kilonova: Lanthanide-Free Ejecta }
\label{sec:blue}

In addition to the highly neutron-rich ejecta ($Y_e \lesssim 0.30$), some of the matter unbound from a NS-NS merger may contain a lower neutron abundance ($Y_e \gtrsim 0.30$) and thus will be free of Lanthanide group elements (e.g.~\citealt{Metzger&Fernandez14,Perego+14,Wanajo+14}).  This low-opacity ejecta can reside either in the polar regions, due to dynamical ejection from the NS-NS merger interface, or in more isotropic outflows from the accretion disk (e.g.~\citealt{Miller+19}).  The quantity of high-$Y_e$ matter will be greatest in cases when BH formation is significantly delayed relative to the lifetime of the accretion disk due to the strong neutrino luminosity of the NS remnant (Fig.~\ref{fig:schematic}, right panel).

The right panel of Fig.~\ref{fig:vanilla} shows a model otherwise identical to that in the left panel, but which assumes a lower opacity $\kappa = 1$ cm$^{2}$ g$^{-1}$ more appropriate to Lanthanide-free ejecta.  The emission now peaks at the visual bands R and I, on a timescale of about 1 day at a level 2\,--\,3 magnitudes brighter than the Lanthanide-rich case.  This luminous, fast-evolving visual signal was key to the discovery of the kilonova counterpart of GW170817 (Sect.~\ref{sec:170817}).

\subsubsection{Mixed Blue + Red Kilonova }

In general, the total kilonova emission can be thought of as a combination of distinct `blue' and `red' components.  This is because both high- and low-$Y_e$ ejecta components can be simultaneously visible following a merger, particularly for viewing angles close to the binary rotation axis (Fig.~\ref{fig:schematic}).  For viewers closer to the equatorial plane, the blue emission may in some cases be blocked by the high-opacity lanthanide-rich tidal ejecta \citep{Kasen+15}.  Thus, although the presence of a days to week-long NIR transient is probably a  generic feature of all mergers, the early blue kilonova phase might only be visible or prominent in a fraction of events.  On the other hand, if the blue component expands faster than the tidal ejecta (or the latter is negligibly low in mass - e.g. for an equal-mass merger), the early blue emission may be visible from a greater range of angles than just pole-on (e.g.~\citealt{Christie+19}).  

It has become common practice following GW170817 to model the total kilonova light curve by adding independent 1D blue (low $\kappa$) and red (high $\kappa$) models on top of one another (e.g.~\citealt{Villar+17}), i.e. neglecting any interaction between the ejecta components.  While extremely useful for obtaining qualitative insight, in detail this assumption will result in quantitative errors in the inferred ejecta properties (e.g.~\citealt{Kasen+17,Wollaeger+18}).  With well-sampled photometry (in both time and frequency), the total ejecta mass should be reasonably well-measured: once the ejecta has become effectively transparent at late times the bolometric luminosity directly traces the radioactive energy input.  

However, at early times when the ejecta is still opaque, the radial and angular structure of the opacity (i.e.~lanthanide abundance $X_{\rm La}$) can couple distinct ejecta components in a way not captured by combining two independent 1D models (e.g.~\citealt{Kasen+15}).  Radial dependence of $X_{\rm La}$ (e.g. due to a $Y_e$ gradient) is straightforward to implement in the toy model through a mass shell-dependent value of $\kappa_v$ (eq.~\ref{eq:tdv}).  If $X_{\rm La}$ increases with radius (i.e. if the ``red'' ejecta resides physically outside of the ``blue''), then in principle even a low amount of red (high-opacity) fast material can `reprocess' the radioactive luminosity generated from a much greater mass of blue (low-opacity) slower material residing behind it.   \citet{Kawaguchi+18} found that this could in principle lead to an over-estimate the quantity of blue ejecta if one models the light curve by simply adding independent red and blue components.  For these reasons, caution must be taken in naively adding `blue' and `red' models and a detailed analysis must take into account not just photometric light curve information, but also spectral features (in the above example, for instance, reprocessing by an outer thin layer of lanthanide-rich matter would generate strong UV line blanketing).  The geometry of the blue and red components of the ejecta, and the observer viewer angle, are also in principle distinguishable by their relative levels of polarization \citep{Covino+17,Matsumoto18,Bulla+19}.

\section{Kilonova Emission from GW170817}
\label{sec:170817}

\begin{table}
\caption{Key Properties of \LIGO}

\begin{tabular}{ccc}

Property & Value &  References  \\

\hline

Chirp mass, $\mathcal{M}_{\rm c}$ (rest frame) & 1.188$^{+0.004}_{-0.002}M_{\odot}$  & 1 \\
First NS mass, $M_{\rm 1}$ & $1.36-1.60M_{\odot}$ (90\%) & 1 \\
Second NS mass, $M_{\rm 2}$ &  $1.17-1.36M_{\odot}$  (90\%) & 1 \\
Total mass, $M_{\rm tot} = M_{1}+M_{2}$ & $\approx 2.74^{0.04}_{-0.01}M_{\odot}$ &  1 \\
Observer angle to orbital axis, $\theta_{\rm obs}$ & $19-42^{\circ}$ (90\%) &  1,2  \\
Blue KN ejecta ($A_{\rm max} \lesssim 140$) & $\approx 0.01-0.02M_{\odot}$ & 3\\
Red KN ejecta ($A_{\rm max} \gtrsim 140$) & $\approx 0.03-0.06M_{\odot}$ & 4 \\ 
Light $r$-process yield ($A \lesssim 140$) & $\approx 0.04-0.07M_{\odot}$ & \\
Heavy $r$-process yield ($A \gtrsim 140$) & $\approx 0.01M_{\odot}$ & \\ 
Energy of GRB jet & $\sim 10^{49}-10^{50}$ erg & 6\\
ISM density & $\sim 10^{-5}-10^{-3}$ cm$^{-3}$ & 7\\
\hline\\
\end{tabular} 
\hspace{2cm}[1] \citealt{LIGO+17PARAMS}; [2] \citet{Finstad+18,Mooley+18b}; [3] e.g.~\citet{Nicholl+17,Kasen+17}; [4] e.g.~\citet{Chornock+17,Kasen+17}; [5] e.g.~\citet{Hallinan+17,Alexander+17}; [6] e.g.~\citet{Nicholl+17,Kasen+17}; [7] e.g.~\citet{Margutti+18b,Mooley+18}
 \label{table:170817}
\end{table}

As introduced in Sect.~\ref{sec:intro}, the termination of the GW inspiral from LIGO/Virgo's first NS-NS merger GW170817 \citep{LIGO+17DISCOVERY} was followed within seconds by a short GRB \citep{Goldstein+17,Savchenko+17,LIGO+17FERMI}.  Roughly 11 hours later, a luminous optical counterpart, dubbed AT2017gfo, was discovered in the galaxy NGC 4993 at a distance of only $\approx 40$ Mpc (\citealt{Coulter+17,Soares-Santos+17,Arcavi+17b,Diaz+17,Hu+17,Lipunov+17,Valenti+17,Troja+17,Kilpatrick+17,Smartt+17,Drout+17,Evans+17,LIGO+17CAPSTONE,McCully+17,Buckley+18,Utsumi+17,Covino+17,Hu+17}).  
Table \ref{table:170817} summaries a few key properties of GW170817 as inferred from its GW/EM emission.

The timeline of the discovery was recounted in the capstone paper written jointly by LIGO/Virgo and astronomers involved in the EM follow-up\citep{LIGO+17CAPSTONE} and will not be recounted here.  In limiting the scope of our discussion, we also do not address the host galaxy and environment of the merger and its implication for NS-NS merger formation channels (\citealt{Blanchard+17,Hjorth+17,Im+17,Levan+17,Pan+17}), nor shall we discuss the plethora of other science opportunities the kilonova enabled (e.g. H0 cosmology; \citealt{LIGO+17H0}).  We also do not touch upon inferences about the GRB jet and its connection to the observed prompt gamma-ray and non-thermal afterglow emission   (e.g.~\citealt{Bromberg+18,Kasliwal+17,Gottlieb+18,MurguiaBerthier+17,Salafia+18,Xiao+17}), though possible connections between the jet and the early kilonova emission will be discussed in Sect.~\ref{sec:cocoon}.  

AT2017gfo started out blue in color, with a featureless thermal spectrum that peaked at UV frequencies (e.g.~\citealt{Nicholl+17,McCully+17,Evans+17}), before rapidly evolving over the course of a few days to become dominated by emission with a spectral peak in the near-infrared (NIR) \citep{Chornock+17,Pian+17,Tanvir+17}.  Although early blue colors are not uncommon among astrophysical transients (most explosions start hot and thereafter cool from expansion), the very fast evolution of AT2017gfo was completely unlike that seen in an previously known extra-galactic event, making its connection to GW170817 of high significance (even before folding in theoretical priors on the expected properties of kilonovae).  Simultaneous optical (e.g.~\citealt{Nicholl+17,Shappee+17}) and NIR \citep{Chornock+17} spectra around day 2.5 appeared to demonstrate the presence of distinct optical and NIR emission components.  \citet{Smartt+17} observed absorption features in the spectra around 1.5 and 1.75 $\mu$m in the spectra which they associated with features of Cs I and Te I (light $r$-process elements).  In broad brush, the properties of the optical/NIR emission agreed remarkably well with those predicted for $r$-process powered kilonova \citep{Li&Paczynski98,Metzger+10,Roberts+11,Barnes&Kasen13,Tanaka&Hotokezaka13,Grossman+14,Martin+15,Tanaka+17,Wollaeger+18,Fontes+17}, a conclusion reached nearly unanimously by the community (e.g.~\citealt{Kasen+17,Drout+17,Tanaka+17,Kasliwal+17,MurguiaBerthier+17,Waxman+18}).  
In discussing the interpretation of AT2017gfo, we start with the most basic and robust inferences that can be made, before moving onto areas where there is less universal agreement.  

\begin{figure}[!t]
\includegraphics[width=1.0\textwidth]{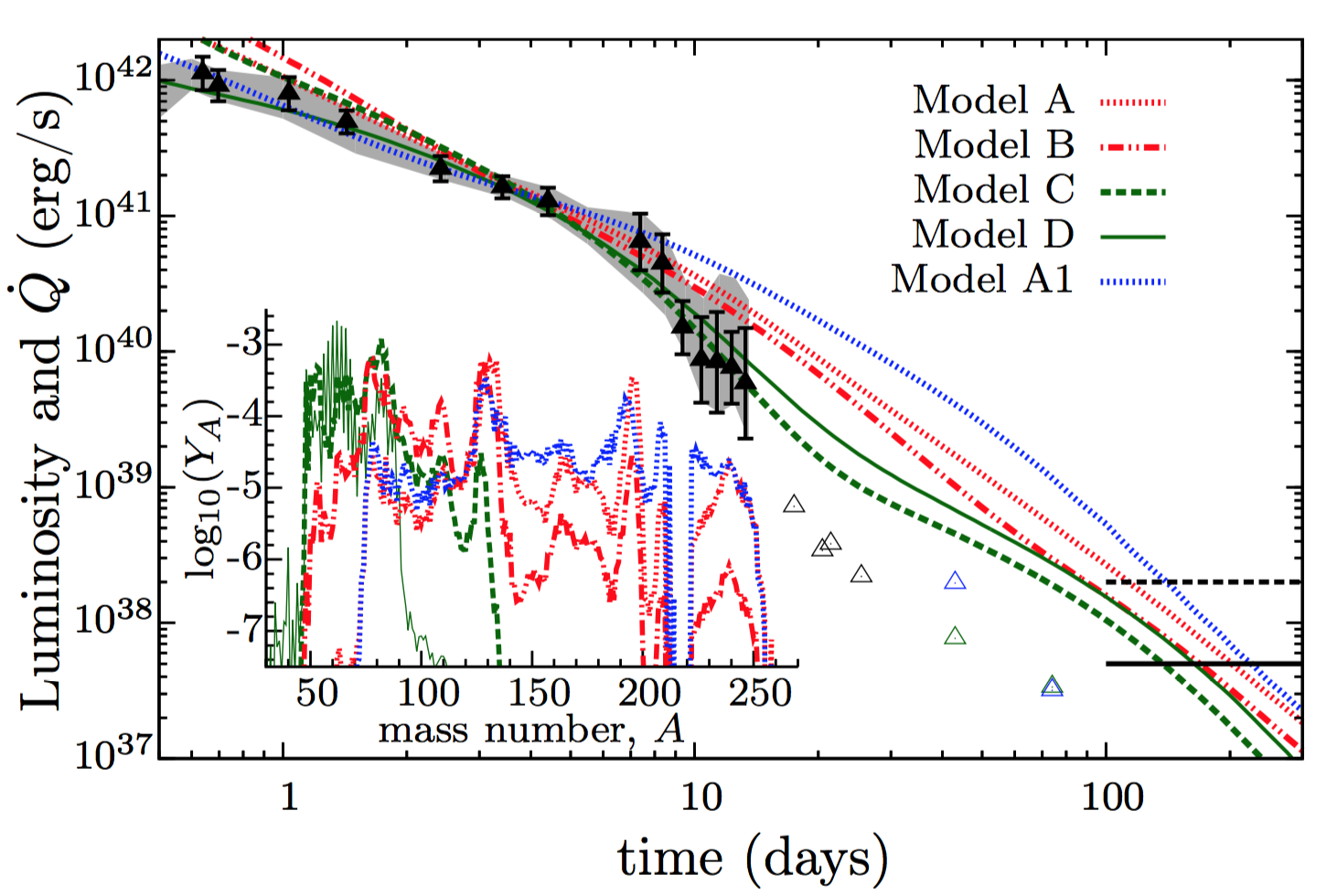}
\caption{Bolometric luminosity of the kilonova AT2017gfo associated with GW170817 from \citet{Smartt+17} with uncertainties derived from the range given in the literature \citep{Smartt+17,Waxman+18,Cowperthwaite+17,Arcavi18}.  Also shown are lower limits (empty triangles) on the late-time luminosity as inferred from the Ks band with VLT/HAWK-I \citep{Tanvir+17} (black) and the 4.5 μm detections by the Spitzer Space Telescope from \citet[green]{Villar+18} and \citep[blue]{Kasliwal+19}. Colored lines show the ejecta heating rate for models with different values for the ejecta mass and average electron fraction as follows: A ($Y_e = 0.15$; $M_{\rm ej} = 0.04M_{\odot}$), B ($Y_e = 0.25$; $M_{\rm ej} = 0.04M_{\odot}$), C ($Y_e = 0.35$; $M_{\rm ej} = 0.055M_{\odot}$), D ($Y_e = 0.45$; $M_{\rm ej} = 0.03M_{\odot}$).  While models $A-D$ assume the FRDM nuclear mass model \citep{Moller+95}, Model A1 ($Y_e = 0.15$; $M_{\rm ej} = 0.02M_{\odot}$) uses the DZ31 nuclear mass model \citep{Duflo&Zuker95}.  Their corresponding $r$-process abundance distributions at t = 1 d are shown in the inset.  Thermalization is calculated following \citep{Kasen&Barnes19} for an assumed ejecta velocity $0.1$ c.  The black solid (dashed) horizontal lines in the lower right corner represent the approximate observation limits of the NIR (MIR) instruments on the James Webb Space Telescope for a merger at 100 Mpc.  Reproduced with permission from \citet{Wu+19a}.}
\label{fig:wu}
\end{figure}

Perhaps the first question one might ask is: {\bf What evidence exists that AT2017gfo was powered by $r$-process heating? and, if so, How much radioactive material was synthesized?}  Figure \ref{fig:wu} from \citet{Wu+19a} shows the bolometric luminosity $L_{\rm bol}(t)$ compiled from observations in the literature \citep{Smartt+17,Cowperthwaite+17,Waxman+18,Arcavi18}compared to several distinct models for the time-dependent heating rate of $r$-process decay $\dot{Q}_r$ (eq.~\ref{eq:qdotr}), in which the authors have varied the mean $Y_e$ of the ejecta contributing to the heating and the nuclear mass model, the latter being one of the biggest nuclear physics uncertainties.

A first takeaway point is the broad similarity between the observed $L_{\rm bol}(t)$ evolution and the power-law-like decay predicted by the decay of a large ensemble of $r$-process isotopes \citep{Metzger+10}.  Furthermore, the total ejecta mass one requires to match the normalization of $L_{\rm bol}$ varies with the assumptions, ranging from $\approx 0.02M_{\odot}$ ($Y_e= 0.15$; DZ31 mass model) to $0.06M_{\odot}$ ($Y_e = 0.35$; FRDM mass model).  This range broadly agrees with that reported by independent groups modeling GW170817 (see \citealt{Cote+18} for a compilation).  It is also entirely consistent with the range of ejecta masses predicted from NS-NS mergers (Sect.~\ref{sec:ejecta}), as we elaborate further below.  Although non-$r$-process powered explanations for AT2017gfo can be constructed (e.g.~invoking magnetar power; Sect.~\ref{sec:magnetar}), they require several additional assumptions and thus are disfavored by Ockham's razor.    

If the yield of $r$-process elements in GW170817 is at all representative of that of NS-NS mergers in the Universe as a whole (as the similarity of its GW-inferred properties compared to the Galactic binary NS population support it being; e.g.~\citealt{Zhao&Lattimer18}), then, even adopting the lowest NS-NS merger rate currently allowed from LIGO/Virgo of $\sim 100$ Gpc$^{-1}$ yr$^{-1}$, an order-of-magnitude estimate (eq.~\ref{eq:Mr}) leads to the conclusion that NS-NS mergers are {\it major} sources of $r$-process elements in the universe (e.g.~\citealt{Kasen+17,Cote+18}).  However, given large current uncertainties on the Galactic rate of NS-NS mergers and the precise abundance distribution synthesized in GW170817, it cannot yet be established that mergers are the exclusive, or even dominant, $r$-process site (see discussion in Sect.~\ref{sec:rprocess}).

\begin{figure}[!t]
\includegraphics[width=1.0\textwidth]{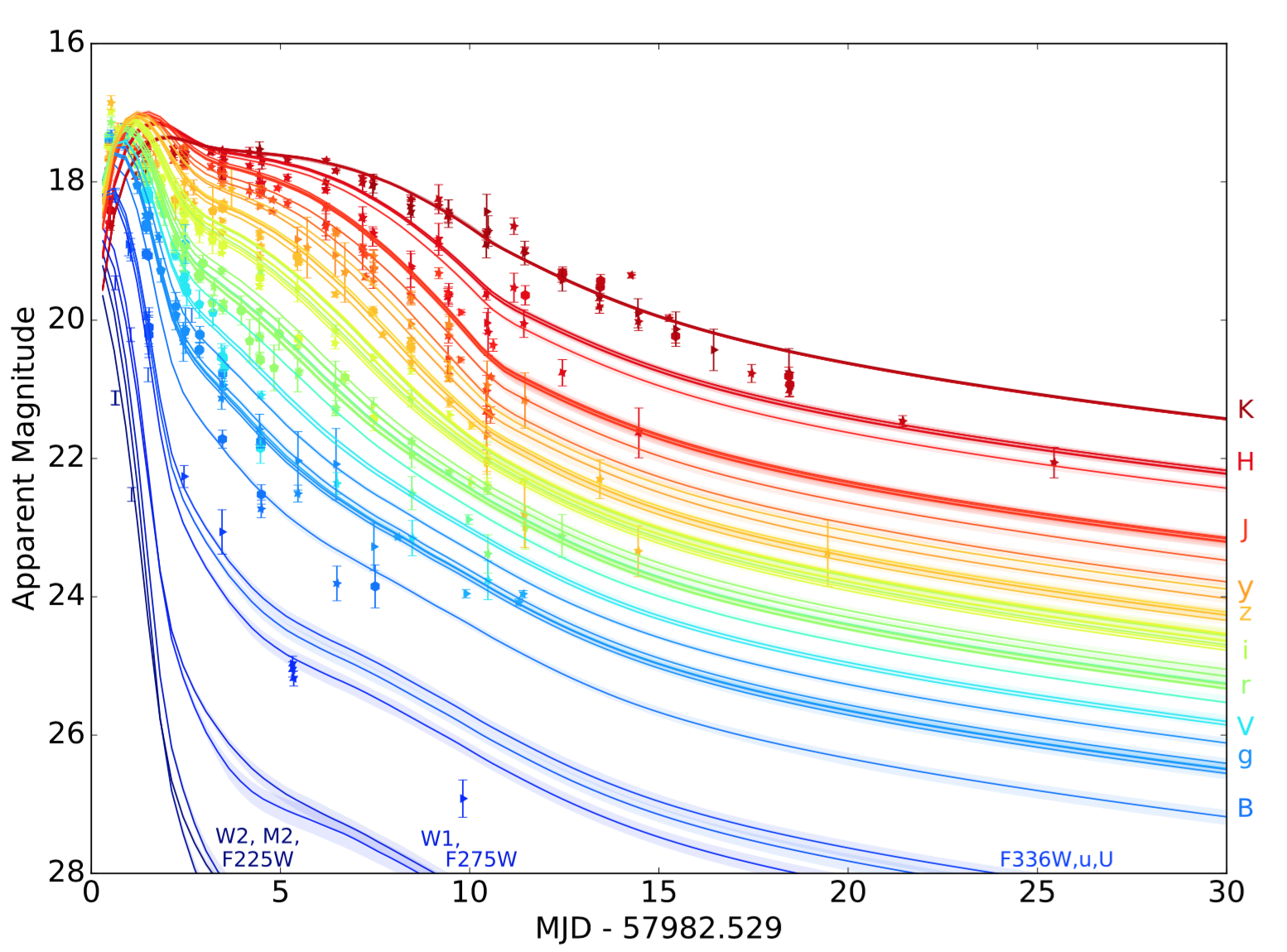}
\caption{UVOIR light curves of AT2017gfo from the data set compiled, along with a best-fit spherically symmetric three-component kilonova model (see text).  The data in this figure was originally presented in \citep{Andreoni+17,Arcavi+17a,Coulter+17,Cowperthwaite+17,Diaz+17,Drout+17,Evans+17,Hu+17, Kasliwal+17,Lipunov+17,Pian+17,Shappee+17,Smartt+17,Tanvir+17,Troja+17,Utsumi+17,Valenti+17}.
Figure reproduced with permission from \citet{Villar+17}.}
\label{fig:Villar}
\end{figure}

With the production and ejection of at least a few hundredths of a solar mass of neutron-rich elements  established, the next question is the detailed nature of its composition.  {\bf Specifically, which $r$-process elements were formed?}  Figure \ref{fig:Villar} shows a compilation of photometric data from the literature on AT2017gfo by \citet{Villar+17}.  The blue/UV bands (e.g. F225W, F275W) fade rapidly from the first observation at 11 hours, while the NIR bands (e.g. JHK) show a much flatter decay over the first week.  The early-time blue emission suggests that the outermost layers of the merger ejecta (at least those dominating the observed emission) are composed of light $r$-process material with a low opacity (blue kilonova; Sect.~\ref{sec:blue}) synthesized from merger ejecta with a relatively high\footnote{\citet{Rosswog+18} argue that the ejecta must have possessed $Y_{e} \lesssim 0.3-0.35$ to produce a smooth radioactive heating rate consistent with the bolometric light curve (due to the fact that discrete isotopes dominate the heating rate for higher $Y_e$; e.g.~\citealt{Lippuner&Roberts15}).  However, this makes the over-restrictive assumption that all the ejecta contains a single precise value of $Y_e$.  A small but finite spread in $Y_e$ about the mean value $\bar{Y}_e$ results in a smooth light curve decay consistent with observations even for $\bar{Y}_e > 0.35$ (\citealt{Wanajo18,Wu+19a}; see Fig.~\ref{fig:wu}).} electron fraction, $Y_{e} \gtrsim 0.25$.  The more persistent late NIR emission instead requires matter with higher opacity, consistent with the inner ejecta layers containing at least a moderate amount of lanthanide or actinide elements (red kilonova; Sect~\ref{sec:red}).   

Motivated by the theoretical prediction of distinct lanthanide-free and lanthanide-rich ejecta components (e.g.~\citealt{Metzger&Fernandez14}), many groups interpreted AT2017gfo using mixed models described in the previous section, comprised of 2 or 3 separate ejecta components with different lanthanide abundances (e.g.~\citealt{Kasen+17,Tanaka+17,Drout+17,Kasliwal+17,Perego+17}; however, see \citealt{Waxman+18}).  As one example, the solid lines in Fig.~\ref{fig:Villar} show a best-fit model from \citet{Villar+18} based on the sum of three spherical gray-opacity kilonova models (``blue'', ``purple'', ``red'') with respective opacities $\kappa = (0.5,3,10)$ cm$^{2}$ g$^{-1}$ (similar to those given in Table \ref{table:opacity}) and from which they infer for the respective components ejecta masses $M_{\rm ej} \approx (0.02,0.047,0.011)M_{\odot}$ and mean velocities $v_{\rm ej} \approx (0.27,0.15,0.14) c$.  Mapping the opacities back to electron fractions using e.g. Table \ref{table:opacity}, one infers that most of the ejecta possessed intermediate values of $Y_e \approx 0.25-0.35$ which generated elements up to the second $r$-process peak.  Smaller quantities of the ejecta had $Y_e \gtrsim 0.4$ or $Y_e \lesssim 0.25$, the latter containing a sufficient neutron abundance to produce some lanthanide elements ($A \gtrsim 140$) if not nuclei extending up to third $r$-process peak ($A \sim 195$) or beyond.  Despite the unprecedented data set available for AT2017gfo, it is unfortunately not possible to reconstruct the detailed abundance pattern synthesized, for instance to test its consistency with that observed in metal-poor stars or in our solar system (e.g.~\citealt{Hotokezaka+18}).

From the inferred ejecta mass, velocity, and $Y_e$ distribution, the next question is: {\bf What phase or phases during or following the merger was this material released?}  One thing is clear: the dynamical ejecta alone is insufficient.  Fig.~\ref{fig:Siegel} shows that the total ejecta mass $\gtrsim 0.02M_{\odot}$ exceeds the predicted dynamical ejecta from essentially all NS-NS merger simulations published to date, while the average velocity of the bulk of the lower-$Y_e$ ejecta $\approx 0.1$ c is also significantly less than predicted for the dynamical ejecta.  The bulk of the ejecta, particularly the redder low-$Y_e$ component, is instead most naturally explained as an outflow from the remnant accretion torus created around the central compact object following the merger (Sect.~\ref{sec:diskejecta}).  GRMHD simulation of the post-merger disk evolution demonstrate that $\approx 40\%$ of the initial mass of the torus (i.e. up to $\sim 0.08M_{\odot}$ in wind ejecta for initial disk masses up to $\approx 0.2M_{\odot}$ predicted by simulations) is unbound at an average velocity of $v \approx 0.1$ c (e.g.~\citealt{Siegel&Metzger17,Fernandez+19}).  The disk wind ejecta can contain a range of electron fractions (and thus produce blue or red emission), depending e.g.~on the lifetime of the central NS remnant prior BH formation (see Fig.~\ref{fig:Lippuner}). 

The physical source of the lanthanide-poor ejecta ($Y_e \gtrsim 0.35$) responsible for powering the early-time blue emission is more open to debate.  Table \ref{table:blueKN} summarizes several possible origins, along with some of their pros and cons.  The high velocities $v_{\rm blue} \approx 0.2-0.3$ c and composition ($Y_{e} \gtrsim 0.25$) broadly agree with predictions for the shock-heated dynamical ejecta (e.g.~ \citealt{Oechslin&Janka06,Sekiguchi+16,Radice+16}).  However, one concern is the large quantity $M_{\rm blue} \gtrsim 10^{-2}M_{\odot}$, which again is higher than the total predicted by most merger simulations (Fig.~\ref{fig:Siegel}), especially considering that only a fraction of the dynamical ejecta will possess a high $Y_e$.  If the blue ejecta is dynamical in origin, this could provide evidence for a small value of the NS radius (\citealt{Nicholl+17}, Sect.~\ref{sec:EOS}) because the quantity of shock-heated ejecta appears to grow with the NS compactness \citep{Bauswein+13}. 

The highest velocity tail of the kilonova ejecta might not be an intrinsic property, but instead the result of shock-heating of an originally slower ejecta cloud by a relativistic jet created following some delay after the merger (e.g.~\citealt{Bucciantini+12,Duffell+15,Gottlieb+18b,Gottlieb+18,Kasliwal+17,Bromberg+17,Piro&Kollmeier18}; Sect.~\ref{sec:cocoon}).  However, even models which invoke an early component of cocoon emission \citep{Kasliwal+17} require a radioactive-powered component of emission which dominates after the first 11 hours that contains a large mass $\sim 10^{-2}M_{\odot}$ of low-opacity (high $Y_e$) matter (stated another way, the observations do not require an early extra emission component beyond radioactivity; see Fig.~\ref{fig:firsthour}).  Further disfavoring a jet-related origin for the blue kilonova ejecta is that the kinetic energy of the latter $M_{\rm blue}v_{\rm blue}^{2}/2 \approx 10^{51}$ ergs exceeds the kinetic energies of cosmological short gamma-ray bursts ($\approx 10^{49}-10^{50}$ erg; \citealt{Nakar07,Berger14}), and that of the off-axis jet specifically required to fit the off-axis afterglow of GW170817 (e.g.~\citealt{Margutti+18}), by a large factor $\sim 10-100$.

\citet*{Metzger+18} proposed an alternative source for the blue kilonova ejecta: a magnetized wind which emerges from the HMNS remnant $\approx 0.1-1$ seconds prior to its collapse to a BH.  While the HMNS remnant was  proposed as a potential ejecta source for GW170817, (e.g.~\citealt{Evans+17}), the velocity and mass-loss rate of such purely neutrino-powered winds \citep{Dessart+09,Perego+14} are insufficient to explain that of the observed kilonova.  \citet*{Metzger+18}  emphasize the role that strong magnetic fields have on increasing the mass-loss rate and velocity of the wind through centrifugal slinging (similar to models of magnetized winds from ordinary stars; \citealt{Belcher&MacGregor76}).  Using a series of 1D wind models, they found that a temporarily-stable magnetar remnant with a surface field strength $B \approx 1-3\times 10^{14}$ G can naturally produce the mass, velocity, and composition of the blue kilonova ejecta in GW170817.  We return to the role that much longer-lived magnetar remnants can play in the kilonova emission in Sect.~\ref{sec:magnetar}.  

Recently, using numerical relativity simulations which include approximate neutrino transport and a treatment of the effects of turbulent viscosity in the disk, \citet{Nedora+19} found that spiral density waves generated in the post-merger accretion disk by the central HMNS remnant can lead to the ejection of $\sim 10^{-2}M_{\odot}$ in matter with $Y_e \gtrsim 0.25$ and velocity $\sim 0.15-0.2$ c.  An open question is whether such spiral waves behave similarly, and can produce ejecta with sufficiently high velocities $\gtrsim 0.2$ c to explain AT2017gfo, even in the physical case in which the magneto-rotational instability operates simultaneously in the disk.  If so, this mechanism would provide an additional promising source of blue kilonova ejecta from a moderately long-lived HMNS.   
  
\begin{table*}
\centering
\caption{Potential Sources of the Fast Blue KN  Ejecta in GW170817
$^{\dagger}$ \label{table:blueKN}}
\begin{tabular}{cccccc}
Ejecta Type & Quantity? & Velocity? & $Y_e$? \\
\hline
Tidal Tail Dynamical & Maybe, if $q \lesssim 0.7^{\ddagger}$ & \checkmark & {\bf Too Low}\\
Shock-Heated Dynamical & Maybe, if $R_{1.6} \lesssim 11$ km$^{\star}$ & \checkmark  & \checkmark  if NS long-lived \\
Accretion Disk Outflow & \checkmark if torus massive & {\bf Too Low} & \checkmark  if NS long-lived \\
HMNS $\nu$-Driven Wind & {\bf Too Low} & {\bf Too Low} & {\bf Too High?} \\
Spiral Wave Wind & \checkmark  & Maybe & \checkmark  \\
Magnetized HMNS Wind & \checkmark  if NS long-lived & \checkmark  & \checkmark   \\
\hline \\
\end{tabular}
\\

$^{\dagger}$Adopted from \citet*{Metzger+18}.  $^{\ddagger}$where $q \equiv M_{1}/M_{2}$ and $M_{1}, M_{2}$ are the individual NS masses (\citealt{Dietrich+17,Dietrich&Ujevic17,Gao+17KILONOVA}).
$^{\star}$However, a small NS radius may be in tension with the creation of a large accretion disk needed to produce the red KN ejecta (\citealt{Radice+18b}). 
\end{table*}

\subsection{Inferences about the Neutron Star Equation of State}
\label{sec:EOS}

GW170817 provided a wealth of information on a wide range of astrophysical topics.  One closely connected to the focus of this review are new constraints it enabled on the equation of state (EOS) of nuclear density matter, that which is responsible for determining the internal structure of the NS and setting its key properties, such as its radius and maximum stable mass $M_{\rm TOV}$ (\citealt{Lattimer&Prakash16,Ozel&Freire16}).    

Even absent a bright EM counterpart, the gravitational waveform can be used to measure or constrain the tidal deformability of the inspiraling stars prior to their disruption resulting from tidal effects on the inspiral phase evolution (e.g.~\citealt{Raithel+18,DeSoumi+18,LIGO+18EOS}).  Assuming two stars with the same EOS, observations of GW170817 were used to place limits on the radius of a 1.6$M_{\odot}$ NS of $R_{1.6} = 10.8^{+2.0}_{-1.7}$ km \citep{LIGO+18EOS}.  Likewise, in BH-NS mergers, measurement of tidal interactions and the cut-off GW frequency at which the NS is tidally disrupted by the BH, provide an alternative method to measure NS radii (e.g.~\citealt{Kyutoku+11,Lackey+14,Pannarale13,Pannarale+15}).  

Unfortunately, current generation of GW detectors are far less sensitive to the post-merger signal and thus of the ultimate fate of the merger remnant, such and whether and when a BH is formed (as was true even for the high signal-to-noise event, GW170817; \citealt{LIGO+17REMNANT}).  Here, EM observations provide a complementary view.  In a BH-NS merger, the presence or absence of an EM counterpart is informative about whether the NS was tidally disrupted and thus can be used to measure its compactness (e.g.~\citealt{Ascenzi+19}).  As discussed in Sect.~\ref{sec:ejecta} and summarized in Table \ref{table:remnants}, the type of compact remnant which is created by a NS-NS merger (prompt collapse, HMNS, SMNS, or stable NS) depends sensitively on the total binary mass $M_{\rm tot}$ relative to various threshold masses, which depend on unknown properties of the EOS, particularly $M_{\rm TOV}$ and $R_{1.6}$ (Fig.~\ref{fig:spindown}).  Thus, if one can infer the type of remnant produced in a given merger from the EM counterpart, e.g. the kilonova or GRB emission, then by combining this with GW measured\footnote{The inspiral waveform most precisely encodes the binary chirp mass $\mathcal{M}_{\rm c}$ (eq.~\ref{eq:Mchirp}).  However, the mapping between $M_{\rm tot}$ and $\mathcal{M}_{\rm c}$ is only weakly dependent on the binary mass ratio $q = M_{1}/M_{2}$ for values of $q \gtrsim 0.7$ characteristic of the known Galactic double NS systems \citep{Margalit&Metzger19}.}  value of $M_{\rm tot}$ one can constrain the values of $M_{\rm TOV}$ and/or $R_{1.6}$.

In GW170817, the large quantity of ejecta $\gtrsim 0.02M_{\odot}$ inferred from the kilonova, and its high electron fraction, strongly disfavored that the merger resulted in a prompt ($\sim$ dynamical timescale) collapse to a BH.  Given that the threshold for prompt collapse depends on the NS compactness \citep{Bauswein+13b}, this enabled \citet{Bauswein+17} to place a lower limit of $R_{1.6} \gtrsim 10.3-10.7$ km (depending on the conservativeness of their assumptions).  \citet{Radice+18b} came to a physically-related conclusion (that GW170817 produced a large ejecta mass not present in the case of prompt BH formation), but expressed their results as a lower limit on tidal deformability instead of $R_{1.6}$ (see also \citealt{Coughlin+18} for a joint Bayesian analysis of the EM and GW data).

Going beyond the inference that GW170817 initially formed a NS remnant instead of a prompt collapse BH to infer the stability and lifetime of the remnant becomes trickier.  Nevertheless, several independent arguments can be made which taken together strongly suggest that remnant was a relatively short-lived HMNS ($t_{\rm collapse} \lesssim 0.1-1$ s), rather than a SMNS or indefinitely-stable NS  \citep{Margalit&Metzger17,Granot+17,Bauswein+17,Perego+17,Rezzolla+18,Ruiz+18,Pooley+18}.  
\begin{itemize}
\item{The presence of a significant quantity of lanthanide-rich disk wind ejecta, as inferred from the presence of red kilonova emission, is in tension with the $Y_e$ distribution predicted for a merger remnant that were to have survived longer than several hundred milliseconds (\citealt{Metzger&Fernandez14,Lippuner+17}; see Fig.~\ref{fig:Lippuner}).}
\item{The kinetic energies of the kilonova ejecta ($\sim 10^{51}$ erg) and the off-axis gamma-ray burst jet inferred from the X-ray/radio afterglow ($\sim 10^{49}-10^{50}$ erg) exceed by a large factor $\gtrsim 10-100$ the rotational energy necessarily released for a SMNS or stable NS remnant to collapse to a BH (Fig.~\ref{fig:spindown}; see further discussion in Sect.~\ref{sec:magnetar}).}
\item{The formation of an ultra-relativistic GRB jet on a timescale of $\lesssim 1$ s after the merger is believed by many to require a clean polar funnel only present above a BH (e.g.~\citealt{Murguia-Berthier+17}; see Sect.~\ref{sec:magnetar} for discussion of this point).   All indications from the GW170817 afterglow point to the presence of an off-axis jet with properties consistent with the cosmological short GRB population (e.g.~\citealt{Wu&MacFadyen19}).}
\item{The magnetar spin-down luminosity could power temporally-extended X-ray emission minutes to days after the merger (Sect.~\ref{sec:magnetar}); however, the observed X-rays from GW170817 are completely explained by the GRB afterglow without excess emission from a long-lived central remnant being required (\citealt{Margutti+17,Pooley+18}; however, see \citealt{Piro+19}).}
\end{itemize}

Taking the exclusion of a SMNS remnant in GW170817 for granted, and combining this inference with the measured binary mass $M_{\rm tot} = 2.74^{+0.04}_{-0.01}M_{\odot}$ \citep{LIGO+17PARAMS} from the GW signal, \citet{Margalit&Metzger17} place an upper limit on the TOV mass of $M_{\rm TOV} \lesssim 2.17M_{\odot}$ (see also \citealt{Shibata+17,Rezzolla+18,Ruiz+18}; however see \citealt{Shibata+19}, who argue for a more conservative constraint of $M_{\rm TOV} \lesssim 2.30M_{\odot}$).  Stated another way, if $M_{\rm TOV}$ were much higher than this limit, one would expect the remnant of GW170817 to have survived longer and produced an EM signal markedly different than the one observed.  If this result holds up to further scrutiny, it provides the most stringent upper limit on $M_{\rm TOV}$ currently available (and one in possible tension with the high NS masses $\approx 2.4M_{\odot}$ suggested for some so-called ``black widow'' pulsars; e.g.~\citealt{Romani+15}).     

The above methods for constraining the NS EOS from pure GW or joint EM/GW data come with systematic uncertainties (most yet to be quantified), albeit different ones than afflict current EM-only methods.  However, there is reason to hope these will improve with additional modeling and observations.  For instance, if our current theoretical understanding of the diversity of possible outcomes of NS-NS mergers  with the in-going binary properties ($M_{\rm tot}$, $q$) is correct (Fig.~\ref{fig:MM}), these predicted trends should be verifiable from a sample of future kilonova/afterglow observations (Sect.~\ref{sec:predictions}).  \citet{Margalit&Metzger19} show that $\sim$ 10 joint EM-GW detections of sufficient quality to accurately ascertain the merger outcome could constrain the values of $R_{1.6}$ and $M_{\rm TOV}$ to the several percent level where systematic effects are certain to dominate the uncertainties.  The future is bright for EM-GW joint studies of the NS EOS in tandem with improvements in our ability to understand and even predict the diverse outcomes of NS-NS or BH-NS merger events (Sect.~\ref{sec:predictions}).

\section{Diversity of Kilonova Signatures}

\label{sec:variations}

The previous section described the most ``vanilla'' models of red/blue kilonovae powered exclusively by $r$-process heating and how the thermal UVOIR emission following GW170817 could be adequately described in this framework.  This section explores additional, sources of emission which are theoretically predicted by some models but have not yet been observed (at least unambiguously).  Either these emission sources were not accessible in GW170817 due to observational limitations, or they could be ruled out in this event but nevertheless may accompany future NS-NS or NS-BH mergers (e.g. for different masses of the in-going binary stars).  Though some of these possibilities remain speculative, their consideration is nevertheless useful to define future observational goals and to inform search strategies regarding just how differently future mergers could appear than GW170817.

\subsection{The First Few Hours}
\label{sec:firsthour}

The UV/optical counterpart of GW170817 was discovered roughly 11 hours after the two stars merged.  This delay was largely due to the event taking place over the Indian Ocean, rendering its sky position initially inaccessible to the majority of ground-based follow-up telescopes.  However, roughly half of future mergers should take place in the northern hemisphere (above the LIGO detectors) which improves the chances of rapid optical follow-up, potentially within hours or less from the time of coalescence (e.g.~\citealt{Kasliwal&Nissanke13}).  A future wide-field UV satellite (or fleet of satellites) able to rapidly cover GW event error regions could revolutionize the early-time frontier.

Kilonovae are powered by the outwards diffusion of thermal radiation.   The earliest time emission therefore probes the fastest, outermost layers of the ejecta.  In the most simple-minded and conservative scenario, these layers are also heated by $r$-process decay, rendering the early-time emission a simple continuation of the $r$-process kilonova to earlier times.  However, due to the higher temperatures when the ejecta is more compact, the ionization states of the outer layers (and hence which elements dominate the line opacity) could differ markedly from those at later times.  Full opacity calculations which extend to thermodynamic conditions appropriate to the first few hours of the transient are a necessary ingredient to making more accurate predictions for this early phase (e.g.~\citealt{Tanaka+19}).  

However, it is also possible that the outermost layers of the ejecta are even hotter$-$and thus more luminous$-$than expected from $r$-process heating alone.  Additional sources of early-time heating include: (1) radioactive decay of {\it free} neutrons which may be preferentially present in the fast outers layers (Sect.~\ref{sec:neutrons}) or (2) the delayed passage through the ejecta by a relativistic jet or wide-angle outflow (Sect.~\ref{sec:cocoon}).  The tail-end of such an extra emission component could in principle have contributed to the earliest epochs of optical/UV emission from GW170817 \citep{Arcavi18}, though the available data is fully consistent with being powered exclusively by $r$-process heating.  

\subsubsection{Neutron Precursor Emission}
\label{sec:neutrons}

\begin{figure}[!t]
\includegraphics[width=0.5\textwidth]{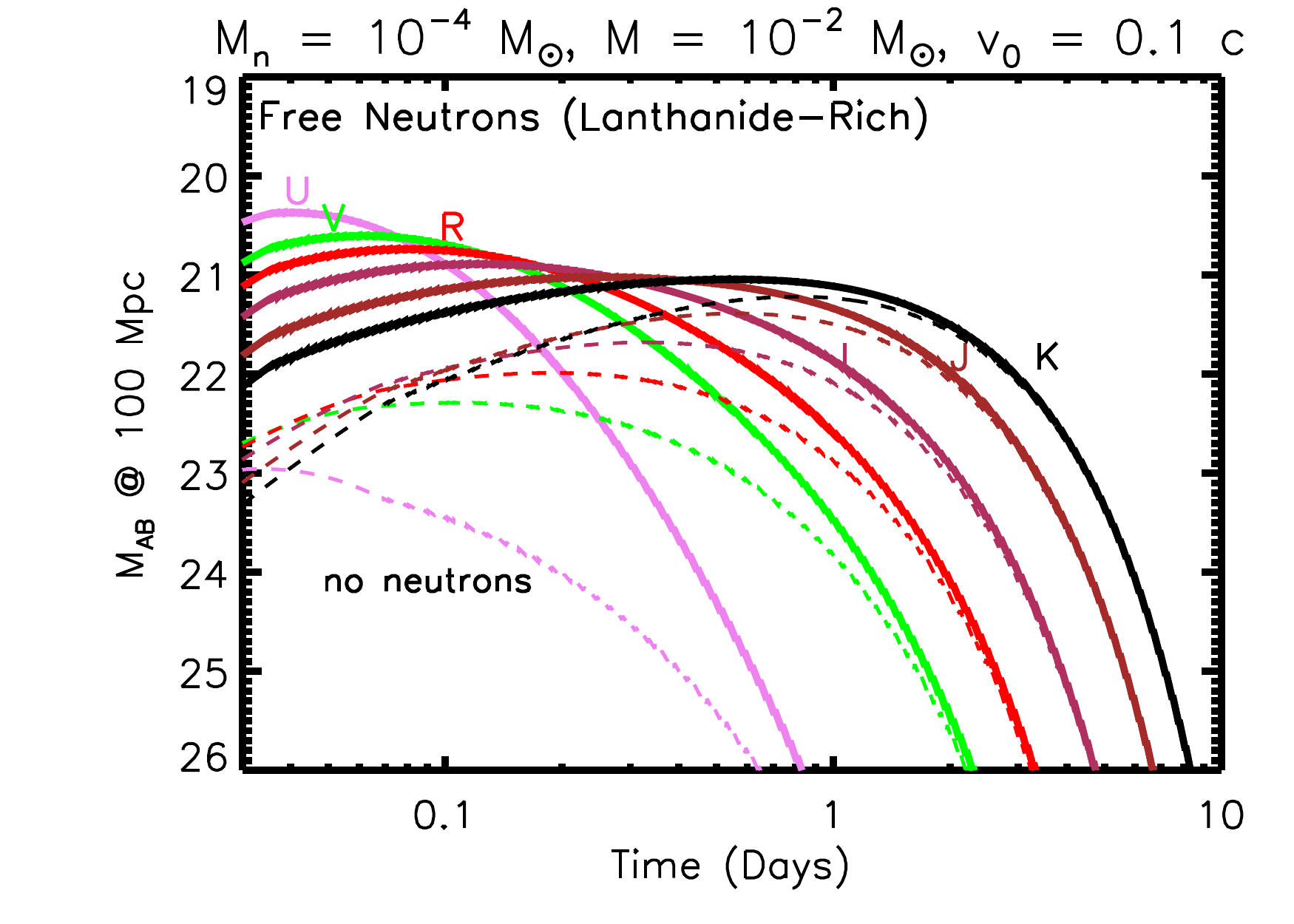}
\includegraphics[width=0.5\textwidth]{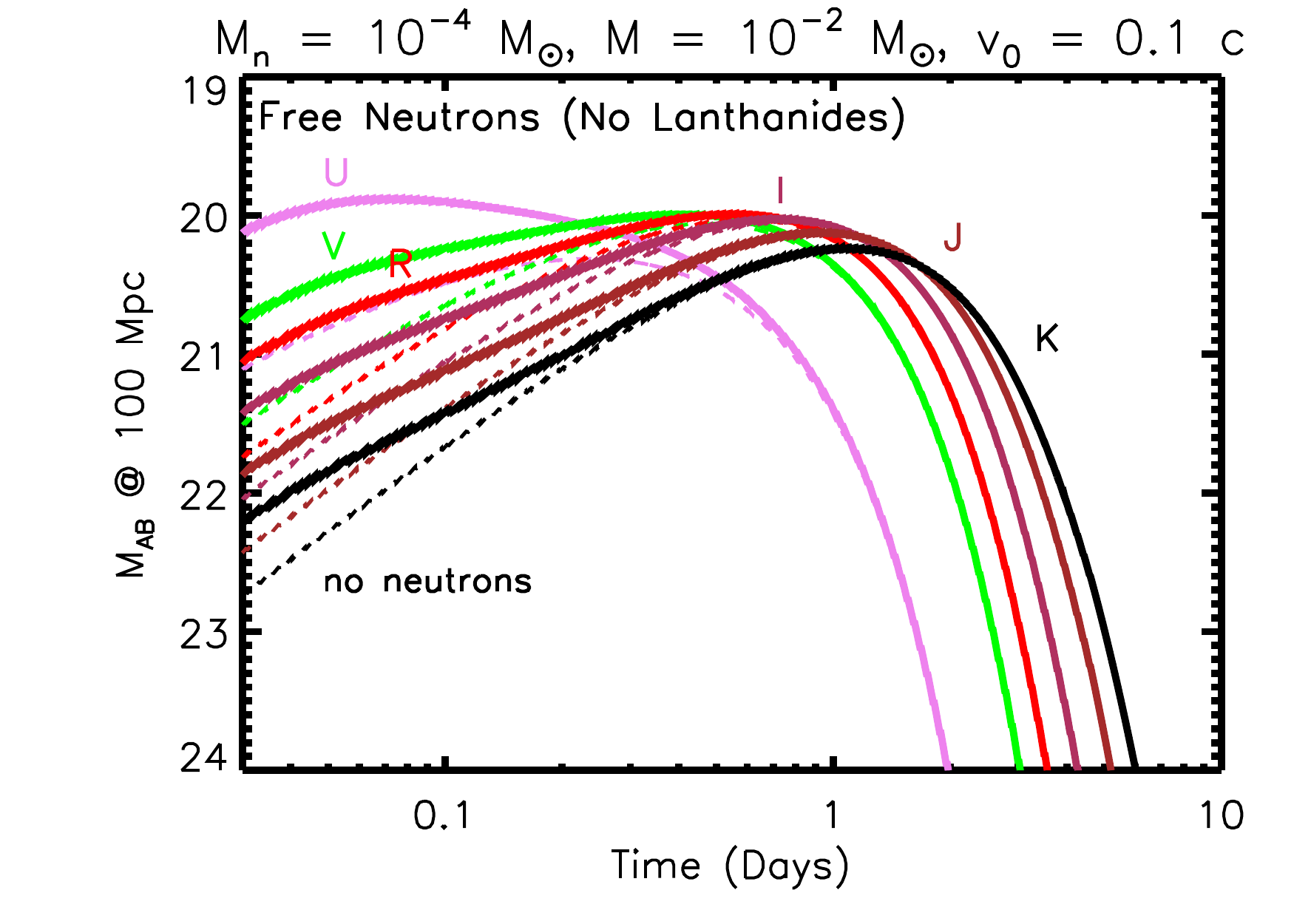}
\caption{Kilonova light curves, including the presence of free neutrons in the outer $M_{\rm n} = 10^{-4}M_{\odot}$ mass layers of the ejecta (``neutron precusor'' emission), calculated for the same parameters of total ejecta mass $M = 10^{-2}M_{\odot}$ and velocity $v_0 = 0.1$c  used in Fig.~\ref{fig:vanilla}.  The left panel shows a calculation with an opacity appropriate to lanthanide-bearing nuclei, while the right panel shows an opacity appropriate to lanthanide-free ejecta.  Models without a free neutron layer ($M_{\rm n} = 0$; Fig.~\ref{fig:vanilla}) are shown for comparison with dashed lines.}
\label{fig:neutrons}
\end{figure}

The majority of the ejecta from a NS-NS merger remain sufficiently dense during its decompression from nuclear densities that all neutrons are captured into nuclei during the $r$-process (which typically takes place seconds after matter is ejected).  However, some NS-NS merger simulations find that a small fraction of the dynamical ejecta (typically a few percent, or $\sim 10^{-4}M_{\odot}$) can expand sufficiently rapidly that the neutrons in the ejecta do not have time to be captured into nuclei \citep{Bauswein+13}, i.e.~the $r$-process ``freezes out".  In the simulations of \citet{Bauswein+13} this fast-expanding matter, which reaches asymptotic velocities $v \gtrsim 0.5$ c, originates from the shocked interface between the merging stars and resides on the outermost layers of the polar ejecta (see also \citealt{Ishii+18}).  Equally fast-expanding material could in principle be produced via other mechanisms which take place after the dynamical phase, e.g. the passage through the ejecta by a GRB jet or in a magnetized wind from the HMNS remnant (see next section).  

Free neutrons, if present in the outer ejecta layers, provide an order of magnitude greater specific heating rate than produced by $r$-process nuclei on timescales of tens of minutes to hours (Fig.~\ref{fig:heating}).  \citet{Metzger+15} emphasized that such super-heating by such a free neutron layer could substantially enhancing the early kilonova emission (see also \citealt{Kulkarni05}).  

An ejecta layer $\delta M_v$ containing free neutrons experiences a radioactive heating rate of
\be
\dot{Q}_{r,v} = \delta M_v X_{n,v}\dot{e}_n(t),
\ee
where the initial mass fraction of neutrons,
\be 
X_{n,v} = \frac{2}{\pi} (1-Y_e)\arctan\left(\frac{M_{n}}{M_v}\right),
\label{eq:Xnv}
\ee
is interpolated in a smooth (but otherwise ad-hoc) manner between the neutron-free inner layers at $M \gg M_n$ and the neutron-rich outer layers $M \ll M_n$, which have a maximum mass fraction of $1- 2Y_e$.  The specific heating rate due to neutron $\beta$-decay (accounting for energy loss to neutrinos) is given by
\be
\dot{e}_n = 3.2\times 10^{14}\exp[-t/\tau_{n}]\,{\rm erg\,s^{-1}\,g^{-1}},
\label{eq:edotn}
\ee
where $\tau_n \approx 900$ s is the neutron half-life.  The rising fraction of free neutrons in the outermost layers produces a corresponding decreasing fraction of $r$-process nuclei in the outermost layers, i.e., $X_{r,v} = 1-X_{n,v}$ in calculating the $r$-process heating rate from Eq.~(\ref{eq:qdotr}).  

Figure \ref{fig:neutrons} shows kilonova light curves, including an outer layer of neutrons of mass $M_n = 10^{-4}M_{\odot}$  and electron fraction $Y_e = 0.1$.  In the left panel, we have assumed that the $r$-process nuclei which co-exist with the neutrons contain lanthanides, and hence would otherwise (absent the neutrons) produce a ``red'' kilonova.  Neutron heating boosts the UVR luminosities on timescales of hours after the merger by a large factor compared to the otherwise identical case without free neutrons (shown for comparison with dashed lines).  Even compared to the early emission predicted from otherwise lanthanide-free ejecta (``blue kilonova''), neutron decay increases the luminosity during the first few hours by a magnitude or more, as shown in the right panel of Fig.~\ref{fig:neutrons}.  

How can such a small layer of neutrons have such a large impact on the light curve?  The specific heating rate due to free neutrons $\dot{e}_n$ (Eq.~\ref{eq:edotn}) exceeds that due to $r$-process nuclei $\dot{e}_r$ (Eq.~\ref{eq:edotr}) by over an order of magnitude on timescales $\sim 0.1-1$ hr after the merger.  This timescale is also, coincidentally, comparable to the photon diffusion depth of the inner edge of the neutron mass layer if $M_{\rm n} \gtrsim 10^{-5}M_{\odot}$.  Indeed, setting $t_{\rm d,v} = t$ in Eq.~(\ref{eq:tdv}), the emission from mass layer $M_v$ peaks on a timescale
\begin{eqnarray}
&&t_{\rm peak,v} \approx \left(\frac{M_{v}^{4/3}\kappa_{v}}{4\pi M^{1/3} v_{0}  c}\right)^{1/2} \nonumber \\
 &\approx& 1.2\,{\rm hr}\left(\frac{M_v}{10^{-5}M_{\odot}}\right)^{2/3}\left(\frac{\kappa_v}{10\,{\rm cm^{2}\,g^{-1}}}\right)^{1/2}\left(\frac{v_0}{0.1\, \rm c}\right)^{-1/2}\left(\frac{M}{10^{-2}M_{\odot}}\right)^{-1/6} 
\label{eq:tpeakv}
\end{eqnarray}The total energy energy released by neutron-decay is $E_n \simeq \int \dot{e}_n M_{\rm n} dt \approx 6\times 10^{45}(M_{\rm n}/10^{-5}M_{\odot})\mathrm{\ erg}$ for $Y_e \ll 0.5$.  Following adiabatic losses, a fraction $\tau_{\rm n}/t_{\rm peak,v} \sim 0.01-0.1$ of this energy is available to be radiated over a timescale $\sim t_{\rm peak,v}$.  The peak luminosity of the neutron layer is thus approximately
\begin{eqnarray}
&& L_{\rm peak,n} \approx  \frac{E_n \tau_n}{t_{\rm peak,v}^{2}} \nonumber \\
&\approx& 3\times 10^{42}\,{\rm erg\,s^{-1}}\left(\frac{M_{v}}{10^{-5}M_{\odot}}\right)^{-1/3}\left(\frac{\kappa_v}{10\,{\rm cm^{2}\,g^{-1}}}\right)^{-1}\left(\frac{v_0}{0.1\, \rm c}\right)\left(\frac{M}{10^{-2}M_{\odot}}\right)^{1/3}, \nonumber\\
\label{eq:Lpeakn}
\end{eqnarray}
and hence is relatively insensitive to the mass of the neutron layer, $M_{v} = M_{\rm n}$.  This peak luminosity can be $\sim 10-100$ times higher than that of the main $r$-process powered kilonova peak.  The high temperature of the ejecta during the first hours of the merger will typically place the spectral peak in the UV, potentially even in cases when the free neutron-enriched outer layers contain lanthanide elements.    

Additional theoretical and numerical work is needed to assess the robustness of the fast-moving ejecta and its abundance of free neutrons, which thus far has been seen in a single numerical code \citep{Bauswein+13}.  The freeze-out of the $r$-process, and the resulting abundance of free neutrons, is also sensitive to the expansion rate of the ejecta \citep{Lippuner&Roberts15}, which must currently be extrapolated from the merger simulations (running at most tens of milliseconds) to the much longer timescales of $\sim$ 1 second over which neutrons would nominally be captured into nuclei.  Figure \ref{fig:neutrons} and Eq.~(\ref{eq:Lpeakn}) make clear that the neutron emission is sensitive to the opacity of the ejecta at early stages, when the temperatures and ionization states of the ejecta are higher than those employed in extant kilonova opacity calculations. 

\subsubsection{Shock Re-Heating (Short-Lived Engine Power)}
\label{sec:cocoon}

\begin{figure}[!t]
\includegraphics[width=1.\textwidth]{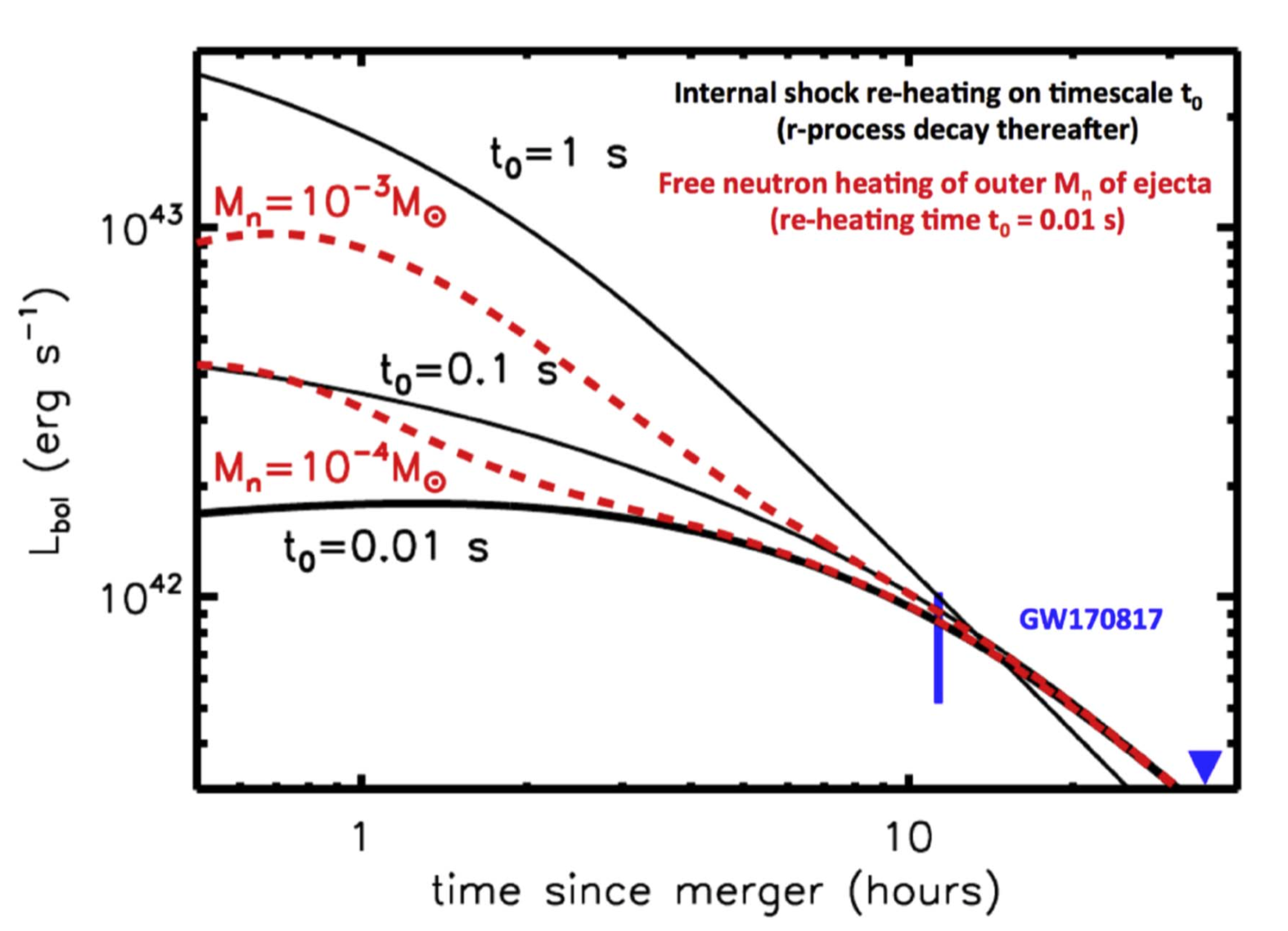}
\caption{Bolometric kilonova light curve during the first few hours of a NS-NS merger, calculated for several model assumptions that can reproduce the measured luminosity $L_{\rm bol} \approx 10^{42}$ erg s$^{-1}$ of AT2017gfo at $t \approx 11$ hr (blue uncertainty bar; e.g.,~\citealt{Arcavi+17a,Cowperthwaite+17,Drout+17}).  Black solid lines show how $r$-process only models change with the assumed timescale $t_0 = 0.01, 0.1, 1$ s at which the outer ejecta was last ``thermalized", i.e. endowed with an internal thermal energy comparable to its asymptotic kinetic energy (at $t \gtrsim t_0$, the ejecta is heated is solely by $r$-process radioactivity in these models).  A small value of $t_0 \sim 0.01$ s corresponds to a dynamical ejecta origin with no additional heating, while a large value of $t_0 \sim 0.01-1$ s represents the case of a long-lived engine (GRB jet, magnetar wind or accretion disk outflow) which re-heats the ejecta on a timescale $\sim t_0$.  We adopt parameters $\beta = 3$, $v_0 = 0.25c$, $M = 0.025M_{\odot}$, $\kappa = 0.5$ cm$^{2}$ g$^{-1}$ (except for the $t_0 = 1$ s case, for which $M = 0.015M_{\odot}$). Red dashed lines show models with $t_0 = 0.01$ s but for which the outer layer of mass $M_n$ is assumed to contains free neutrons instead of r-process nuclei (a model similar to those shown in Fig.~\ref{fig:neutrons}).  Note that the early-time signatures of neutron decay are largely degenerate with late-time shock re-heating of the ejecta.    Figure reproduced with permission from \citet{Metzger+18}, copyright of the authors.}
\label{fig:firsthour}
\end{figure}

The ejecta from a NS merger is extremely hot $\gg 10^{10}$ K immediately after becoming unbound from the central remnant or accretion disk.  However, due to $PdV$ losses, the temperature drops rapidly $\propto  1/R$ as the ejecta radius $R = vt$ expands.  Absent additional sources of heating, the internal energy decays in time as
\begin{eqnarray}
e_0(t) &\simeq& 0.68\frac{\rho^{1/3}s^{4/3}}{a^{1/3}} \nonumber \\
&\approx& 4\times 10^{12}\,{\rm erg\,g^{-1}}\left(\frac{s}{20{\rm k_b\,b^{-1}}}\right)^{4/3}\left(\frac{M}{10^{-2}M_{\odot}}\right)^{1/3}\left(\frac{t}{1\,\rm day}\right)^{-1}\left(\frac{v}{0.1c}\right)^{-1},
\label{eq:e0}
\end{eqnarray}
where the ejecta density has been set to its mean value $\rho = 3M/(4\pi R^{3})$ and the entropy $s$ normalized to a value 20 $k_b$ per baryon typical of the shock-heated polar dynamical or disk wind ejecta (the unshocked tidal tail material can be even colder).  As the ejecta expands, it receives heating from the decay of $r$-process nuclei at the rate given by equation (\ref{eq:rshort}).  The thermal energy input from $r$-process decay on timescales $\sim t$ is thus approximately given by
\be
e_r \sim \dot{e}_r t \approx 9\times 10^{14}\left(\frac{t}{1\,{\rm day}}\right)^{-0.3}\,{\rm erg\,g}^{-1}
\ee
The key point to note is that $e_r \gg e_0$ within minutes after the merger.  This demonstrates why the initial thermal energy of the ejecta can be neglected in calculating the kilonova emission on timescales of follow-up observations of several hours to days (or, specifically, why the toy model light curves calculations presented thus far are insensitive to the precise initial value of $E_v$). 

However, the early-time luminosity of the kilonova could be substantially boosted from this naive expectation if the ejecta is {\it re-heated} at large radii, well after its initial release (i.e. if the ejecta entropy is substantially larger than assumed in eq.~\ref{eq:e0}).  

One way such re-heating could take place is by the passage of a relativistic GRB jet through the polar ejecta, which generates a shocked ``cocoon'' of hot gas (e.g.~\citealt{Gottlieb+18,Kasliwal+17,Piro&Kollmeier18}).  However, the efficiency of this heating process is debated.  \citet{Duffell+18} found, using a large parameter study of jet parameters (jet energies $\sim 10^{48}-10^{51}$ erg and opening angles $\theta \sim 0.07-0.4$ covering the range thought to characterize GRBs) that the thermal energy deposited into the ejecta by the jet falls short of that produced by the $r$-process heating on the same timescale by an order of magnitude or more.  Jet heating is particularly suppressed when the relativistic jet successfully escapes from the ejecta, evidence in GW170817 by late-time afterglow observations \citep{Margutti+17,Margutti+18,Mooley+18}.  

An alternative means to shock-heat the ejecta is by a wind from the magnetized central NS remnant prior to its collapse into a BH \citep{Metzger+18}.  Such a wind is expected to have a wide opening angle and to accelerate to trans-relativistic speeds over a characteristic timescale of seconds (e.g.~\citealt{Metzger+08b}).\footnote{This secular acceleration of the wind is driven by its diminishing baryon-loading rate due to neutrino-driven mass ablation from the neutron star surface following its Kelvin-Helmholtz cooling evolution.}  This delay in the wind acceleration, set by the Kelvin-Helmholtz cooling of the remnant, would naturally allow the dynamical ejecta time to reach large radii before being hit and shocked by the wind.  Although this has not yet been explored in the literature, even time variability in the accretion disk outflows (Sect.~\ref{sec:diskejecta}) could generate internal shocks and re-heat the wind ejecta over timescales comparable to the disk lifetime $\lesssim$ seconds (e.g.~\citealt{Fernandez+19}).  

In all of these mechanisms, re-setting of the ejecta thermal energy at large radii is key to producing luminous emission, because otherwise the freshly-deposited energy is degraded by $PdV$ expansion before being radiated.  This is illustrated by Fig.~\ref{fig:firsthour}, where black lines show how the early-time kilonova light curve is enhanced when the ejecta is re-thermalized (its thermal energy re-set to a value comparable to its kinetic energy, i.e. $E_v \sim \delta M_v v^2/2$) at different times, $t_0$, following its initial ejection.  As expected, larger $t_0$ (later re-thermalization) results in more luminous emission over the first few hours.  

But also note that the light curve enhancement from jet/wind re-heating looks broadly similar that resulting from the outer layers being composed of free neutrons (shown for comparison as red lines in Fig.~\ref{fig:firsthour}).  This degeneracy between free neutrons and delayed shock-heating makes the two physical processes challenging to observationally distinguish \citep{Arcavi18,Metzger+18}.  Well-sampled early-time light curves, e.g.~to search for a subtle bump in the light curve on the neutron half-live $\tau_{\beta} \approx 10^{3}$ s, could be necessary to make progress on the interpretation.  Regardless, the first few hours of the kilonova is an important frontier for future EM follow-up efforts: the signal during this time is sensitive to the origin of the ejecta and how it interacts with the central engine (the details which are largely washed out at later times when $r$-process heating takes over).

\subsection{Long-Lived Engine Power}
\label{sec:engine}

\begin{figure}[!t]
\includegraphics[width=1.\textwidth]{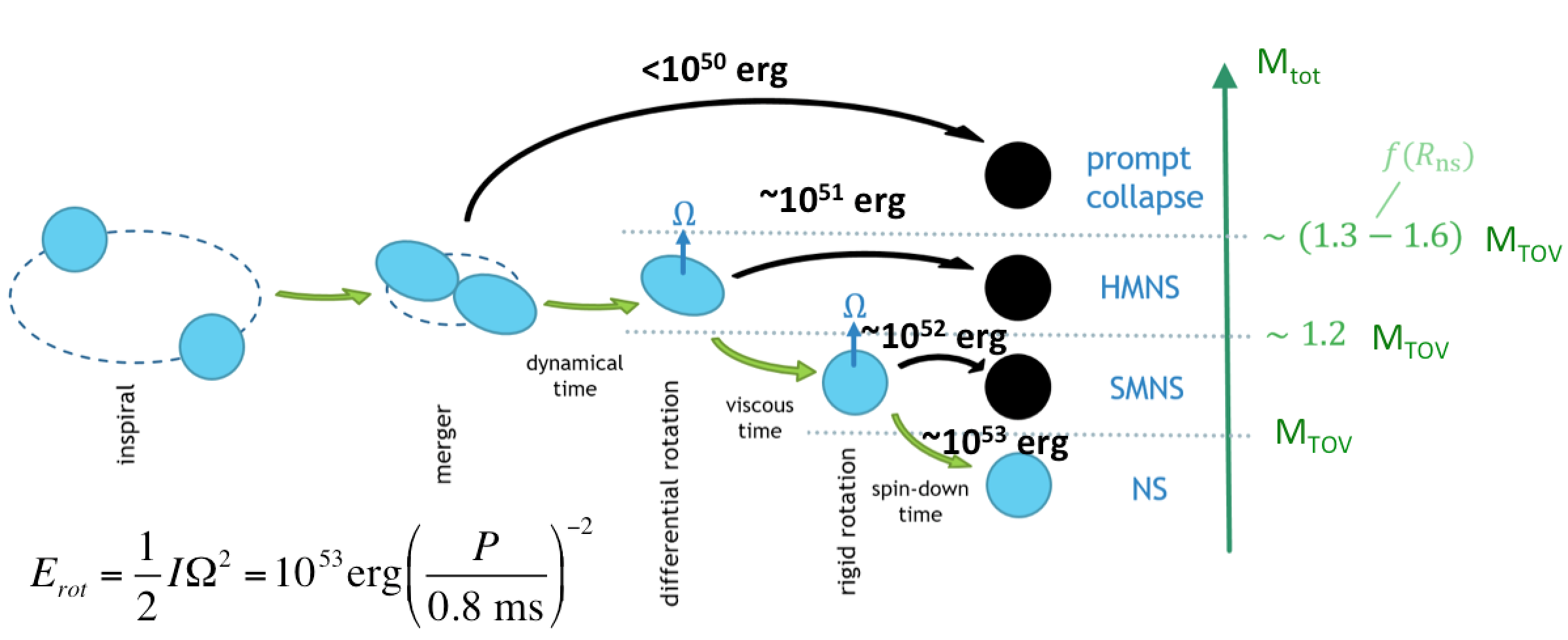}
\caption{The four possible outcomes of a NS-NS merger depend on the total binary mass relative to various threshold masses, each of which is proportional to the maximum mass, $M_{\rm TOV}$, of a non-rotating NS (Table \ref{table:remnants}).  Prompt BH formation or a short-lived HMNS results generates ejecta with a relatively low kinetic energy $\sim 10^{50}-10^{51}$ erg (energy stored in the differential rotation of the HMNS remnant can largely be dissipated as heat and thus lost to neutrinos).  By contrast, the delayed formation of a BH through spin-down of a SMNS or stable remnant takes place over longer, secular timescales and must be accompanied by the release of substantial rotational energy $\sim 10^{52}-10^{53}$ erg.  Unless effectively ``hidden" through GW emission, a large fraction of this energy will be transferred to the ejecta kinetic energy (and, ultimately, the ISM forward shock), thus producing a more luminous kilonova and synchrotron afterglow than for a short-lived remnant.  Figure credit: Ben Margalit.  }
\label{fig:spindown}
\end{figure}

The end product of a NS-NS or BH-NS merger is a central compact remnant, either a BH or a massive NS (Sect.~\ref{sec:ejecta}).  Sustained energy input from this remnant can produce an additional  source of ejecta heating in excess of the minimal contribution from radioactivity, thereby substantially altering the kilonova properties (e.g.~\citealt{Yu+13,Metzger&Piro14,Wollaeger+19}).    

Evidence exists for late-time central engine activity following short GRBs, on timescales from minutes to days.  A fraction $\approx 15-25\%$ of \textit{Swift} short bursts are followed by a distinct hump of hard X-ray emission lasting for tens to hundreds of seconds following the initial prompt spike (e.g.~\citealt{nb06,Perley+09,Kagawa+15}).   The isotropic X-ray light curve of such temporally extended emission in GRB 080503 is shown in the bottom panel of Fig.~\ref{fig:heating} (\citealt{Perley+09}).  Other GRBs exhibit a temporary flattening or ``plateau'' in their X-ray afterglows lasting $\approx 10^2-10^3$ seconds \citep{nkg+06}, which in some cases abruptly ceases \citep{Rowlinson+10}.  X-ray flares are also observed at even later timescales of $\sim$few days \citep{Perley+09,fbm+14}.  The power output of the central engine required to explain this emission is uncertain by several orders of magnitude because it depends on the radiative efficiency and beaming fraction of the (likely jetted) X-ray emission.  Nevertheless, comparison of the left and right panels of Fig.~\ref{fig:heating} makes clear that the energy input of a central remnant can compete with, or even dominate, that of radioactivity on timescales from minutes to weeks after the merger (e.g.~\citealt{Kisaka+17}).

\subsubsection{Fall-Back Accretion}
\label{sec:fallback}

\begin{figure}[!t]
\includegraphics[width=0.5\textwidth]{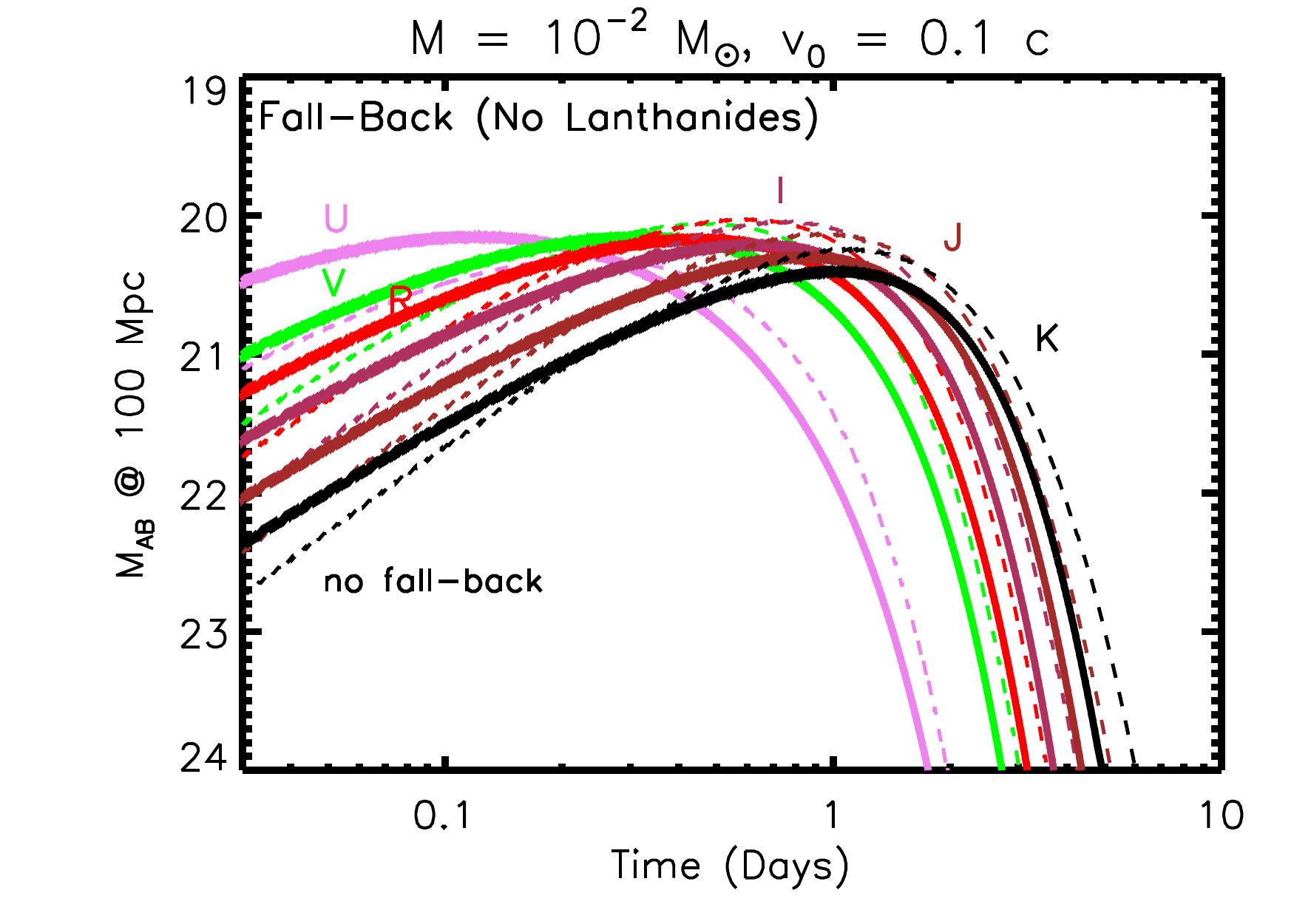}
\includegraphics[width=0.5\textwidth]{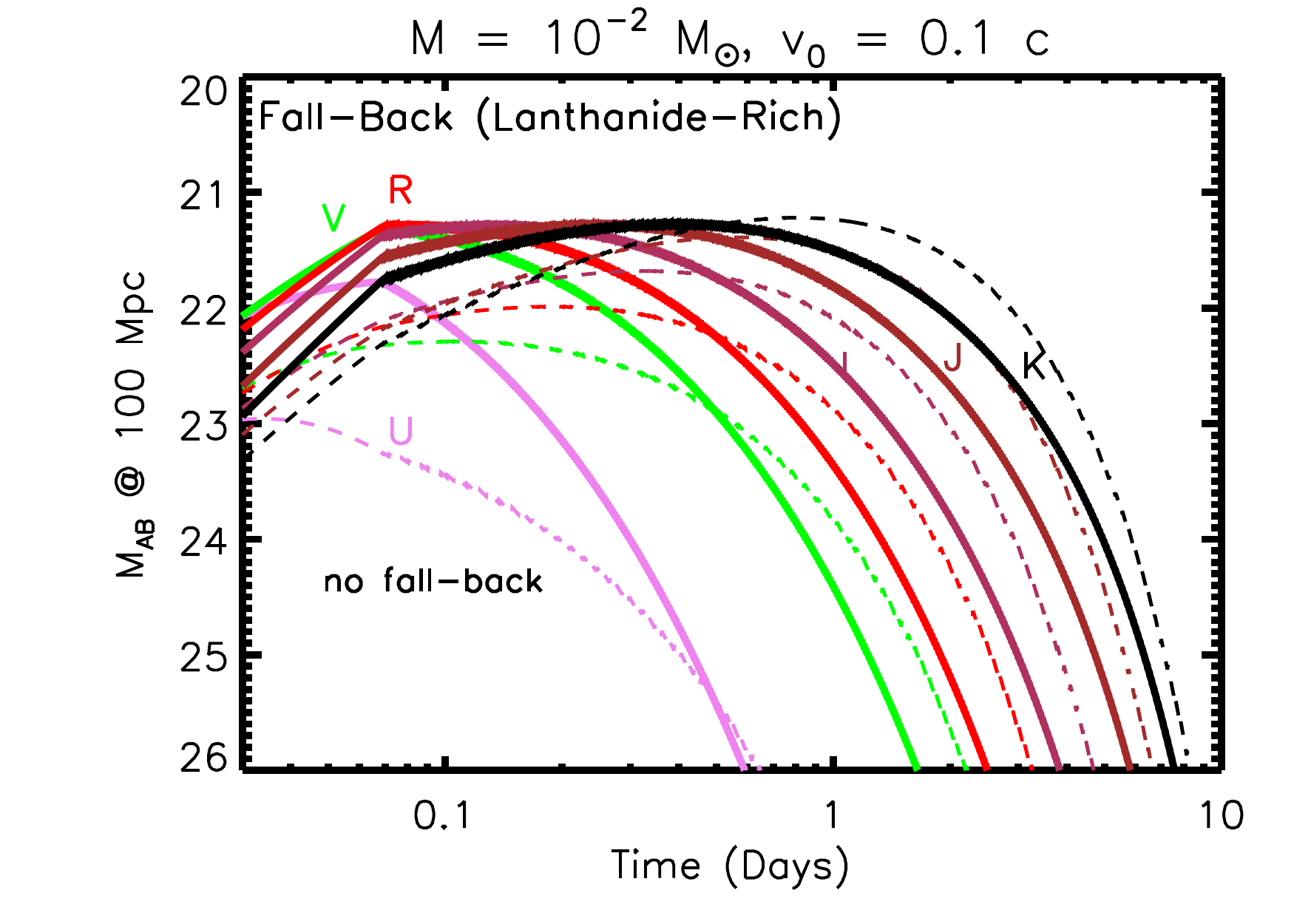}
\caption{Kilonova light curves powered by fall-back accretion, calculated for the same parameters of total ejecta mass $M = 10^{-2}M_{\odot}$ and velocity $v_0 = 0.1$ c used in Fig.~\ref{fig:vanilla}, shown separately assuming opacities appropriate to lanthanide-free ($\kappa = 1$ cm$^{2}$ g$^{-1}$; left panel) and lanthanide-bearing ($\kappa = 20$ cm$^{2}$ g$^{-1}$; right panel) ejecta.   We adopt ejecta heating rate following Eq.~(\ref{eq:Lxfb}) for a constant efficiency $\epsilon_{\rm j} = 0.1$ and have normalized the fall-back mass to an optimistic value $\dot{M}_{\rm fb}(t = 0.1) = 10^{-2}M_{\odot}$ s$^{-1}$.}
\label{fig:fallback}
\end{figure}

In addition to the unbound ejecta during a NS-NS/BH-NS merger (or the accretion disk after the merger), a comparable mass could remain gravitationally bound to the central remnant.  Depending on the energy distribution of this matter, it will fall back to the center and enter the accretion disk over timescales ranging from seconds to days or longer after the coalescence event \citep{Rosswog07,Rossi&Begelman09,Chawla+10,Kyutoku+15}.  At late times $t \gg 0.1$ s, the mass fall-back rate decays as a power-law
\be
\dot{M}_{\rm fb} \approx \left(\frac{\dot{M}_{\rm fb}(t = 0.1\,{\rm s})}{10^{-3}M_{\odot}\,s^{-1}}\right)\left(\frac{t}{0.1\,{\rm s}}\right)^{-5/3},
\label{eq:mdotfb}
\ee
where the normalization $\dot{M}_{\rm fb}(t = 0.1)$ at the reference time $t = 0.1$ s can vary from $\sim 10^{-3}M_{\odot}\,{\rm s}^{-1}$ in NS-NS mergers, to values up to an order of magnitude larger in BH-NS mergers \citep{Rosswog07,Foucart+15}.  

There are several caveats to the presence of fall-back accretion.  Simulations show that disk outflows from the inner accretion flow in BH-NS mergers can stifle the fall-back of material, preventing it from reaching the BH on timescales $t \gtrsim 100$ ms \citep{Fernandez+15}.  Heating due to the $r$-process over the first $\sim 1$ second can also unbind matter that was originally marginally-bound, generating a cut-off in the fall-back rate after a timescale of seconds or minutes \citep{Metzger+10b,Desai+19}.  Furthermore, matter which does return to the central remnant is only tenuously bound and unable to cool through neutrinos, which may drastically reduce the accretion efficiency (the fraction of $\dot{M}_{\rm fb}$ that remains bound in the disk; \citealt{Rossi&Begelman09}).  Despite these concerns, some fall-back and accretion by the central remnant is likely over the days-weeks timescales of the observed kilonovae.

If matter reaches the central compact object at the rate $\dot{M}_{\rm fb}$ (Eq.~\ref{eq:mdotfb}), then a fraction of the resulting accretion power $L_{\rm acc} \propto \dot{M}_{\rm fb}c^{2}$ would be available to heat the ejecta, contributing to the kilonova luminosity.  The accretion flow is still highly super-Eddington throughout this epoch ($L_{\rm acc} \gg L_{\rm Edd} \sim 10^{39}$ erg s$^{-1}$) and might be expected power a collimated ultra-relativistic jet, similar but weaker than that responsible for generating the earlier GRB.  At early times, the jet has sufficient power to propagate through the ejecta, producing high energy emission at larger radii (e.g.~powering the short GRB or temporally-extended X-ray emission following the burst).  However, as the jet power decreases in time it is more likely to become unstable (e.g.~\citealt{Bromberg&Tchekhovskoy16}), in which case its Poynting flux or bulk kinetic energy would be deposited as heat behind the ejecta.  A mildly-relativistic wind could be driven from the inner fall-back-fed accretion disk, which would emerge into the surroundings and collide/shock against the (potentially slower, but higher mass) ejecta shell, thermalizing the wind's kinetic energy and providing a heat source behind the ejecta \citep{Dexter&Kasen13}.  

Heating by fall-back accretion can be crudely parametrized as follows,
\be
\dot{Q}_{\rm fb}= \epsilon_{j} \dot{M}_{\rm fb} c^{2} \approx 2\times 10^{51}\,{\rm erg\,s^{-1}}\left(\frac{\epsilon_{j}}{0.1}\right)\left(\frac{\dot{M}_{\rm fb}(0.1{\rm s})}{10^{-3}M_{\odot}\,s^{-1}}\right)\left(\frac{t}{\, \rm 0.1 s}\right)^{-5/3},
\label{eq:Lxfb}
\ee
where $\epsilon_{j}$ is a jet/disk wind efficiency factor.\footnote{If the jet derives its power from the Blandford--Znajek process, then the jet luminosity actually depends on the magnetic flux threading the BH rather than its accretion rate, at least up to fluxes for which the jet power saturates at  $\epsilon_j \approx 1$ (\citealt{Tchekhovskoy+11}).  \citet{Kisaka&Ioka15} suggest that topology of the accreted magnetic field from fall-back could give rise to a complex temporal evolution of the jet power, which differs from the $\propto t^{-5/3}$ decay predicted by Eq.~(\ref{eq:Lxfb}) for $\epsilon_j = constant$.}
For optimistic, but not physically unreasonable, values of $\epsilon_j \sim 0.01-0.1$ and $\dot{M}_{\rm fb}(0.1{\rm s}) \sim 10^{-3}M_{\odot}$ s$^{-1}$, Fig.~\ref{fig:heating} shows that $\dot{Q}_{\rm fb}$ can be comparable to radioactive heating on timescales of days to weeks.  

Figure \ref{fig:fallback} shows toy model light curves calculated assuming the ejecta (mass $M = 10^{-2}M_{\odot}$ and velocity $v_0 = 0.1$ c) is heated by fall-back accretion according to Eq.~(\ref{eq:Lxfb}) under the optimistic assumption that $\epsilon_{\rm j}\dot{M}_{\rm fb}(t = 0.1) = 10^{-3}M_{\odot}$ s$^{-1}$ on the very high end of the values suggested by merger simulations \citep{Rosswog07} and expected for BH-powered outflows ($\epsilon_j \sim 1$).  The left and right panels show the results separately in the case of ejecta with a low ($\kappa = 1$ cm$^{2}$ g$^{-1}$) and high ($\kappa = 20$ cm$^{2}$ g$^{-1}$) opacity, respectively.  Fall-back enhance the peak brightness, particularly those bands which peak during the first $\lesssim 1$ day, by up to a magnitude or more, compared to the otherwise similar case with pure $r$-process heating (reproduced from Fig.~\ref{fig:vanilla} and shown for comparison with dashed lines).  If the amount of fall-back, and the jet/accretion disk wind efficiency is high, we conclude that accretion power could in principle provide a moderate boost to the observed kilonova emission, particularly in cases where the ejecta mass (and thus intrinsic $r$-process decay power) is particularly low.

Based on the high observed X-ray luminosity from GRB 130603B simultaneous with the excess NIR emission, \cite{Kisaka+16} argue that the latter was powered by reprocessed X-ray emission rather than radioactive heating \citet{Tanvir+13,Berger+13}.  However, the validity of using the observed X-rays as a proxy for the ejecta heating rely on the assumption that the X-ray emission is instrinsically isotropic (i.e.~we are peering through a hole in the ejecta shell), as opposed to being geometrically or relativistically-beamed as a part of a jet-like outflow from the central engine (a substantial beaming-correction to the observed isotropic X-ray luminosity would render it too low to power the observed NIR emission).  \citet{Matsumoto+18} and \citet{Li+18} made a similar argument that AT2017gfo was powered by a central engine.  However, unlike in 130603B, no X-ray emission in excess of the afterglow from the external shocked ISM was observed following GW170817 at the time of the kilonova \citep{Margutti+17}.  Furthermore, the observed bolometric light curve is well explained by $r$-process radioactive decay without the need for an additional central energy source (Fig.~\ref{fig:wu}).

\subsubsection{Long-Lived Magnetar Remnants}
\label{sec:magnetar}

As described in Sect.~\ref{sec:ejecta}, the type of compact remnant produced by a NS-NS merger prompt BH formation, hypermassive NS, supramassive NS, or indefinitely stable NS) depends sensitively on the total mass of the binary relative to the poorly constrained TOV mass, $M_{\rm TOV}$.  A lower limit of $M_{\rm TOV} \gtrsim 2-2.1M_{\odot}$ is set by measured pulsar masses \citep{Demorest+10,Antoniadis+13,Cromartie+19}, while an upper limit of $M_{\rm TOV} \lesssim 2.16M_{\odot}$ is suggested for GW170817 (Sect.~\ref{sec:EOS}).  Taken as granted, these limits, combined with the assumption that the measured mass distribution of the Galactic population of binary neutron stars is representative of those in the universe as a whole, leads to the inference that $\approx 18-65\%$ of mergers will result in a long-lived SMNS \citep{Margalit&Metzger19} instead of the short-lived HMNS most believe formed in GW170817 (Table \ref{table:remnants}).   

The massive NS remnant created by a NS-NS merger will in general have more than sufficient angular momentum to be rotating near break-up (\citealt{Radice+18c}; however, see \citealt{Shibata+19}).  A NS of mass $M_{\rm ns}$ rotating near its mass-shedding limit possesses a rotational energy,
\be
E_{\rm rot} = \frac{1}{2}I\Omega^{2} \simeq 1\times 10^{53}\left(\frac{I}{I_{\rm LS}}\right)\left(\frac{M_{\rm ns}}{2.3 M_{\odot}}\right)^{3/2}\left(\frac{P}{\rm 0.7\rm ms}\right)^{-2}\,{\rm erg},
\label{eq:Erot}
\ee
where $P = 2\pi/\Omega$ is the rotational period and $I$ is the NS moment of inertia, which we have normalized to an approximate value for a relatively wide class of nuclear equations of state $I_{\rm LS} \approx  1.3\times 10^{45}(M_{\rm ns}/1.4M_{\odot})^{3/2}\mathrm{\ g\ cm}^{2}$, motivated by  Fig.~1 of \citet{Lattimer&Schutz05}.  This energy reservoir is enormous, both compared to the kinetic energy of the merger ejecta ($\approx 10^{50}-10^{51}\mathrm{\ erg}$) and to that released by its radioactive decay.  Even if only a modest fraction of $E_{\rm rot}$ were to be extracted from the remnant hours to years after the merger by its electromagnetic spin-down, this would substantially enhance the EM luminosity of the merger counterparts \citep{Yu+13,Gao+13,Metzger&Piro14,Gao+15,Siegel&Ciolfi16a}.

This brings us back to a crucial qualitative difference between the formation of a HMNS and a SMNS or stable NS remnant.  A HMNS can be brought to the point of collapse by the accretion of mass and redistribution of its internal angular momentum.  However, energy dissipated by removing internal differential rotational support can largely be released as heat and thus will escape as neutrino emission (effectively unobservable at typical merger distances).  Thus, the angular momentum of the binary can largely be trapped in the spin of the newly-formed BH upon its collapse, rendering most of $E_{\rm rot}$ unavailable to power EM emission.  

By contrast, a SMNS is (by definition) supported by its solid-body rotation, even once all forms of differential rotation have been removed.  Thus, {\it angular momentum must physically be removed from the system to allow the collapse, and the removal of angular momentum brings with it a concomitant amount of rotational energy.}  The left panel of Fig.~\ref{fig:MM} shows the ``extractable" rotational energy  for mergers which leave SMNS remnants and how quickly it grows with decreasing binary chirp mass (proxy for the mass $M_{\rm tot}$).\footnote{The extractable energy is defined as the difference between the rotational energy at break-up and the rotational energy after the object has spun down to the point of becoming unstable and collapsing into a BH.}  This energy budget increases from $\lesssim 10^{51}$ erg for remnants near the HMNS-SMNS boundary at $M_{\rm tot} \approx 1.2M_{\rm TOV}$ to the full rotational energy $E_{\rm rot} \approx 10^{53}$ erg (eq.~\ref{eq:Erot}) for the lowest mass, indefinitely stable remnants $M_{\rm tot} \lesssim M_{\rm TOV}$.

A strong magnetic field provides an agent for extracting rotational energy from the NS remnant via electromagnetic spin-down (the same mechanism at work in ordinary pulsars).  MHD simulations show that the original magnetic fields of the NS in a NS-NS merger are amplified to very large values, similar or exceeding the field strengths of $10^{15}-10^{16}$ G of Galactic magnetars \citep{Price&Rosswog06,Zrake&MacFadyen13,Kiuchi+14}.  However, most of this amplification occurs on small spatial scales, and at early times when the NS is still differentially-rotating, resulting in a complex and time-dependent field geometry \citep{Siegel+14}.  Nevertheless, by the time the NS enters into a state of solid body rotation (typically within hundreds of milliseconds following the merger), there are reasons to believe the remnant could possess an ordered dipole magnetic field of comparable strength, $B \sim 10^{15}-10^{16}$ G.  For instance, an ordered magnetic field can be generated by an $\alpha-\Omega$ dynamo, driven by the combined action of its rapid millisecond rotation period and thermal and lepton gradient-driven convection in the cooling remnant \citep{Thompson&Duncan93}.  

The spin-down luminosity of an aligned dipole\footnote{Unlike vacuum dipole spin-down, the spin-down rate is not zero for an aligned rotator in the force-free case, which is of greatest relevance to the plasma-dense, post-merger environment.} rotator is given by (e.g.~\citealt{Philippov+15})
\be
L_{\rm sd}   = \left\{
\begin{array}{lr}
\frac{\mu^{2}\Omega^{4}}{c^{3}} =  7\times 10^{50}\,{\rm erg\,s^{-1}}\left(\frac{I}{I_{\rm LS}}\right)\left(\frac{B}{10^{15}\,{\rm G}}\right)^{2}\left(\frac{P_{\rm 0}}{\rm 0.7\,ms}\right)^{-4}\left(1 + \frac{t}{t_{\rm sd}}\right)^{-2}
, &
t < t_{\rm collapse}\\
0 &
t > t_{\rm collapse} \\
\end{array}
\label{eq:Lsd}
\right. ,
\ee
where $\mu = B R_{\rm ns}^{3}$ is the dipole moment, $R_{\rm ns} = 12\,{\rm km}$ is the NS radius, $B$ is the surface equatorial dipole field, 
\be
t_{\rm sd} = \left.\frac{E_{\rm rot}}{L_{\rm sd}}\right|_{t = 0}\simeq 150\,{\rm s}\left(\frac{I}{I_{\rm LS}}\right)\left(\frac{B}{10^{15}\,{\rm G}}\right)^{-2}\left(\frac{P_{\rm 0}}{\rm 0.7\,ms}\right)^{2}
\label{eq:tsd}
\ee
is the characteristic spin-down time over which an order unity fraction of the rotational energy is removed, where $P_{0}$ is the initial spin-period and we have assumed a remnant mass of $M = 2.3M_{\odot}$.  The latter is typically close to, or slightly exceeding, the mass-shedding limit of $P = 0.7$ ms.\footnote{If the remnant is born with a shorter period, mass shedding or non-axisymmetric instabilities set in which will result in much more rapid loss of angular momentum to GWs \citep{Shibata+00}, until the NS rotates at a rate close to $P_0 \gtrsim 0.7$ ms.}


The spin-down luminosity in Eq.~(\ref{eq:Lsd}) goes to zero\footnote{The collapse event itself has been speculated to produce a brief (sub-millisecond) electromagnetic flare \citep{Palenzuela+13} or a fast radio burst \citep{Falcke&Rezzolla13} from the detaching magnetosphere; however, no accretion disk, and hence long-lived transient, is likely to be produced \citep{Margalit+15}.} when the NS collapses to the BH at time $t_{\rm collapse}$.  For a stable remnant, $t_{\rm collapse} \rightarrow \infty$, but for supramassive remnants, the NS will collapse to a black hole after a finite time which can be estimated\footnote{We also assume that dipole spin-down exceeds gravitational wave losses, as is likely valid if the non-axisymmetric components of the interior field are $\lesssim 100$ times weaker than the external dipole field \citep{DallOsso+09}.  We have also neglected angular momentum losses due to $f$-mode instabilities \citep{Doneva+15}.} by equating $\int_0^{t_{\rm collapse}}L_{\rm sd}dt$ to the extractable rotational energy.  

\cite{Yu+13} suggested\footnote{In fact, \cite{Kulkarni05} earlier had suggested energy input from a central pulsar as a power source.} that magnetic spin-down power, injected by the magnetar behind the merger ejecta over a timescale of days, could enhance the kilonova emission (the termed such events ``merger-novae''; see also \citealp{Gao+15}).  Their model was motivated by similar ideas applied to super-luminous supernovae \citep{Kasen&Bildsten10,Woosley10,Metzger+14} and is similar in spirit to the `fall-back powered' emission described in Sect.~\ref{sec:fallback}.  Although the spin-down luminosity implied by Eq.~(\ref{eq:Lsd}) is substantial on timescales of hours to days, the fraction of this energy which will actually be thermalized by the ejecta, and hence available to power kilonova emission, may be much smaller.

As in the Crab Nebula, pulsar winds inject a relativistic wind of electron/positron pairs.  This wind is generally assumed to undergo shock dissipation or magnetic reconnection near or outside a termination shock, inflating a nebula of relativistic particles \citep{Kennel&Coroniti84}.  Given the high energy densities of the post-NS-NS merger environment, these heated pairs cool extremely rapidly via synchrotron and inverse Compton emission inside the nebula \citep{Metzger+14, Siegel&Ciolfi16a, Siegel&Ciolfi16b}, producing broadband radiation from the radio to gamma-rays (again similar to conventional pulsar wind nebulae; e.g., \citealp{Gaensler&Slane06}).  A fraction of this non-thermal radiation, in particular that at UV and soft X-ray frequencies, will be absorbed by the neutral ejecta walls and reprocessed to lower, optical/IR frequencies \citep{Metzger+14}, where the lower opacity allows the energy to escape, powering luminous kilonova-like emission.

On the other hand, this non-thermal nebular radiation may also escape directly from the ejecta without being thermalized, e.g.~through spectral windows in the opacity.  This can occur for hard X-ray energies above the bound-free opacity or (within days or less) for high energy $ \gg$ MeV gamma-rays between the decreasing Klein--Nishina cross section and the rising photo-nuclear and $\gamma-\gamma$ opacities  (Fig.~\ref{fig:opacities}).  Furthermore, if the engine is very luminous and the ejecta mass sufficiently low, the engine can photo-ionize the ejecta shell, allowing radiation to freely escape even from the far UV and softer X-ray bands (where bound-free opacity normally dominates).  While such leakage from the nebula provides a potential isotropic high energy counterpart to the merger at X-ray wavelengths \citep{Metzger&Piro14,Siegel&Ciolfi16a, Siegel&Ciolfi16b,Wang+16}, it also reduces the fraction of the magnetar spin-down luminosity which is thermalized and available to power optical-band radiation.  

\begin{figure}[!t]
\includegraphics[width=0.5\textwidth]{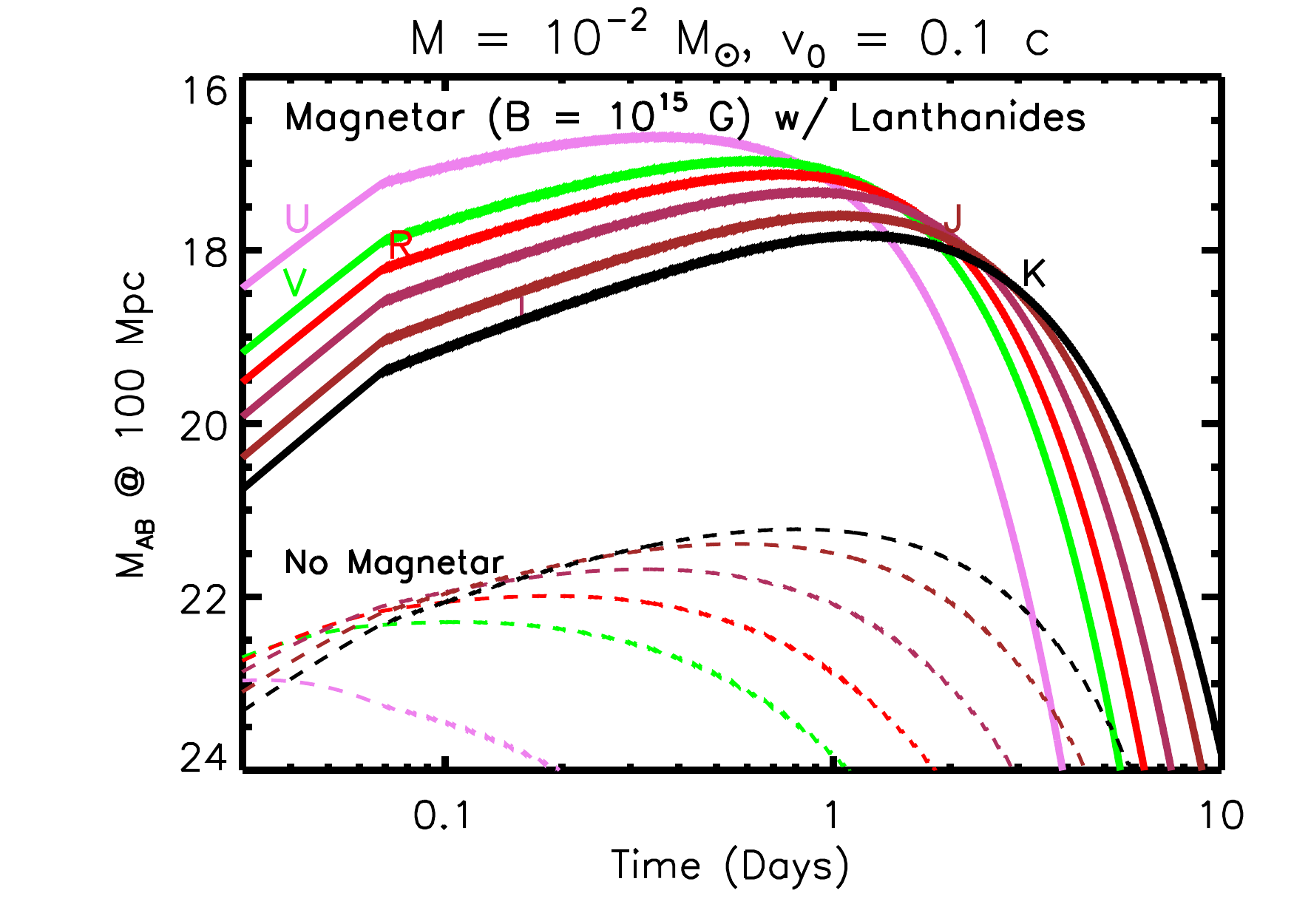}
\includegraphics[width=0.5\textwidth]{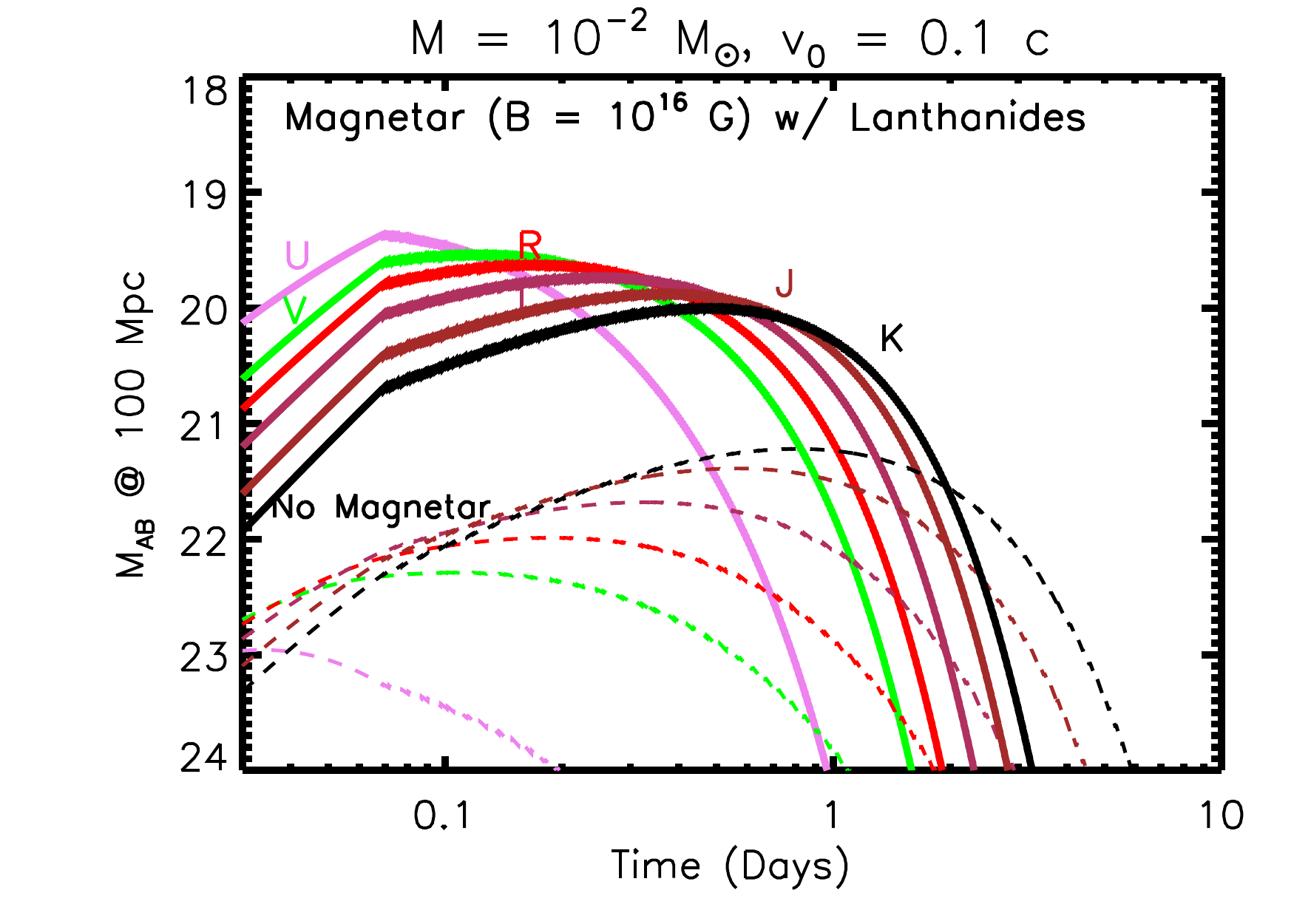}
\caption{Kilonova light curves, boosted by spin-down energy from an indefinitely stable magnetar ($t_{\rm collapse} = \infty$), and taking an opacity $\kappa = 20$ cm$^{2}$ g$^{-1}$ appropriate to lanthanide-rich matter.  We assume an ejecta mass $M = 0.1M_{\odot}$ \citep{Metzger&Fernandez14}, initial magnetar spin period $P_0 = 0.7$ ms,  thermalization efficiency $\epsilon_{\rm th} = 1$ and magnetic dipole field strength of $10^{15}$ G (left panel) or $10^{16}$ G (right panel).  Dashed lines show for comparison the purely $r$-process powered case.}
\label{fig:magnetar15}
\end{figure}

\begin{figure}[!t]
\includegraphics[width=0.5\textwidth]{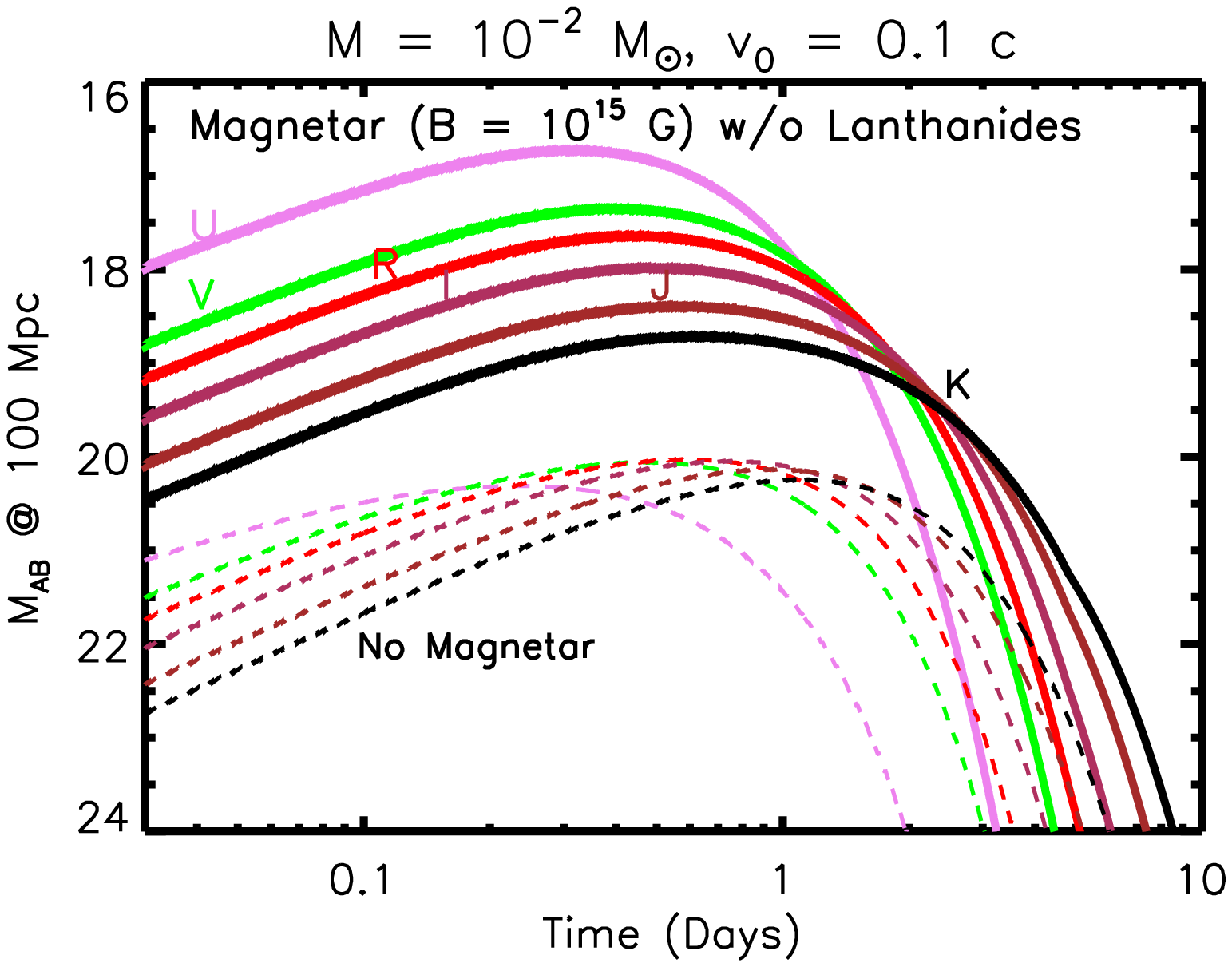}
\includegraphics[width=0.5\textwidth]{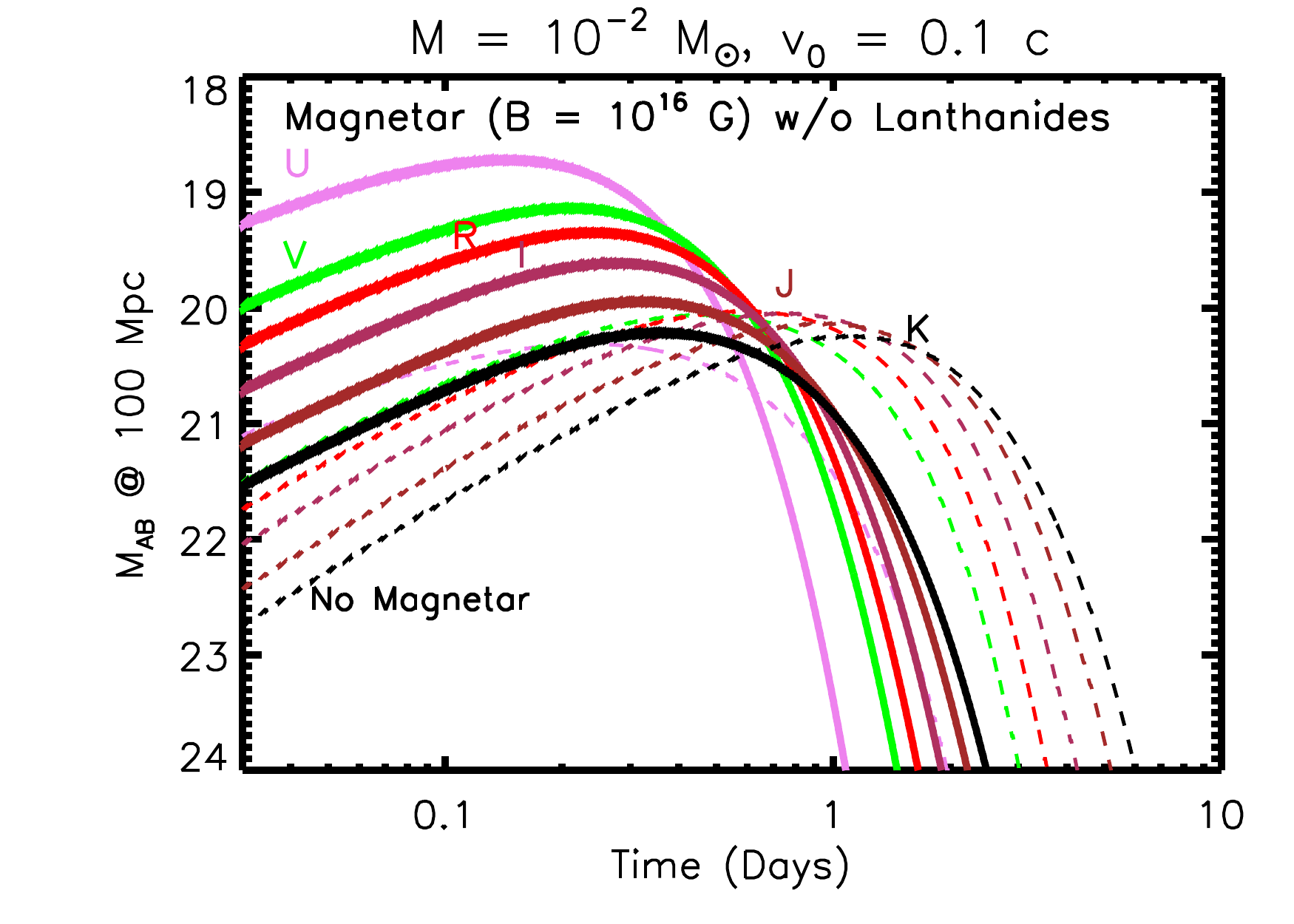}
\caption{Same as Fig.~\ref{fig:magnetar15}, but calculated for an ejecta opacity $\kappa = 1$ cm$^{2}$ g$^{-1}$ relevant to lanthanide-free matter.}
\label{fig:magnetar16}
\end{figure}

We parameterize the magnetar contribution to the ejecta heating as
\be
\dot{Q}_{\rm sd} = \epsilon_{\rm th}L_{\rm sd},
\label{eq:qdotsd}
\ee  
where, as in the fall-back case (Eq.~\ref{eq:Lxfb}), $\epsilon_{\rm th}$ is the thermalization efficiency.  We expect $\epsilon_{\rm th} \sim 1$ at early times when the ejecta is opaque (unless significant energy escapes in a jet), but the value of $\epsilon_{\rm th}$ will decrease as the optical depth of the expanding ejecta decreases, especially if the ejecta becomes ionized by the central engine.

\cite{Metzger&Piro14} point out another inefficiency, which, unlike radiation leakage, is most severe at early times.  High energy $\gtrsim$ MeV gamma-rays in the nebula behind the ejecta produce copious electron/positron pairs when the compactness is high.  These pairs in turn are created with enough energy to Compton upscatter additional seed photons to sufficient energies to produce another generation of pairs (and so on...).  For high compactness $\ell \gg 1$, this process repeats multiple times, resulting in a `pair cascade' which acts to transform a significant fraction $Y \sim 0.01-0.1$ of the pulsar spin-down power $L_{\rm sd}$ into the rest mass of electron/positron pairs \citep{Svensson87,Lightman+87}.  Crucially, in order for non-thermal radiation from the central nebula to reach the ejecta and thermalize, it must diffuse radially through this pair cloud, during which time it experiences adiabatic PdV losses.  Because at early times the Thomson optical depth of the pair cloud, $\tau_{\rm n}$, actually exceeds the optical depth through the ejecta itself, this suppresses the fraction of the magnetar spin-down power which is available to thermalize and power the emission.

Following \cite{Metzger&Piro14} and \cite{Kasen+16}, we account in an approximate manner for the effect of the pair cloud by suppressing the observed luminosity according to,
\be
L_{\rm obs} = \frac{L}{1 + (t_{\rm life}/t)}
\label{eq:Lobs}
\ee
where $L$ is the luminosity of the kilonova, calculated as usual from the energy equation (\ref{eq:dEdt}) using the magnetar heat source (Eq.~$\ref{eq:qdotsd}$), and
\be
\frac{t_{\rm life}}{t} = \frac{\tau_{\rm n}v}{c(1-A)} \approx \frac{0.6}{1-A}\left(\frac{Y}{0.1}\right)^{1/2}\left(\frac{L_{\rm sd}}{10^{45}\, {\rm erg\,s^{-1}}}\right)^{1/2}\left(\frac{v}{0.3\,\rm c}\right)^{1/2}\left(\frac{t}{\rm 1\, day}\right)^{-1/2}
\ee
is the characteristic `lifetime' of a non-thermal photon in the nebula relative to the ejecta expansion timescale, where $A$ is the (frequency-averaged) albedo of the ejecta.  In what follows we assume $A = 0.5$.

For high spin-down power and early times ($t_{\rm life} \gg t$), pair trapping acts to reduce the thermalization efficiency of nebular photons, reducing the effective luminosity of the magnetar-powered kilonova by several orders of magnitude compared to its value were this effect neglected.  The bottom panel of Fig.~\ref{fig:heating} shows the spin-down luminosity $L_{\rm sd}$ for stable magnetars with $P_0 = 0.7$ ms and $B = 10^{15}, 10^{16}$ G.  We also show the spin-down power, `corrected' by the factor $(1 + t_{\rm life}/t)^{-1}$, as in Eq.~(\ref{eq:Lobs}) for $Y = 0.1$.\footnote{We emphasize, however, that when one is actually calculating the light curve, the pair suppression (Eq.~\ref{eq:Lobs}) should be applied \emph{after} the luminosity has been calculated using the full spin-down power as the heating source (Eq.~\ref{eq:qdotsd}).  This is because the non-thermal radiation trapped by pairs is also available to do PdV work on the ejecta, accelerating it according to Eq.~(\ref{eq:dvdt}).}


Figures \ref{fig:magnetar15} and \ref{fig:magnetar16} shows kilonova light curves, calculated from our toy model, but including additional heating of the ejecta due to rotational energy input from an indefinitely stable magnetar with assumed dipole field strengths of $B = 10^{15}$ G and $10^{16}$ G, respectively.  We again show separately cases in which we assume a high value for the ejecta opacity $\kappa = 20$ cm$^{2}$ g$^{-1}$ appropriate for lanthanide-rich ejecta compared to low opacity $\kappa = 1$ cm$^{2}$ g$^{-1}$ appropriate to Lanthanide-free (light $r$-process) elements.  The latter case is probably the most physical one, because the majority of the disk wind ejecta (which typically dominates the total) is expected to possess a high-$Y_e$ in the presence of a long-lived stable merger remnant ($t_{\rm life} \approx \infty$ in Fig.~\ref{fig:Lippuner}).  

A long-lived magnetar engine has three main effects on the light curve relative to the normal (pure $r$-process-powered) case: (1) increase in the peak luminosity, by up to $4-5$ magnitudes; (2) more rapid evolution, i.e. an earlier time of peak light; (3) substantially bluer colors.  Feature (1) is simply the result of additional heating from magnetar spin-down, while feature (2) results from the greater ejecta velocity due to the kinetic energy added to the ejecta by the portion of the spin-down energy released before the ejecta has become transparent (that portion going into $PdV$ rather than escaping as radiation).  Feature (3) is a simple result of the fact that the much higher luminosity of the transient increases its effective temperature, for an otherwise similar photosphere radius near peak light. 

Figs.~\ref{fig:magnetar15} and \ref{fig:magnetar16} represent close to the most ``optimistic" effects a long-lived magnetar could have on the kilonova light curves.  The effects will be more subtle, and closer to the radioactivity-only models, if the magnetar is a less-stable SMNS (that collapses into a BH before all of $E_{\rm rot}$ is released) or if a substantial fraction of its rotational energy escapes as gamma-rays or is radiated by the magnetar through GW emission (e.g.~\citealt{DallOsso&Stella07,Corsi&Meszaros09}), rather than being transferred to the ejecta through its magnetic dipole spin-down.  Such effects are easy to incorporate into the toy model by cutting off the magnetar spin-down heating for  $t \ge t_{\rm collapse}$ in equation (\ref{eq:Lsd}), or by including additional losses due to GW radiation into the magnetar spin-down evolution (e.g.~ \citealt{Li+18} and references therein).


\section{Observational Prospects and Strategies}
\label{sec:discussion}

\begin{figure}[!t]
\includegraphics[width=1.0\textwidth]{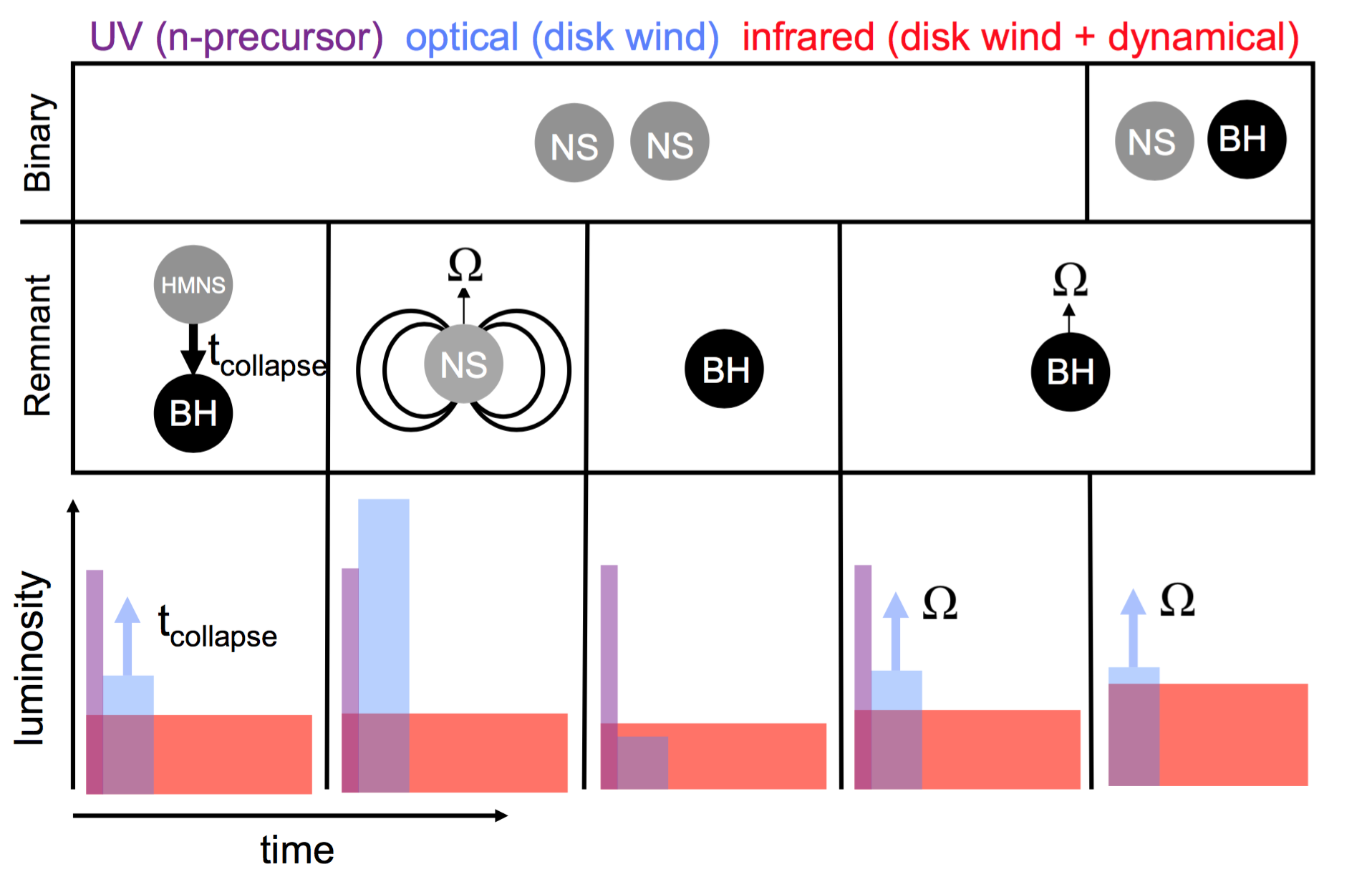}
\caption{Schematic illustration mapping different types of mergers and their outcomes to trends in their kilonova light curves. The top panel shows the progenitor system, either an NS-NS or an NS-BH binary, while the middle plane shows the final merger remnant (from left to right: an HMNS that collapses to a BH after time $t_{\rm collapse}$, a spinning magnetized NS, a non-spinning BH and a rapidly spinning BH). The bottom panel illustrates the relative amount of UV/blue emission from an neutron precursor (purple), optical emission from lanthanide-free material (blue) and IR emission from lanthanide containing ejecta (red).  Note: the case of a NS-NS merger leading to a slowly spinning black hole is unlikely, given that at a minimum the remnant will acquire the angular momentum of the original binary orbit.  Modified from a figure originally presented in \citet{Kasen+15}, copyright by the authors.}
\label{fig:outcomes}
\end{figure}

With the basic theory of kilonova in place, we now discuss several implications for past and present kilonova observations.  Table \ref{table:range} provides rough estimates for the expected range in kilonova luminosities, timescales, isotropy to accompany NS-NS and BH-NS mergers for different assumptions about the merger remnant/outcome.  Figure \ref{fig:outcomes} illustrates some of this diversity graphically.

\begin{table}[!t]
\caption{Range of Kilonova Properties \label{table:range}}

\begin{tabular}{ccccccc}

Event & Remnant/Outcome & $L_{\rm pk}$$^{(a)}$ (erg s$^{-1}$) & $t_{\rm pk}$$^{(b)}$ & Color & Isotropic?$^{(c)}$  \\

\hline
NS-NS & Prompt BH & $\sim 10^{40}-10^{41}$ & $\sim 3$ day & Mostly Red & N \\
 - 	& HMNS $\Rightarrow$ BH & $\sim 10^{41}-10^{42}$ & $\sim 1$ day & Blue \& Red & $\sim$ Y \\
- & SMNS  $\Rightarrow$  BH & $\sim 10^{42}-10^{44}$ & $\lesssim 1$ day & Mostly Blue & $\sim$ Y \\
- & Stable NS & $\sim 10^{43}-10^{44}$ & $\lesssim $ 1 day & Mostly Blue & $\sim$ Y \\
BH-NS & $R_{\rm t} \gtrsim R_{\rm isco}$$^{(d)}$ & $\sim 0$ & N/A  & N/A & N/A \\
- & $R_{\rm t} \lesssim R_{\rm isco}$ & $\sim 10^{41}-10^{42}$ & $\sim $ 1 week & Mostly Red & N \\
\hline \\
\end{tabular}
\\
$^{(a)}$Estimated range in peak luminosity.  Does not account for extra sources of early-time heating from free neutrons or shocks, which could enhance the peak luminosity in the first hours (Sect.~\ref{sec:firsthour}).  $^{(b)}$Estimated peak timescale.  $^{(c)}$Whether to expect large pole-equatorial (or azimuthal, in the NS-BH case) isotropy in the kilonova properties. $^{(d)}$Whether a BH-NS merger is accompanied by mass ejection depends on whether the NS is tidally disrupted sufficiently far outside of the BH event horizon to generate unbound tidal material and the formation of an accretion disk.  Very roughly, this condition translates into a comparison between the tidal radius of the NS, $R_{\rm t}$ (which depends on the BH-NS mass ratio and the NS radius), and the radius of the innermost stable circular orbit of the BH, $R_{\rm isco}$ (which depends on the BH mass and spin);  see futher discussion in Sect.~\ref{sec:ejecta}.  
\end{table}

\subsection{Kilonova Candidates Following Short GRBs}
\label{sec:candidates}

If short duration GRBs originate from NS-NS or NS-BH mergers, then one way to constrain kilonova models is via optical and NIR follow-up observations of nearby short bursts on timescales of hours to a week.  All else being equal, the closest GRBs provide the most stringent constraints; however, the non-thermal afterglow emission---the strength of which can vary from burst to burst---must also be relatively weak, so that it does not outshine the thermal kilonova.  Blue kilonova emission similar in luminosity to GW170817 would have been outshone by the afterglow emission in all but a small handful of observed short GRBs, but the longer-lived red kilonova emission has a better chance of sticking out above the fading, relatively blue afterglow.  

The NIR excess observed following GRB 130603B \citep{Berger+13,Tanvir+13}, if powered by the radioactive decay of $r$-process nuclei, required a total ejecta mass of lanthanide-bearing matter of $\gtrsim 0.1M_{\odot}$ \citep{Barnes+16}.  This is $\sim 3-5$ times greater than the ejecta mass inferred for GW170817 (\citealt{Fong+17}; Sect.~\ref{sec:170817}).  As with GW170817, the ejecta implied by kilonova models of 130603B is too high to be explained by the dynamical ejecta from a NS-NS merger, possibly implicating a BH-NS merger in which the NS was tidally disrupted well outside the BH horizon \citep{Hotokezaka+13b,Tanaka+14,Kawaguchi+16}.  However, NS-NS mergers can also produce such a high ejecta mass if a large fraction of the remnant accretion disk (which possess masses up to $\sim 0.2M_{\odot}$) is unbound in disk winds (\citealt{Siegel&Metzger17} found that $\approx 40\%$ of the disk mass could be unbound).  Alternatively, the unexpectedly high luminosity of this event could attributed to energy input from a central engine rather than radioactivity \citep{Kisaka+16}, which for fall-back accretion indeed produces the correct luminosity to within an order-of-magnitude (Fig.~\ref{fig:fallback}).  

\cite{Yang+15,Jin+15,Jin+16} found evidence for NIR emission in excess of the expected afterglow following the short GRBs 050709 and 060614, indicative of possible kilonova emission.  The short GRB 080503 \citep{Perley+09} showed an optical peak on a timescale of $\sim 1$ day, which could be explainded as blue kilonova powered by $r$-process heating \citep{Metzger&Fernandez14,Kasen+15} or a central engine \citep{Metzger&Piro14,Gao+15}.  These possibilities unfortunately could not be distinguished because the host galaxy of GRB080503 was not identified, resulting in its distance and thus luminosity being unconstrained.\footnote{Rebrightening in the X-ray luminosity, coincident with the optical brightening, was also observed following GRB 080503 \citep{Perley+09}.  Whether the optical emission is powered exclusively by $r$-process heating or not, this could potentially be consistent with non-thermal emission from a central engine \citep{Metzger&Piro14,Gao+15,Siegel&Ciolfi16a,Siegel&Ciolfi16b}.}

\citet{Gompertz+18} found three short bursts (GRBs 050509b, 061201, and 080905A) where, if the reported redshifts were correct, deep upper limits rule out the presence of a kilonova similar to AT2017gfo by several magnitudes (see also \citealt{Fong+17}). Given the diverse outcomes of NS-NS mergers, and how the properties of the remnant can dramatically effect the quantity and composition of the kilonova outflows, variation in the ejecta properties by an order of magnitude or more would not be unexpected.  For instance, high-mass mergers that undergo a prompt collapse to a BH eject substantially less mass (particularly of the high-$Y_e$ kind capable of producing blue kilonova emission), but still could generate an accretion disk of sufficient mass to power a GRB jet.\footnote{If the jet is powered by the Blandford-Znajek mechanism, the disk mass $M_{\rm t}$ only need be sufficiently massive to hold a magnetic field of the requisite strength in place.  This is not a very stringent constraint: GRB jets are weak in energy $E_{\rm GRB} \sim 10^{49}$ erg compared to the maximum accretion power available, $\sim M_{\rm t}c^{2} \sim 10^{52}(M_{\rm t}/10^{-2}M_{\odot})$ erg.}  

Even with deep observations of a particularly nearby burst, \cite{Fong+16b} emphasize the challenges to constrain vanilla blue/red kilonova models with ground follow-up of GRBs.  This highlights the crucial role played by the Hubble Space Telescope, and in the future by the James Webb Space Telescope (JSWT) and Wide Field Infrared Survey Telescope (WFIRST), in such efforts.  Fortunately, NS-NS mergers detected by Advanced LIGO at distances $<200$ Mpc (redshift $z < 0.045$) are at least three times closer ($>2.5$ mags brighter) than the nearest cosmological short GRBs.  Nevertheless, the detection rate of well-localized short GRBs is currently higher than GW events.  For this reason among others (e.g.~the information obtained on GRB jet opening angles from afterglow jet breaks), we advocate for continued space-based observations performing late-time follow-up of short GRB afterglows to search for kilonova signatures.

\subsection{Gravitational Wave Follow-Up}
\label{sec:detection}

This review has hopefully made clear that kilonovae are not likely to be homogeneous in their properties, with potentially significant differences expected in their colors and luminosities, depending on the type of merging system, the properties of the in-going binary, and, potentially, our viewing angle relative to the binary inclination (see, Figs.~\ref{fig:MM}, \ref{fig:outcomes} and Table \ref{table:range}).  Here we discuss prospects and strategies of GW follow-up in the case of different merger outcomes.   

\noindent{\bf Prompt Collapse to BH.} In a NS-NS merger the observed emission is expected to depend sensitively on the lifetime of the central NS remnant, which in turn will depend on the in-going binary mass (Table \ref{table:remnants}).  When BH formation is prompt, the ejecta mass will in most cases be low $\ll 10^{-2}M_{\odot}$ and radioactivity (and, potentially, fall-back accretion of the tidal tail) will provide the only heating sources.  The lack of a HMNS remnant will also result in a greater fraction of ejecta being lanthanide-rich and thus generating red kilonova emission.  While some blue ejecta could still originate from the accretion disks, its relatively low velocity $\sim 0.1$ c could result in its emission being blocked for equatorial viewing angles.  Little or no neutron precursor emission is expected.   For a merger at $\sim 100$ Mpc, a purely red $r$-process powered kilonova of ejecta mass $\sim 3\times 10^{-3}M_{\odot}$ would peak over a timescale of a few days in the NIR at $IJK \sim 23-24$ (scaling from Fig.~\ref{fig:vanilla}, right panel).   Given such relatively dim emission, only the largest aperture telescopes (e.g.~ DECam, Subaru HSC, or LSST) are capable of detecting the low-$M_{\rm ej}$ red kilonova of a prompt collapse (Fig.~\ref{fig:vanilla}, right panel).

\noindent{\bf HMNS Remnant (GW170817-like).} The situation is more promising for lower-mass mergers in which at least a moderately long-lived HMNS remnant forms.  Shock-heated matter from the merger interface can generate high-$Y_e$ dynamical ejecta comparable of exceeding that of the lanthanide-rich tidal tail.  Likewise, outflows from the magnetized HMNS,  or from the accretion disk prior to BH formtion (Fig.~\ref{fig:Lippuner}), will produce a greater quantity of high-$Y_e$ lanthanide free material than in the prompt collapse case.  For a merger generating $\sim 10^{-2}M_{\odot}$ of blue kilonova ejecta  (similar to GW170817) at $\sim 100$ Mpc, the resulting blue kilonova emission could peak at UVR $\sim$ 19-20  (Fig.~\ref{fig:vanilla}, right panel) on a timescale of several hours to days.  Even if this blue emission is somehow not present or is blocked by lanthanide-rich matter, a source at 100 Mpc could still reach $U \sim 20$ on a timescale of hours if the outer layers of the ejecta contain free neutrons (Fig.~\ref{fig:neutrons}) or if the ejecta has been shock heated within a second after being first ejected (Fig.~\ref{fig:firsthour}).  Although not much brighter in magnitude than the NIR peak at later times, the blue kilonova may be the most promising counterpart for the majority of follow-up telescopes, for which the greatest sensitivity at optical wavelengths (a fact that the discovery of AT2017gfo made obvious).  It is thus essential that follow-up begin within hours to one day following the GW trigger.

\noindent{\bf SMNS/Stable Remnant.}  Although potentially rare, the most promising cases are likely the mergers of low-mass NS-NS binaries which generate long-lived SMNS or stable NS remnants ($t_{\rm collapse} \gg 300$ ms).  Even ignoring the possibility of additional energy input from magnetar spin-down, the quantity of blue disk wind ejecta in this case is substantially enhanced ($t_{\rm collapse} = \infty$ in Fig.~\ref{fig:Lippuner}) and could approach $\sim 0.1M_{\odot}$, boosting the peak luminosity of the blue kilonova by a magnitude or more from what was observed in GW170817.  Allowing also for energy input from the magnetar rotational energy (in addition to radioactivity), the transient at 100 Mpc could reach $UVI \approx 17-19$ (Figs.~\ref{fig:magnetar15}, \ref{fig:magnetar16}).  However, the latter values are highly uncertain as they depend on several unknown factors: the dipole magnetic field of the magnetar remnant, the thermalization efficiency of the magnetar nebula by the ejecta (Eq.~\ref{eq:qdotsd}), and the precise NS collapse time (which in turn depends on the binary mass and the magnetic field strength).  Shallower follow-up observations, such as those used in the discovery and follow-up of GW170817, are thus still relevant to kilonova follow-up even for more distant events (they could also be sufficient to detect the on-axis GRB afterglow in the rare case of a face-on merger).

\noindent{\bf BH-NS Mergers.} BH-NS mergers are ``all or nothing" events.  If the NS is swallowed whole prior to being tidally disrupted, then little or no kilonova emission is expected.  However, in the potentially rare cases when the BH is low in mass and rapidly spinning (in the prograde orbital sense), then the NS is tidally disrupted well outside of the horizon and the quantity of dynamical ejecta can be larger than in NS-NS mergers, by a typical factor of $\sim 10$ (Sect.~\ref{sec:ejecta}).  All else being equal, this results in the kilonova peaking one magnitude brighter in BH-NS mergers.  Likewise, the mass fall-back rate in BH-NS mergers can be up to $\sim 10$ times higher than in NS-NS mergers \citep{Rosswog07}, enhancing potential accretion-powered contributions to the kilonova emission (Fig.~\ref{fig:fallback}, bottom panel).  However the amount of high-$Y_e$ ejecta is potentially less than in NS-NS mergers due to the lack in BH-NS mergers of shock-heated ejecta or a magnetar remnant, and for the same reason no neutron precursor is anticipated, unless it can be somehow generated by the GRB jet.  The accretion disk outflows could still produce a small quantity of blue ejecta, but its velocity is likely to be sufficiently low $\sim 0.1$ c that it will be blocked by the (faster, more massive) tidal tail, at least for equatorial viewing angles.  Taken together, the kilonova emission from BH-NS mergers is more likely to dominated by the red component, although moderate amounts of high-$Y_e$ matter and blue emission could still be produced by the disk winds \citep{Just+15,Fernandez+15}.  Unfortunately for purposes of follow-up, any benefits of the higher dynamical ejecta mass on the light curve luminosity may be more than offset by the larger expected source distance, which will typically be $\approx 2\mbox{\,--\,}3$ times greater than the 200 Mpc horizon characteristic of NS-NS mergers for an otherwise equal GW event detection rates.  See, e.g.~ \citet{Bhattacharya+19} for further discussion of the diverse EM counterparts of BH-NS mergers.

\noindent{\bf Search Strategies.}  Several works have explored the optimal EM follow-up strategies of GW sources, or ways to achieve lower latency GW triggers \citep{Metzger&Berger12,Cowperthwaite&Berger15,Gehrels+16,Ghosh+16,Howell+16,Rana+16}.  Extremely low latency \citep{Cannon+12,Chen&Holz15}, though crucial to searching for a potential low-frequency radio burst \citep{Kaplan+16}, is generally not essential for kilonova follow-up.  One possible exception is the speculative neutron precursor (Sect.~\ref{sec:neutrons}), which peaks hours after the merger.  However, in this case, the greatest advantage is arguably to instead locate the follow-up telescope in North America, producing a better chance of the source being directly overhead of the LIGO detectors where their sensitivity is greatest \citep{Kasliwal&Nissanke13}.  

In the future it would also aid EM follow-up efforts if LIGO were to provide more information on the properties of its binaries to the wider astronomy community at the time of the GW trigger \citep{Margalit&Metzger19}.  The predicted signal from a NS-NS is expected to depend on the binary inclination and the total binary mass, $M_{\rm tot}$.  The inclination cannot be measured with high precision because it is largely degenerate with the (initially unknown) source distance, though it could be determined once the host galaxy was identified.  However, the chirp mass and total binary mass can be reasonably accurately determined in low latency (e.g.~\citealt{Biscoveanu+19}).  Once the mapping between EM counterparts and the binary mass is better established, providing the binary mass in low latency could provide crucial information for informing search strategies, or even prioritizing, EM follow-up.  This is especially important given the cost/scarcity of follow-up resources capable of performing these challenging deep searches over large sky areas.
 
The generally greater sensitivity of telescopes at optical wavelengths, as compared to the infrared, motivates a general strategy by which candidate targets are first identified by wide-field optical telescopes on a timescale of days, and then followed-up with spectroscopy or photometry in the NIR over a longer timescale of $\sim 1$ week.  \cite{Cowperthwaite&Berger15} show that no other known or predicted astrophysical transients are as red and evolve as quickly as kilonovae, thus reducing the number of optical false positives to a manageable level.  Follow-up observations of candidates at wavelengths of a few microns could be accomplished, for instance, by the James Webb Space Telescope \citep{Bartos+16}, WFIRST \citep{Gehrels&Spergel15}, or a dedicated GW follow-up telescope with better target-of-opportunity capabilities.  

Another goal for future kilonova observations would be a spectroscopic measurement of absorption lines from individual $r$-process elements.  Individual lines are challenging to identify for the simple reason that most of their specific wavelengths cannot be predicted theoretically with sufficient precision and have not been measured experimentally.  Furthermore, at early times, the absorption lines are Doppler-broadened near peak light due to the substantial velocities $v \gtrsim 0.1$ c of the ejecta.  Broad absorption features were seen in AT2017gfo (e.g.~\citealt{Chornock+17}), but these likely represented a blend of multiple lines.  Fortunately, line-widths become narrower post-maximum as the photosphere recedes to lower velocity coordinates through the ejecta and nebular lines appear (ejecta velocities as low as $\sim 0.03$c are predicted for the disk wind ejecta).  Unfortunately, emission becomes significantly dimmer at these late times and line blending could remain an issue.  Spectroscopic IR observations of such dim targets is a compelling science case for future 30-meter telescopes.  For instance, the planned Infrared Imaging Spectrograph (IRIS) on the Thirty Meter Telescope \citep{Skidmore+15} will obtain a signal to noise ratio of 10 per wavelength channel (spectral resolution $R = 4000$) for a $K = 25$ mag point source. 

\subsection{Summary of Predictions}
\label{sec:predictions}

We conclude with a summary of predictions for a future large samples of kilonovae.  Once Advanced LIGO/Virgo reach design sensitivity within a few years, NS-NS mergers could be detected as frequently as once per week to once per month.  This rate will increase by roughly another factor of $\sim 8$ with the planned LIGO A+ upgrades in the mid 2020s \citep{Reitze+19}, and yet further with proposed third generation GW detectors, such as Einstein Telescope \citep{Punturo+10} and Cosmic Explorer \citep{Reitze+19b}, which could come online in the 2030s.  

As mentioned earlier in this review, high-fidelity first-principles kilonova models are not currently available.  Reasons for this include uncertainties in: (1) the predicted ejecta properties (mass, velocity, $Y_e$) due to numerical limitations combined with present ignorance about the NS equation of state; (2) the properties of unstable neutron-rich nuclei, which determine the details of the $r$-process and the radioactive heating rate; (3) radiative transfer in face of the complex atomic structure of heavy $r$-process elements and the potential break-down of the approximations which are more safely applied to modeling supernova with less exotic ejecta composition.   
  
Nevertheless, we can still highlight a few {\it trends} which a future sample of joint GW/EM events should bear out if our understanding of these events is even qualitatively correct.  These predictions include:
\begin{itemize} 
\item{For a total binary mass ($\sim$ chirp mass) significantly higher than GW170817, the remnants of NS-NS  merger will undergo a prompt collapse to a BH, {\it resulting in a kilonova which is dimmer and redder than AT2017gfo}.  The blue component of the ejecta, if present at all, will arise from the relatively low-velocity disk wind and is likely be blocked for equatorial viewers by the lanthanide-rich tidal tail ejecta.  A GRB jet can still be produced because a low-mass accretion disk can still form, but it could be less energetic than the off-axis jet inferred for GW170817.  Such events could be relatively rare if the prompt collapse threshold mass is high. }
\item{For binary masses below the (uncertain) prompt collapse threshold mass, the merger will form a HMNS remnant, perhaps similar to that believed to be generated in GW170817.  {\it The strength of the blue kilonova relative to the red kilonova emission will increase with decreasing total binary mass}, as the ejecta is dominated by the disk wind and the average $Y_e$ rises with the collapse time (Fig.~\ref{fig:Lippuner}).  BH formation is likely to still be relatively prompt, providing no hindrance to the production of an ultra-relativistic GRB jet and afterglow emission.}
\item{Below another uncertain critical binary mass threshold, the merger remnant will survive for minutes or longer as a quasi-stable SMNS or indefinitely stable NS remnant.  The disk wind ejecta will be almost entirely blue but a weak red component could still be present from the tidal tail ejecta (even though such an emission component was likely swamped in GW170817). {\it The mean velocity of the ejecta will increase with decreasing remnant mass, as the amount of rotational energy extracted from the remnant prior to BH formation increases} (Fig.~\ref{fig:MM}).  Less certain is whether a powerful ultra-relativistic GRB jet will form in this case, due to baryon pollution from the wind of the remnant (though an exception to this expectation would be instructive).  However, the afterglow produced by the interaction of the jet/kilonova ejecta with the ISM (particularly the radio emission, which is roughly isotropic and generated by even trans-relativistic ejecta) will be even more luminous than in the HMNS case due to the addition of magnetar rotational energy \citep{Metzger&Bower14, Horesh+16,Fong+16}.}
\item{For BH-NS mergers, in the (possibly rare) subset of low-mass/high-spin BHs which disrupt the NS well outside the horizon, {\it the red kilonova will be more luminous, and extend to higher velocities, than in the NS-NS case} due to the greater quantity of tidal tail ejecta.  As in the prompt collapse of a NS-NS, the blue component$-$if present$-$will arise from the relatively low-velocity disk wind and thus could be blocked for equatorial viewers by the lanthanide-rich tidal tail ejecta. }

\end{itemize}

\section{Final Thoughts}
\label{sec:conclusions}

The first edition of this review was written in the year prior to the discovery of GW170817.  In the final section of that article I pondered:  ``Given the rapid evolution of this field in recent years, it is natural to question the robustness of current kilonova models.  What would it mean if kilonova emission is ruled out following a NS-NS merger, even to stringent limits?"  I concluded this was untenable: ``First, it should be recognized that---unlike, for instance, a GRB afterglow---kilonovae are largely thermal phenomena.  The ejection of neutron-rich matter during a NS-NS merger at about ten percent of the speed of light appears to be a robust consequence of the hydrodynamics of such events, which all modern simulations agree upon.  Likewise, the fact that decompressing nuclear-density matter will synthesize heavy neutron rich isotopes is also robust.  The properties of individual nuclei well off of the stable valley are not well understood, although that will improve soon due to measurements with the new Facility for Rare Isotope Beams (e.g.~\citealt{Horowitz+18}).  However, the combined radioactive heating rate from a large ensemble of decaying nuclei is largely statistical in nature and hence is also relatively robust, even if individual isotopes are not; furthermore, most of the isotopes which contribute to the heating on the timescale of days to weeks most relevant to the kilonova peak are stable enough that their masses and half-lives are experimentally measured."  

Despite the success that astrophysics theorists and numerical relativists had in anticipating many of the properties of GW170817, the tables may soon be turned as we struggle to catch up to the rich phenomenology which is likely to explode from the new GW/EM science.  In particular, much additional work is required on the theory side to transform kilonovae into quantitative probes of the diversity of merger outcomes and nuclear physics.  Among the largest remaining uncertainty in kilonova emission relates to the wavelength-dependent opacity of the ejecta, in particular when it includes lanthanide/actinides isotopes with partially-filled f-shell valence shells \citep{Kasen+13,Tanaka&Hotokezaka13,Fontes+16}.  As discussed in Sect.~\ref{sec:opacity}, the wavelengths and strengths of the enormous number of lines of these elements and ionization states are not experimentally measured and are impossible to calculate from first principles from multi-body quantum mechanics with current computational capabilities.  Furthermore, how to handle radiative transport in cases when the density of strong lines becomes so large that the usual expansion opacity formalism breaks down requires further consideration and simulation work.  Another theoretical issue which deserves prompt attention with additional theory work is the robustness of the presence of free neutrons in the outermost layers of the ejecta, given their potentially large impact on the very early-time kilonova optical emission (Sect.~\ref{sec:neutrons}).  The issue of large-scale magnetic field generation, and its impact on the GRB jet and post-merger outflows (e.g. from the remnant NS or accretion disk), is likely to remain a challenging issue for man years to come, one which is hardly unique to this field of astrophysics.  Here, I suspect that nature will teach us more than we can deduce ourselves.   

With ongoing dedicated effort, as more detections or constraints on kilonovae become possible over the next few years, we will be in an excellent position to use these observations to probe the physics of binary NS mergers, their remnants, and their role as an origin of the $r$-process. 

\begin{acknowledgements}

I want to thank Gabriel Mart{\'{\i}}nez-Pinedo and Almudena Arcones, my colleagues who developed the nuclear reaction network and assembled the microphysics needed to calculate the late-time radioactive heating, and who were enthusiastic back in 2009 about reviving the relic idea of \cite{Burbidge+56} of an ``$r$-process-powered supernova".  I am also indebted to Rodrigo Fernandez, Tony Piro, Eliot Quataert, and Daniel Siegel, with whom I worked over many years to make the case to a skeptical community that accretion disk outflows, rather than dynamical material, was the dominant source of ejecta in NS mergers (a prediction which most believe was borne out by the large ejecta masses GW170817).  I want to thank Todd Thompson, who taught me much of what I know about the $r$-process and the possibly important role of magnetar winds.  I also want to thank Edo Berger, with who I worked out the practical aspects of what it would take to detect and characterize kilonovae.

I also want to acknowledge my many other collaborators on binary neutron star mergers, who helped shape many of the ideas expressed in this article.  These include, but are not limited to, Jennifer Barnes, Andrei Beloborodov, Josh Bloom, Geoff Bower, Niccolo Bucciantini, Phil Cowperthwaite, Alessandro Drago, Wen-Fai Fong, Daniel Kasen, Ben Margalit, Raffaella Margutti, Daniel Perley, Eliot Quataert, Antonia Rowlinson, and Meng-Ru Wu.  I gratefully acknowledge support from NASA (grant number NNX16AB30G) and from the Simons Foundation (grant number 606260).
\end{acknowledgements}


\bibliographystyle{spbasic}      

\end{document}